\pdfoutput = 1

\documentclass[12pt,a4paper]{article}
\usepackage{multirow}
\usepackage[dvips]{graphicx}
\usepackage{latexsym,amsmath,amsfonts,amscd}
\usepackage{epsfig}
\usepackage{indentfirst}
\usepackage{verbatim}
\usepackage{color}
\usepackage{colordvi}
\usepackage{appendix}
\usepackage{tikz}
\usepackage{epstopdf}

\begin{document}

\baselineskip=2pc

\begin{center}

{\Large \bf A  hybrid WENO method with modified ghost fluid method for compressible two-medium flow problems\footnote{The research is partly supported by Science Challenge Project, No. TZ2016002  and NSAF grant U1630247.
}}
\end{center}

\centerline{Zhuang Zhao\footnote{School of Mathematical Sciences, Xiamen University,
Xiamen, Fujian 361005, P.R. China. E-mail: zzhao@stu.xmu.edu.cn.}, Yibing Chen\footnote{Institute of Applied Physics and Computational Mathematics, Beijing 100094, China. E-mail: chen\_yibing@iapcm.ac.cn.} and Jianxian
Qiu\footnote{School of Mathematical Sciences and Fujian Provincial
Key Laboratory of Mathematical Modeling and High-Performance
Scientific Computing, Xiamen University,
Xiamen, Fujian 361005, P.R. China. E-mail: jxqiu@xmu.edu.cn.}
}

\vspace{.1in}

\baselineskip=1.8pc

\centerline{\bf Abstract}

\bigskip

In this paper, we develop a simplified hybrid  weighted essentially non-oscillatory (WENO) method  combined with the modified ghost fluid method (MGFM) \cite{LKY1} to simulate the compressible two-medium flow problems. The MGFM can turn the two-medium flow problems into two single-medium cases by defining the ghost fluids status in terms of the predicted the interface status, which makes the material interface ``invisible". For the single medium flow case, we adapt between the linear upwind scheme and the WENO scheme automatically by identifying the regions of the extreme points for the reconstruction polynomial as same as the hybrid WENO scheme \cite{mzq}. Instead of calculating their exact locations, we only need to know the regions of the extreme points  based on the zero point existence theorem, which is simpler for implementation and saves computation time. Meanwhile, it still keeps the robustness and has high efficiency.  Extensive numerical results for both one and two dimensional two-medium flow problems are performed to demonstrate the good performances of the proposed method.

\vfill {\bf Key Words:} hybrid WENO scheme,
two-medium flow problems, modified ghost fluid method, zero point existence theorem

{\bf AMS(MOS) subject classification:} 65M60, 35L65

\pagenumbering{arabic}

\newpage

\baselineskip=2pc

\section{Introduction}
\label{sec1}
\setcounter{equation}{0}
\setcounter{figure}{0}
\setcounter{table}{0}

In this paper, we propose a simplified hybrid weighted essentially non-oscillatory (WENO) method with modified ghost fluid method (MGFM) \cite{LKY1} for simulating compressible two-medium flow problems. For two-medium flow problems, the equation of state (EOS) would switch between the different medium, which may cause numerical oscillations or inaccuracies near the material interface. Hence, many researchers have used various additional works and techniques to overcome this difficulty, and there are two major options to simulate the compressible two-medium flow problems.

One is the front capturing method, where the high resolution methods are applied  to suppress the non-physical oscillations near discontinuities by bringing the numerical diffusion or viscosity, which inherently exists in the method itself or is given artificially. The front capturing method can deal with large deformation problems and relatively easy to extend to high dimension. However, the numerical inaccuracies and oscillations are inevitable near the interface, therefore, various techniques were introduced by Larrouturou \cite{mix1}, Karni \cite{mix2}, Abgrall et al. \cite{mix3,mix5}, Shyue et al. \cite{mix4}, Saurel et al. \cite{mix6} and  Chen et al. \cite{mix7} to resolve this difficulty. The other one is the front tracking method, which terms the discontinuities between the two-medium flows as internal moving interfaces. It works well at multi-material interfaces, but it would have difficulties  about the entanglement of the Lagrangian meshes and the extension to high dimension, and there are some typical methods about the front tracking method, such as volume of fluid (VOF) method \cite{VOF}, level set method \cite{Lset} and other front tracking methods \cite{Fronttracking1,Fronttracking2}.

To combine the best properties of the front capturing and tracking methods, Fedikw et al. \cite{FAMO} proposed a new numerical method for
treating interfaces using a level set function  in Eulerian schemes named as the ghost fluid method (GFM), which makes the interface ``invisible". In the framework of the GFM \cite{FAMO}, the pressure and velocity at the ghost fluid nodes near the interface are defined as the local real pressure and velocity, while the density is obtained by isobaric fixing. It can easily turn the two-medium flow problems into two single-medium flow cases, and for the single-medium flow problems, many classical and mature schemes can be applied. Hence, it provides an alternative and flexible way to simulate the two-medium flow problems, and the extension to high dimension becomes fairly straightforward. However,  it may cause numerical inaccuracies in the case of a strong shock impacting on the interface, and the reason may be that the statuses near the interface are affected by the wave interaction and the material properties on both sides. Therefore, Liu et al. \cite{LKY1} developed a modified ghost fluid method (MGFM), in which a multi-material Riemann problem is defined and solved approximately or exactly to predict the interfacial status, then, the predicted interfacial status is applied to define the fluid values at the ghost points. The MGFM combines the advantages of the GFM \cite{FAMO} and
the implicit characteristic methods \cite{LKY,LKY2}, and it takes the interaction of shock with the interface into consideration. Later, the interface interaction GFM (IGFM) \cite{IGFM}, the real GFM (RGFM) \cite{RGFM} and the practical GFM (PGFM) \cite{PGFM} were developed following the idea of the Riemann problem-based technique in the MGFM \cite{LKY1}. The MGFM is robust and less problem related, and it has been applied in various situations as in \cite{MGFM1,MGFM2,QLH,MGFM3,MGFM4}, and the accuracy analysis and errors estimation can be seen in \cite{MGFM_errors1,MGFM_errors2}. The GFM \cite{FAMO} and its relevant ghost fluid methods \cite{LKY1,IGFM,RGFM,PGFM} are non-conservative near the interface, and the conservative scheme can be seen in \cite{QLH0}.

Here, we would use the MGFM to define the ghost fluid status for the two-medium flow problems considering its great performances, and for the single-medium flow problems, many successful numerical schemes can  be applied for it. Among them, the finite difference or finite volume weighted essentially non-oscillatory (WENO) schemes have been widely applied for the single-medium flow problems which usually contain shock, contact discontinuities and  sophisticated smooth structures simultaneously. And in 1994, the first WENO scheme was constructed by Liu, Osher and Chan \cite{loc} on the basis of ENO schemes \cite{ho,h1,heoc}, where all candidate stencils were used with a nonlinear convex methodology to obtain higher order accuracy in the smooth regions, then, Jiang and Shu \cite{js}  proposed  the third and fifth-order finite difference WENO schemes in multi-space dimension, in which a general definition  for smoothness indicators and nonlinear weights was presented. After this, WENO schemes have been further developed in \cite{hs,lpr,SHSN,zs,ccd,ZQd,ZS2}. However, the cost of computing the nonlinear weights and local characteristic decompositions is still very high. Hence,  Hill and Pullin \cite{Hdp} combined the tuned center-difference schemes with WENO schemes to expect that the nonlinear weights would be achieved automatically in the smooth regions away from shocks, but a switching principle was still significant. Later, Li and Qiu \cite{Glj} studied the hybrid WENO scheme using different switching principles \cite{TJS}, which shows different principles would have different influences for the hybrid WENO scheme \cite{Glj}, and the majorities of the troubled-cell indicators need to adjust parameters for different problems to balance better non-oscillations near discontinuities and less computation cost, simultaneously. Hence, Qiu et al. \cite{mzq,ZZCQ} used a new simple switching principle, which employed  different reconstruction method automatically based on the locations of all extreme points of the big reconstruction polynomial for numerical flux. Then, we develop this methodology in this paper, in which we  only need to know the regions of the  extreme points, rather than calculating their exact locations as in \cite{mzq,ZZCQ}.

In this paper, to keep the robustness of the MGFM \cite{LKY1} and high efficiency of the  hybrid WENO schemes \cite{mzq,ZZCQ}, we first use the methodology introduced in \cite{LKY1} to predict the interfacial status based on a multi-material Riemann problem, then, the predicted interfacial status is applied to define the ghost fluid values, by which it turns a two-medium flow problems into two single-medium cases. For the single-medium problems, we would solve it by the hybrid WENO method, where we reconstruct the numerical flux by upwind linear approximation directly if all extreme points of the big reconstruction polynomial for numerical flux  are located outside of the big stencil, otherwise we use the classical WENO procedure \cite{js}. But we only need to know the regions of the  extreme points in terms of the  zero point existence theorem, instead of calculating their exact locations as in \cite{mzq,ZZCQ}, and it is more easy for implementation and saves computation time. Meanwhile, it still keeps the robustness of the WENO scheme \cite{js} and the MGFM \cite{LKY1} to simulate the two-medium flow problems. In addition, it has higher efficiency with less computation costs than the WENO scheme \cite{js} for employing linear approximation straightforwardly in the smooth regions.

The organization of the paper is as follows: in Section 2, the detailed implementation procedures of  the  finite difference hybrid WENO scheme combined with the MGFM are presented for two-medium flow problems. In Section 3, Extensive numerical results for gas-gas and gas-water interaction problems in one and two dimensions are presented to illustrate
good performances of the proposed scheme. Concluding remarks are given in Section 4.

\section{Numerical Methods}

We first introduce the governing equations for
the compressible two-medium flow problems, then, we give a brief review  about the finite difference hybrid WENO method \cite{mzq} for single-medium flow problems, but we have an improvement about the identification technique for the regions of the extreme points of the big reconstruction polynomial.
Next, we introduce the level set technique to track the moving interface, then, we briefly introduce the modified ghost fluid method (MGFM) \cite{LKY1} to define the status of ghost fluids. Finally,  we give the summary of the implementation procedures.

\subsection{Governing equations}

We consider the hyperbolic conservations laws given as follows
  \begin{equation}
\label{EQ0} \left\{
\begin{array}
{ll}
U_t+ \nabla \cdot F(U)=0, \\
U(x,0)=U_0(x), \\
\end{array}
\right.
\end{equation}
where $U$ is $(\rho,\rho \mu,E)^T$ and $F(U)$ is $(\rho \mu,\rho \mu^{2}+p,\mu(E+p))^T$ for one dimensional Euler equations, while   for two dimensional Euler equations, $U$ is $(\rho,\rho \mu, \rho \nu, E)^T$, and $F(U)$ is $(F_1(U),F_2(U))$ with $F_1(U)=(\rho \mu,\rho \mu^{2}+p, \rho \mu \nu, \mu(E+p))^T$, $F_2(U)=(\rho \nu,\rho \mu \nu, \rho \nu^{2}+p, \nu(E+p))^T$. In order to close the systems, the equations of state (EOS) is still required. The $\gamma$-law for gas is
\begin{equation*}
  \rho e= p/(\gamma -1),
\end{equation*}
and Tait EOS  for the water medium \cite{ccg,FAMO,LKY} is given as
\begin{equation*}
  \rho e= (p+N\bar B)/(N -1),
\end{equation*}
in which $\bar B=B-A$, $N=7.15$, $A=1.0\times 10^5$ Pa, $A=3.31\times 10^8$ Pa, and $\rho_0=1000.0 kg/m^3$.

\subsection{Hybrid WENO scheme for single-medium flow}
\label{sec22}
\setcounter{equation}{0}
\setcounter{figure}{0}
\setcounter{table}{0}

We first consider one dimensional scalar hyperbolic conservation laws
  \begin{equation}
\label{EQ} \left\{
\begin{array}
{ll}
u_t+ f_x(u)=0, \\
u(x,0)=u_0(x). \\
\end{array}
\right.
\end{equation}
We  divide the computing domain by uniform grid points $\{x_i\}$, and $h$ is denoted as
$x_{i+1}-x_{i}$. The cell $I_i$ is defined as $[x_{i-1/2},x_{i+1/2}]$, where $x_{i+1/2}$ is set as $x_{i+1/2}=x_i+h/2$, then, the semi-discrete finite difference scheme of (\ref{EQ}) is written as
\begin{equation}
\label{ode} \frac{du_i(t)}{dt}=- \frac 1 {h} \left ( \hat f_{i+1/2}-\hat f_{i-1/2}\right ),
\end{equation}
in which $u_i(t)$ is represented as $u(x_i,t)$, and the numerical flux $\hat f_{i+1/2}$  is a fifth order approximation of $v_{i+1/2}=v(x_{i+1/2})$, in which $v(x)$ is  defined implicitly as in \cite{js}:
\begin{equation*}
\label{impli} f(u(x))=\frac 1 h \int_{x-h/2}^{x+h/2} v(x) dx,
\end{equation*}
then, the right hand item of (\ref{ode}) is the  fifth order approximation for  $-f_x(u)$ at $x_i$. For the stability of the finite difference scheme, we split the flux $f(u)$ into two parts:  $f(u)=f^+(u)+f^-(u)$, in which $\frac{df^+(u)}{du}\geq 0$ and
$\frac{df^-(u)}{du}\leq 0$, and the Lax-Friedrichs flux splitting method is applied here as
\begin{equation*}
\label{lfsplit0} f^{\pm}(u)=\frac 1 2(f(u)\pm\alpha u),
\end{equation*}
where $\alpha=$ $\max_{u}|f'(u)|$.

Next, we introduce the detailed procedures for the reconstruction of the numerical flux $\hat f^+_{i+1/2}$, which is the fifth order approximation of $f^+(u(x_{i+1/2}))$, and the reconstruction formulas for $\hat f^-_{i+1/2}$ are mirror symmetric with respect to $x_{i+1/2}$ of
that for  $\hat f^+_{i+1/2}$.  $\hat f_{i+1/2}$
is finally taken as $\hat f^+_{i+1/2}+\hat f^-_{i+1/2}$. Now, we first give a big stencil: $S_0=\{I_{i-2},...,I_{i+2}\}$, then we can easily obtain the fourth degree polynomial $p_0(x)$ in terms of the following requirements as
\begin{eqnarray*}
\label{ver1} \frac 1 h \int_{I_j}p_0(x)dx= f^+(u_j),\
j=i-2,...,i+2.
\end{eqnarray*}
For simplicity, $\frac{(x-x_i)}h$  is set as $\xi$, then we have
\begin{equation*}
\begin{split}
p_0(x)=&\frac1{1920}[(-116f^+_{i-1}+9f^+_{i-2}+2134f^+_{i}-116f^+_{i+1}+9f^+_{i+2})-\\
&40(34f^+_{i-1}-5f^+_{i-2}-34f^+_{i+1}+5f^+_{i+2})\xi+120(12f^+_{i-1}-\\
&f^+_{i-2}-22f^+_{i}+12f^+_{i+1}-f^+_{i+2})\xi^2+160(2f^+_{i-1}-f^+_{i-2}-\\
&2f^+_{i+1}+f^+_{i+2})\xi^3-80(4f^+_{i-1}-f^+_{i-2}-6f^+_{i}+4f^+_{i+1}-f^+_{i+2})\xi^4)].
\end{split}
\end{equation*}

To increase the efficiency, we also use the thought of the hybrid WENO schemes \cite{mzq,ZZCQ}, in which the linear upwind approximation or WENO reconstruction is applied automatically based on the locations of the extreme points of the big polynomial $p_0(x)$. More explicitly, if all extreme points are located outside of the big spatial stencil $S_0$, $\hat{f}^{+}_{i+1/2}$ is taken as $p_0(x_{i+\frac 1 2})$ directly, otherwise the  classical WENO procedures \cite{js} would be used to reconstruct it. Unlike the hybrid WENO schemes \cite{mzq,ZZCQ} solving the real zero points of $p_0^\prime(x)$ exactly, we identify the regions of the extreme points of $p_0(x)$ in terms of the zero point existence theorem as that if the endpoint values of $p'_0(x)$ and the extreme values of $p'_0(x)$ have same signs on the big stencil $S_0$, it means there is no single zero points of $p_0^\prime(x)$ on $S_0$, that is to say, there is no extreme points of $p_0(x)$ located in $S_0$. Also, we present their performances in the numerical examples, which shows the new identification skill can catch the regions for  the extreme points of the big polynomial $p_0(x)$ as the old one in \cite{mzq,ZZCQ}, but it has higher efficiency. In addition, the new one has simpler implementation procedure as it only needs to solve the zero points of the quadratic polynomial $p_0^{\prime \prime}(x)$, while the old one has to calculate the roots of the cubic polynomial $p_0^\prime(x)$.

Then, we review the classical WENO procedure \cite{js} for the reconstruction of $\hat{f}^{+}_{i+1/2}$. Firstly, the big stencil $S_0$ is divided
into three smaller stencils: $S_1=\{I_{i-2},I_{i-1},I_{i}\}$, $S_2=\{I_{i-1},I_{i},I_{i+1}\}$ and $S_3=\{I_{i},I_{i+1},I_{i+2}\}$, then, the polynomials $p_l(x)$ are obtained by the following requirements as
\begin{eqnarray*}
\label{ver} \frac 1 h \int_{I_j}p_l(x)dx= f^+(u_j),\
j=i-3+l,...,i-1+l, \ l=1,2,3.
\end{eqnarray*}
The explicit values of $p_l(x)$ at the point $x_{i+1/2}$ can be seen in \cite{js}, and the linear weights are computed by $p_0(x_{i+1/2})=\sum_{l=1}^3\gamma_lp_l(x_{i+1/2})$, in which $\gamma_1=\frac{1}{10}$, $\gamma_1=\frac{3}{5}$ and $\gamma_2=\frac{3}{10}$. To measure how smooth these small polynomials $p_l(x)$ are in the target cell $I_i$, we use the same definition of smoothness indicators $\beta_l$ seen in \cite{js,s3} as
\begin{equation*}
\beta_l=\sum_{\alpha=1}^2\int_{I_i}h^{2\alpha-1}(\frac{d ^\alpha
p_l(x)}{d x^\alpha})^2dx,
\end{equation*}
then, the nonlinear weights are
\begin{equation*}
\label{omega} \omega_l=\frac {\overline
\omega_l}{\sum_{k=0}^{r}\overline \omega_k}, \,\, \,\overline
\omega_l=\frac {\gamma_l}{(\beta_l+\varepsilon)^2},\ l=1,2,3,
\end{equation*}
where $\varepsilon=10^{-6}$, and the explicit values of $\omega_l$ also can be seen in \cite{js}. Finally, the WENO reconstruction of $\hat f^+_{i{+}1/2}$ is
\begin{equation*}
\label{reconeq}\hat{f}^+_{i{+}1/2}=\sum_{l=1}^3\omega_lp_l(x_{i+ 1/2}).
\end{equation*}

After the spatial discretization, the semi-discrete scheme  (\ref{ode})  is discretized by the  third order TVD Runge-Kutta method  \cite{so1} in time as
\begin{eqnarray}
\label{RK}\left \{
\begin{array}{lll}
     u^{(1)} & = & u^n + \Delta t L(u^n),\\
     u^{(2)} & = & \frac 3 4u^n + \frac 1 4u^{(1)}+\frac 1 4\Delta t L(u^{(1)}),\\
     u^{(n+1)} & = &\frac 1 3 u^n +  \frac 2 3u^{(2)} +\frac 2 3\Delta t L(u^{(2)}).
\end{array}
\right.
\end{eqnarray}

{\bf \em Remark 1:} For the systems, such as the one dimensional compressible Euler equations, WENO procedure is performed in the local characteristic directions to overcome the oscillations nearby discontinuities as in \cite{js}, while the linear approximation is directly computed in each component. For two dimensional problems, the spatial reconstruction is performed by dimension by dimension.

\subsection{Level set equation}
\label{sec23}

We choose the next level set technique to track the moving interface. For one dimensional problems, the level set equation  is
\begin{equation}\label{Levelset1}
  \phi_t+u\phi_x=0,
\end{equation}
while for two dimensional case  it is written as
\begin{equation}\label{Levelset2}
  \phi_t+u\phi_x+v\phi_x=0,
\end{equation}
where $\phi$ is a signed distance function. $u$ and $v$ are the velocity of the flow in the $x$ and $y$ directions, respectively. We would solve the equations (\ref{Levelset1}) and (\ref{Levelset2}) by the  fifth order  finite difference hybrid WENO method introduced in Appendix A. However, if the velocity field has a large gradient in the vicinity of the interface, the level set method may cause seriously distorted contours. Therefore, the re-initialization technique is needed to remedy this influence. For one dimensional problems, we can obtain the position of the interface exactly by Newton's iteration method, then we  re-distribute the signed distance function $\phi$. For two dimensional case, the interface is a curve, so we need to use other ways for re-initialization, in which we solve the re-initialization equation as:
\begin{equation}\label{Leve_reinitial}
  \phi_t+ S(\phi_0)(\sqrt{\phi_x^2+\phi_y^2}-1)=0,
\end{equation}
where $S$ is the sign function of $\phi_0$, and the equation (\ref{Leve_reinitial}) is also calculated by the hybrid WENO method shown in Appendix A.

\subsection{Modified Ghost Fluid Method}
\label{sec24}
We use the modified ghost fluid method (MGFM) \cite{LKY1} to define the information of the ghost cells as it considers the interaction of shock with the interface correctly. The main procedures of the MGFM are that we first predict the interface status by solving a two-medium Riemann problem exactly or approximately, then, the predicted interface status is  used to define the ghost fluid status for each fluid, by which it turns a two-medium  flow problem into two single-medium flow problems.
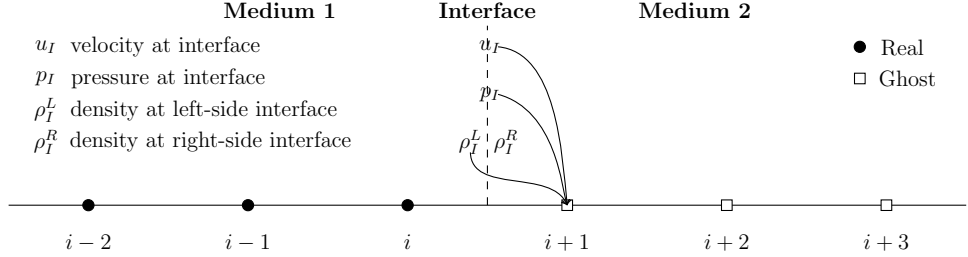
\begin{figure}
\tikzset{global scale/.style={
 scale=#1,every node/.append style={scale=#1}}}
  \centering
 \begin{tikzpicture}[global scale = 0.7]
\draw(-9,0)--(9,0); 
\draw[dashed](0,0)--(0,3.4)node[above]{\bf Interface};
\draw[fill=black](-1.5,0)circle(3pt);
\draw(-1.5,-0.7)node{$i$};
\draw[fill=black](-4.5,0)circle(3pt);
\draw(-4.5,-0.7)node{$i-1$};
\draw[fill=black](-7.5,0)circle(3pt);
\draw(-7.5,-0.7)node{$i-2$};
\draw(1.5,-0.7)node{$i+1$};
\draw(4.5,-0.7)node{$i+2$};
\draw(7.5,-0.7)node{$i+3$};
\draw[fill=white](1.4,-0.1)rectangle(1.6,0.1); 
\draw[fill=white](4.4,-0.1)rectangle(4.6,0.1);
\draw[fill=white](7.4,-0.1)rectangle(7.6,0.1);
\draw(0.05,3.0)node{$u_{I}$};
\draw [->] (0.2,3.0) .. controls (1.3,2.9) and (1.4,1) .. (1.5,0);
\draw(0.05,2.1)node{$p_{I}$};
\draw [->] (0.2,2.1) .. controls (1.3,1.8) and (1.4,0.3) .. (1.5,0);
\draw(-0.32,1.2)node{$\rho_I^L$};
\draw(0.35,1.2)node{$\rho_I^R$};
\draw [->] (-0.32,1.0) .. controls (-0.3,0.2) and (1.2,0.7) .. (1.5,0);
\draw(-3.9,3.7)node{\bf Medium 1};
\draw(3.9,3.7)node{\bf Medium 2};
\draw[fill=black](7,3.0)circle(3pt);
\draw(7.8,3.0)node{Real};
\draw[fill=white](6.9,2.3)rectangle(7.1,2.5); 
\draw(7.9,2.4)node{Ghost};
\draw(-8.3,3.0)node{$u_{I}$};
\draw(-6.05,3.0)node{velocity at interface};
\draw(-8.3,2.4)node{$p_{I}$};
\draw(-6,2.4)node{pressure at interface};
\draw(-8.3,1.8)node{$\rho_I^L$};
\draw(-5.33,1.8)node{density at left-side interface};
\draw(-8.3,1.2)node{$\rho_I^R$};
\draw(-5.2,1.2)node{density at right-side interface};
\end{tikzpicture}
\caption{Isentropic fixing for 1D two-medium flow problems.}
\label{If1}
\end{figure}

For one dimensional case, we only introduce how to define the ghost fluid status for Medium 1 in detail, and the definition of the ghost fluid status for Medium 2 is similar. Let's assume that  the interface is located between $i$ and $i+1$ seen in Figure \ref{If1}, then, we use the status of $U_{i-1}$ and $U_{i+2}$ to define the two-medium Riemann problem suggested in \cite{LKY1}, and
obtain the interfacial status: $u_i$ (velocity), $p_i$ (pressure), $\rho_i^L$ (density at left-side) and $\rho_i^R$ (density at right-side). Later, we take the predicted $u_i$, $p_i$, $\rho_i^L$ as the velocity, pressure and density at the ghost point $i+1$, but at these points $i$, $i+2$ and $i+3$,
the pressure and velocity are those on the  real local fluid, and the density at these points is replaced by the isentropic fixing \cite{FAMO,LKY1}.

\begin{figure}
\tikzset{global scale/.style={
 scale=#1,every node/.append style={scale=#1}}}
  \centering
  \begin{tikzpicture}[global scale = 0.7]
\draw (-4,0) -- (4,0);
\draw (0,-4) -- (0,4);
\draw[step=1.5cm] (-4,-4) grid (4,4);
\draw [dashed] (-3.8,3.8) .. controls (3,0) and (1,-2) .. (-2,-3.8);
\draw[fill=black](0,0)circle(3pt);
\draw(0.2,-0.2)node{$A$};
\draw[fill=black](-1.5,0)circle(3pt);
\draw[fill=black](-1.5,1.5)circle(3pt);
\draw[fill=black](0,-1.5)circle(3pt);
\draw[fill=black](-1.5,-1.5)circle(3pt);
\draw[fill=black](-1.5,-3)circle(3pt);
\draw[fill=black](-3,-3)circle(3pt);
\draw[fill=black](-3,-1.5)circle(3pt);
\draw[fill=black](-3,-0)circle(3pt);
\draw[fill=black](-3,1.5)circle(3pt);
\draw[fill=black](-3,3)circle(3pt);
\draw[fill=white](1.4,1.4)rectangle(1.6,1.6); 
\draw(1.7,1.3)node{$B$};
\draw[fill=white](1.4,1.4)rectangle(1.6,1.6); 
\draw[fill=white](2.9,2.9)rectangle(3.1,3.1); 
\draw[fill=white](1.4,2.9)rectangle(1.6,3.1); 
\draw[fill=white](-0.1,2.9)rectangle(0.1,3.1); 
\draw[fill=white](-1.6,2.9)rectangle(-1.4,3.1); 
\draw[fill=white](-0.1,1.4)rectangle(0.1,1.6); 
\draw[fill=white](2.9,1.4)rectangle(3.1,1.6); 
\draw[fill=white](1.4,-0.1)rectangle(1.6,0.1); 
\draw[fill=white](2.9,-0.1)rectangle(3.1,0.1); 
\draw[fill=white](1.4,-1.6)rectangle(1.6,-1.4); 
\draw[fill=white](2.9,-1.6)rectangle(3.1,-1.4); 
\draw[fill=white](-0.1,-3.1)rectangle(0.1,-2.9); 
\draw[fill=white](1.4,-3.1)rectangle(1.6,-2.9); 
\draw[fill=white](2.9,-3.1)rectangle(3.1,-2.9); 
\draw(-3,0.75)node{\bf Medium 1};
\draw(2.7,2.25)node{\bf Medium 2};
\draw(2.6,-2.4)node{\bf Interface};
\draw [->] (1.4,-2.4)--(0.6,-1.8);
\draw[fill=black](4.4,3.0)circle(3pt);
\draw(5.1,3.0)node{Real};
\draw[fill=white](4.3,2.3)rectangle(4.5,2.5); 
\draw(5.2,2.4)node{Ghost};
\end{tikzpicture}
\caption{Isentropic fixing for 2D two-medium flow problems.}
\label{If2}
\end{figure}
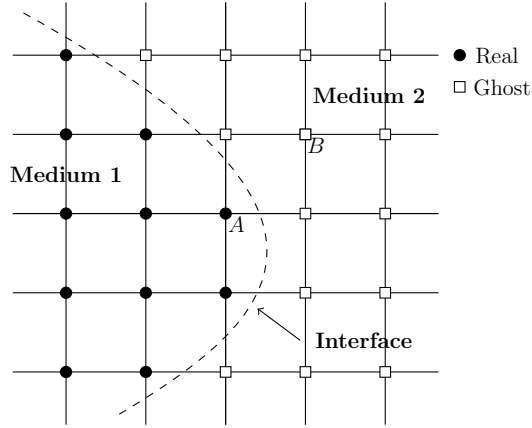

For two dimensional case, it would have one difficulty about the definition of the two-medium
Riemann problem for there is  two velocity components. However, we can know the normal direction $\overrightarrow{n}$ near the interface employing the level set function ($ \overrightarrow{n}=\nabla \phi/|\nabla \phi|$), then, we  obtain the  normal velocity $u_N$ and  tangential velocity $u_T$, in which $u_N$ is defined as $(\mu,\nu)\cdot\overrightarrow{n}$, then, we apply the normal velocity $u_N$, the pressure $p$ and the density $\rho$ to define the two-medium
Riemann problem like one dimensional case. In terms of  the MGFM \cite{LKY1}, we need to define a computation domain for each medium that includes boundary points and grid points in the interfacial regions by $|\phi|<\epsilon$, where $\epsilon$ is set to be  3$\max(\Delta x, \Delta y)$ for the fifth order hybrid WENO scheme. Later, we would only introduce the definition of the status for Medium 1 at the points $A$ and $B$ (seen in Figure \ref{If2}) in detail. To define the status at the point $A$ in Medium 1, we need to find other point next to the interface ($|\phi|<\epsilon$) located in the Medium 2, and let's assume that $B$ is the target point as the angle made by  the normal of  $B$ and $A$ is the minimum, then we define Riemann problem in the normal direction as
\begin{equation*}
  U|_{t=t_n}=\left\{
\begin{array}
{ll}
U_A, \\
U_B, \\
\end{array}
\right.
\end{equation*}
in which  $U_A$ = $(\rho^A,u_N^A,p^A)$ and $U_B$ = $(\rho^B,u_N^B,p^B)$, then we can solve it approximately or exactly to predict the status $u_I$ (velocity), $p_I$ (pressure), $\rho_I^L$ (density at left-side) and $\rho_I^R$ (density at right-side). Notice that node $A$ is located in Medium 1, then, we only need to define the density by isentropic fixing \cite{FAMO,LKY1}, but node $B$ is located in Medium 2, therefore, we need to define the status at node $B$ by  $u_I$, $p_I$ and $\rho_I^L$, and its tangential velocity is still the original one. In addition, the definition of the ghost fluid status for Medium 2 is similar.

\subsection{Summary of the Procedures}

 Now, we give a brief summary of the procedures for simulating two-medium flow problems. Let's assume  that the flow status at $t_n$  has been obtained, then we can advance the respective quantities to $t_{n+1}$ following as:

 \textbf{Step 1.} Calculate the time step $\Delta t$, satisfying the stability condition over the whole range.

\textbf{Step 2.} Solve the level set function $\phi$, and obtain the locations of the interface in the next intermediate time step introduced in Section \ref{sec23}.

\textbf{Step 3.} Define the Riemann problem near the interface and predict the interface status, then use it to define the fluid values at the ghost points for Mediums 1 and 2, respectively, given in Section \ref{sec24}.

 \textbf{Step 4.} Solve the governing equations for Mediums 1 and 2, respectively, advancing the solution to the next intermediate time level, shown in Section \ref{sec22}.

\textbf{Step 5.} Repeat Steps 2, 3 and 4 at each intermediate time step of the third order
TVD Runge-Kutta method, and advance the solution from $U_n$ to $U_{n+1}$, then, re-initialize the level set function $\phi$.

\section{Numerical Results}
\label{sec3}
\setcounter{equation}{0}
\setcounter{figure}{0}
\setcounter{table}{0}

In this section, we perform the numerical results   of the new simplified hybrid WENO scheme and classical WENO scheme \cite{js} with the modified ghost fluid method for two-medium flow problems which are outlined in  the previous section, and the CFL number is set as 0.6. In addition, we also make a comparison between the new identification skill for the regions of the extreme points introduced in the previous section and the old one used in the hybrid WENO schemes \cite{mzq,ZZCQ}. The units for the density, velocity, pressure, length and time are kg$/\text{m}^3$, m/s, Pa, m, and s, respectively.

Here, we use ``New/simplified hybrid WENO method'' to denote the new simplified finite difference hybrid WENO scheme with  the modified ghost fluid method developed in this paper, and use ``Classical WENO method'' to represent the classical finite difference WENO scheme \cite{js} with  the modified ghost fluid method. In addition, we use  ``Old hybrid WENO method'' to denote the  finite difference hybrid WENO scheme \cite{mzq} with  the modified ghost fluid method.

\begin{figure}
 \centerline{\psfig{file=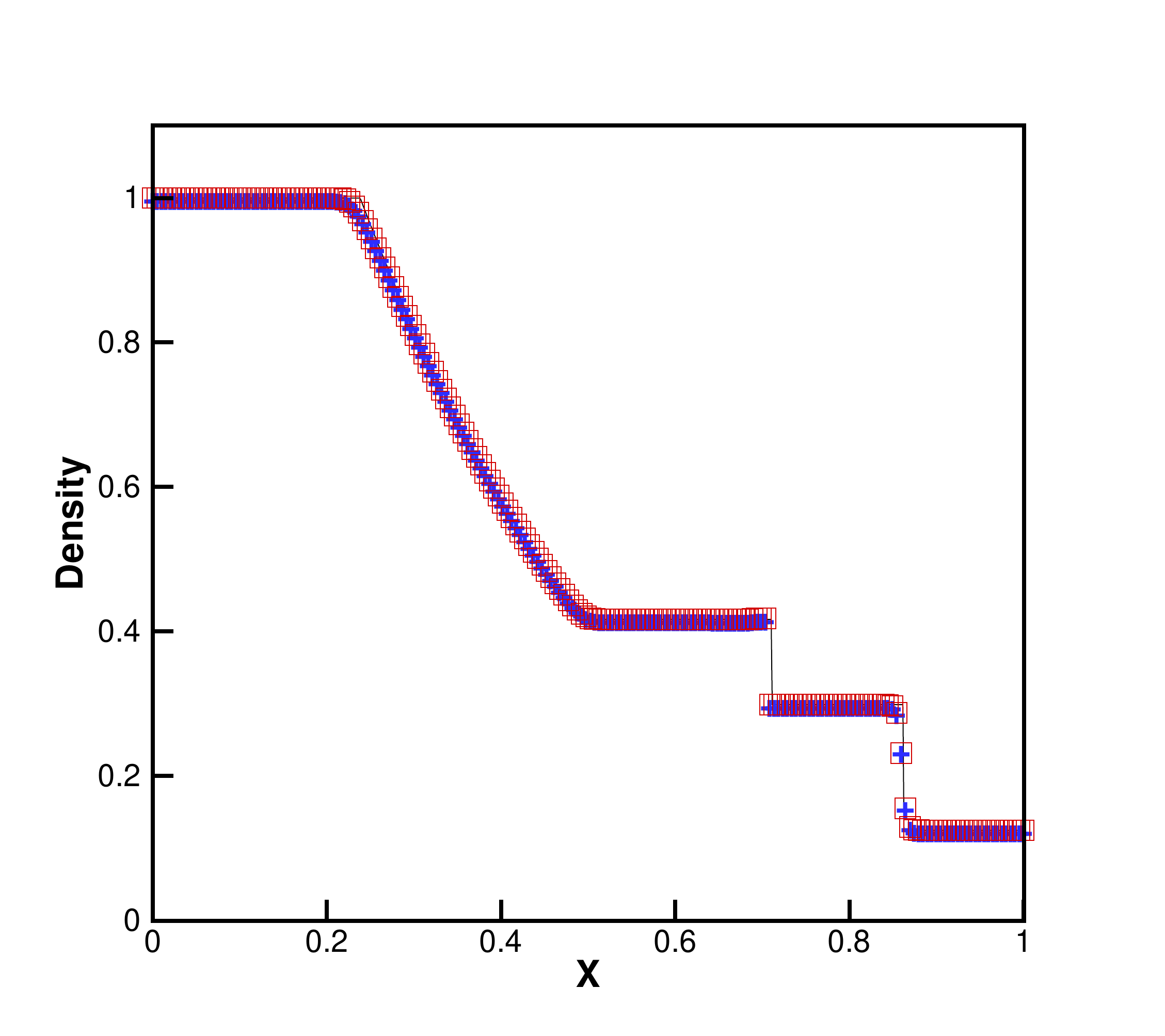,width=2 in}
 \psfig{file=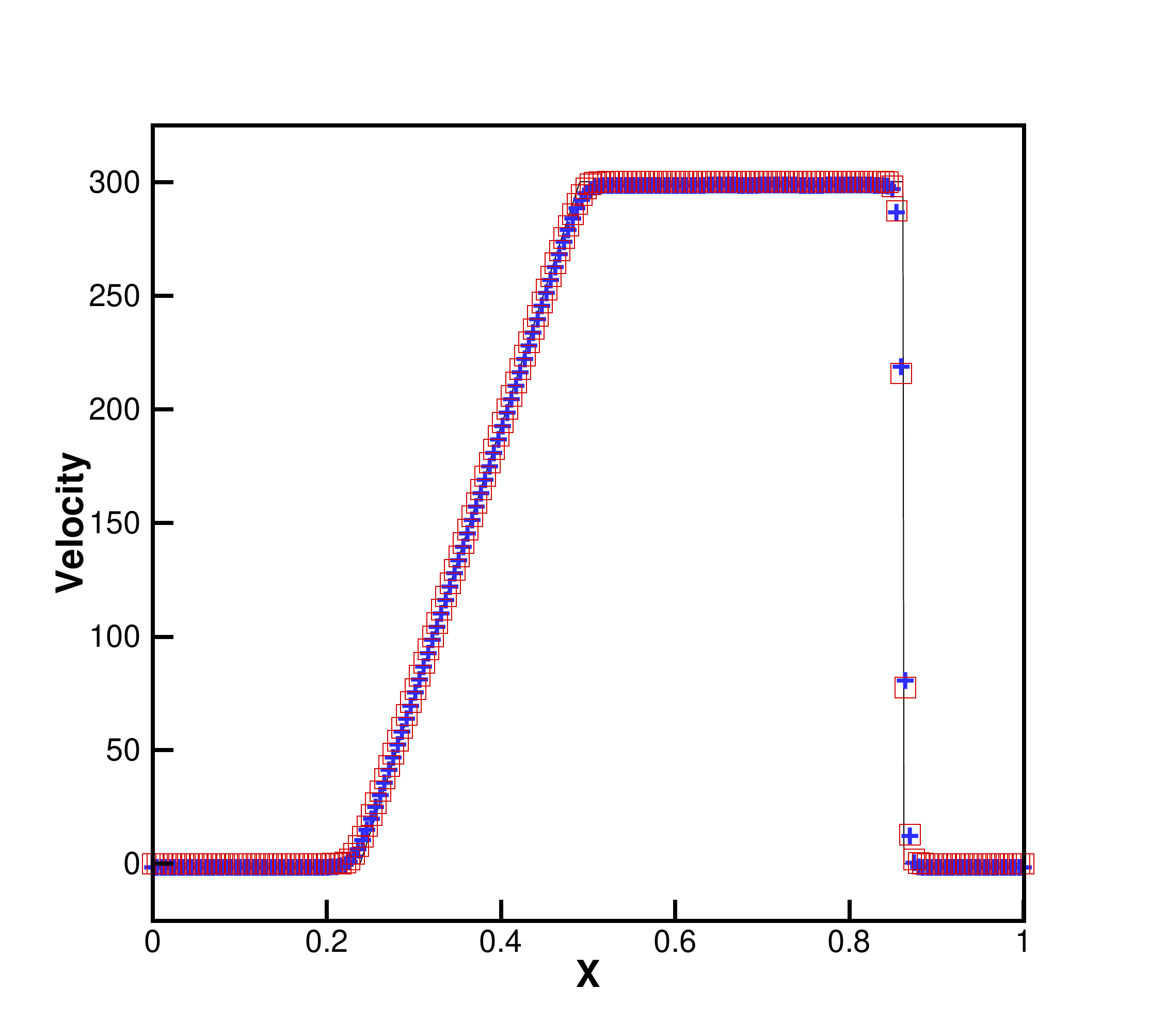,width=2 in}\psfig{file=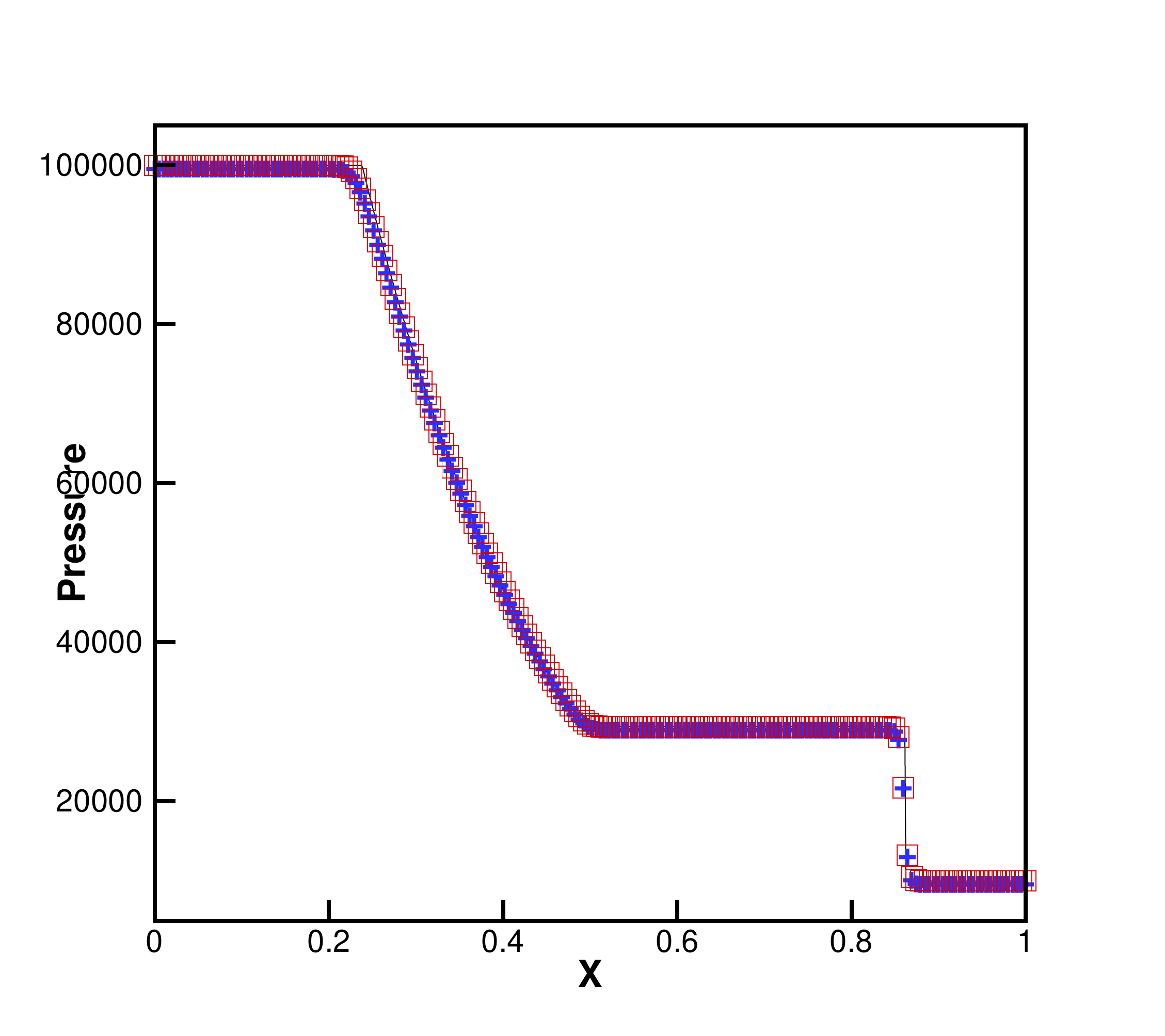,width=2 in}}
\caption{Example 3.1. t=0.0007. From left to right: density; velocity; pressure. Solid line: the exact solution; plus signs:  the results of Classical WENO method; squares:  the results of New/simplified hybrid WENO method. Grid points: 200.}
\label{Ex1}
\end{figure}

\begin{figure}
  \centerline{\psfig{file=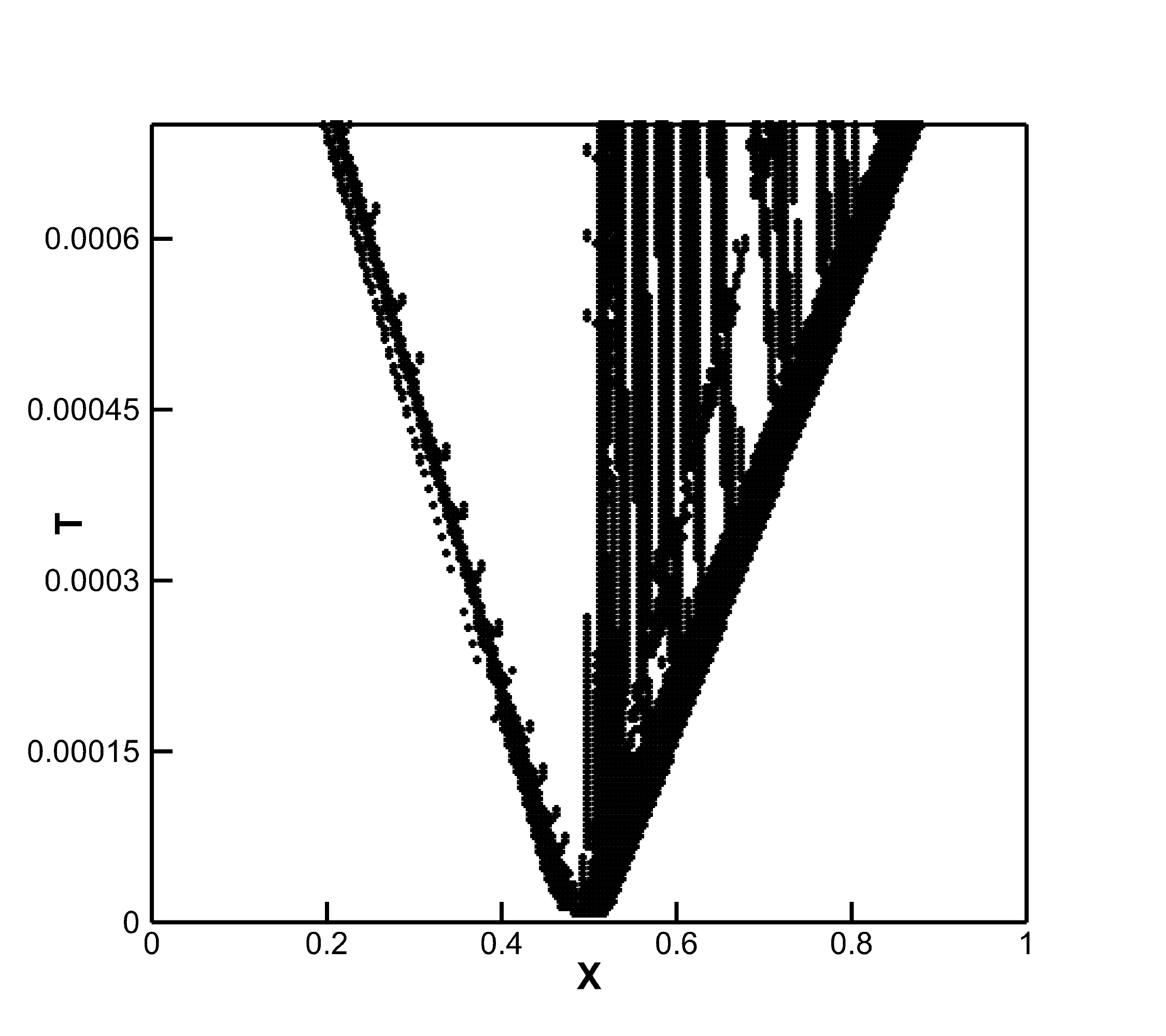,width=2 in}
 \psfig{file= 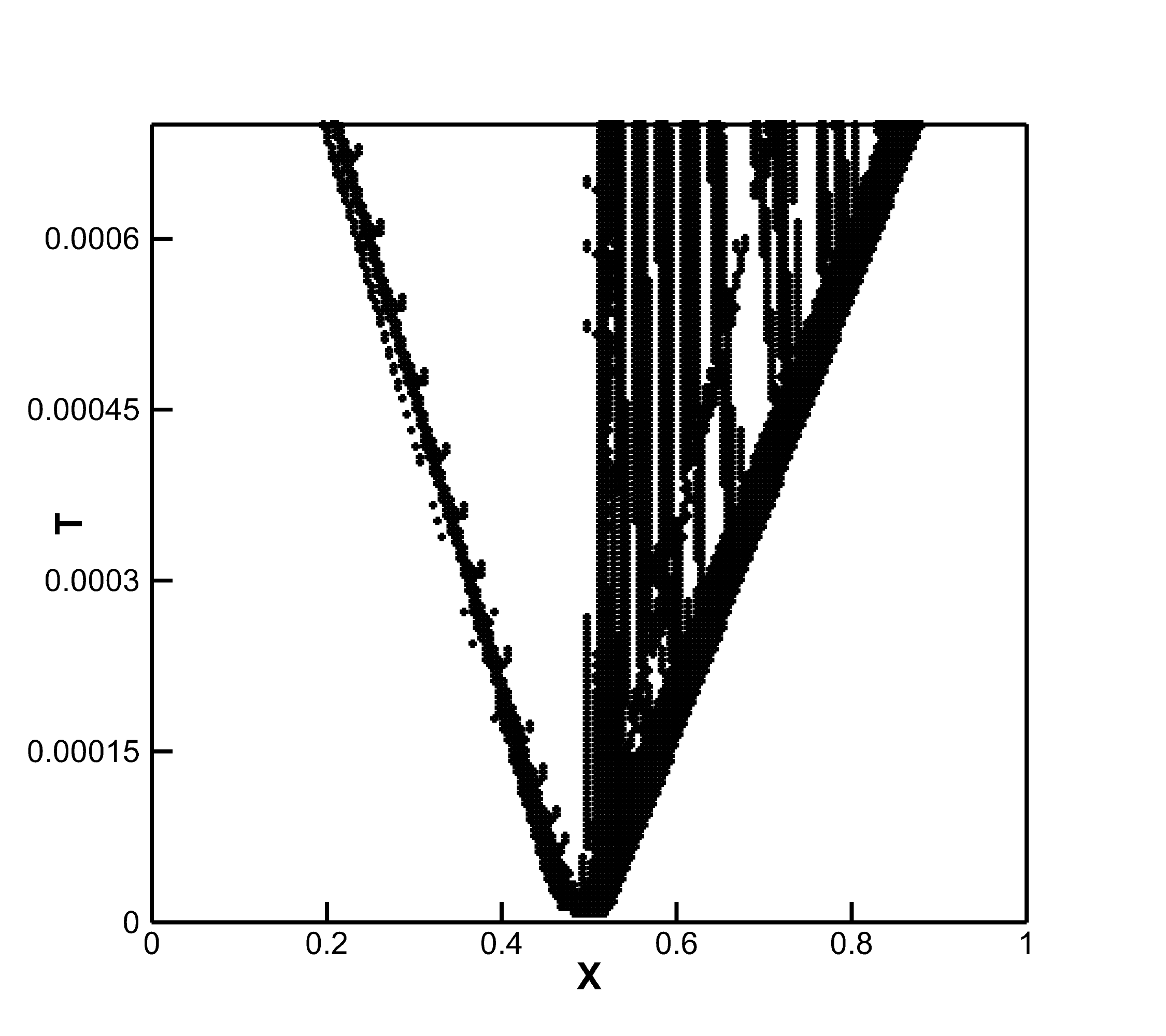,width=2 in}}
  \centerline{\psfig{file=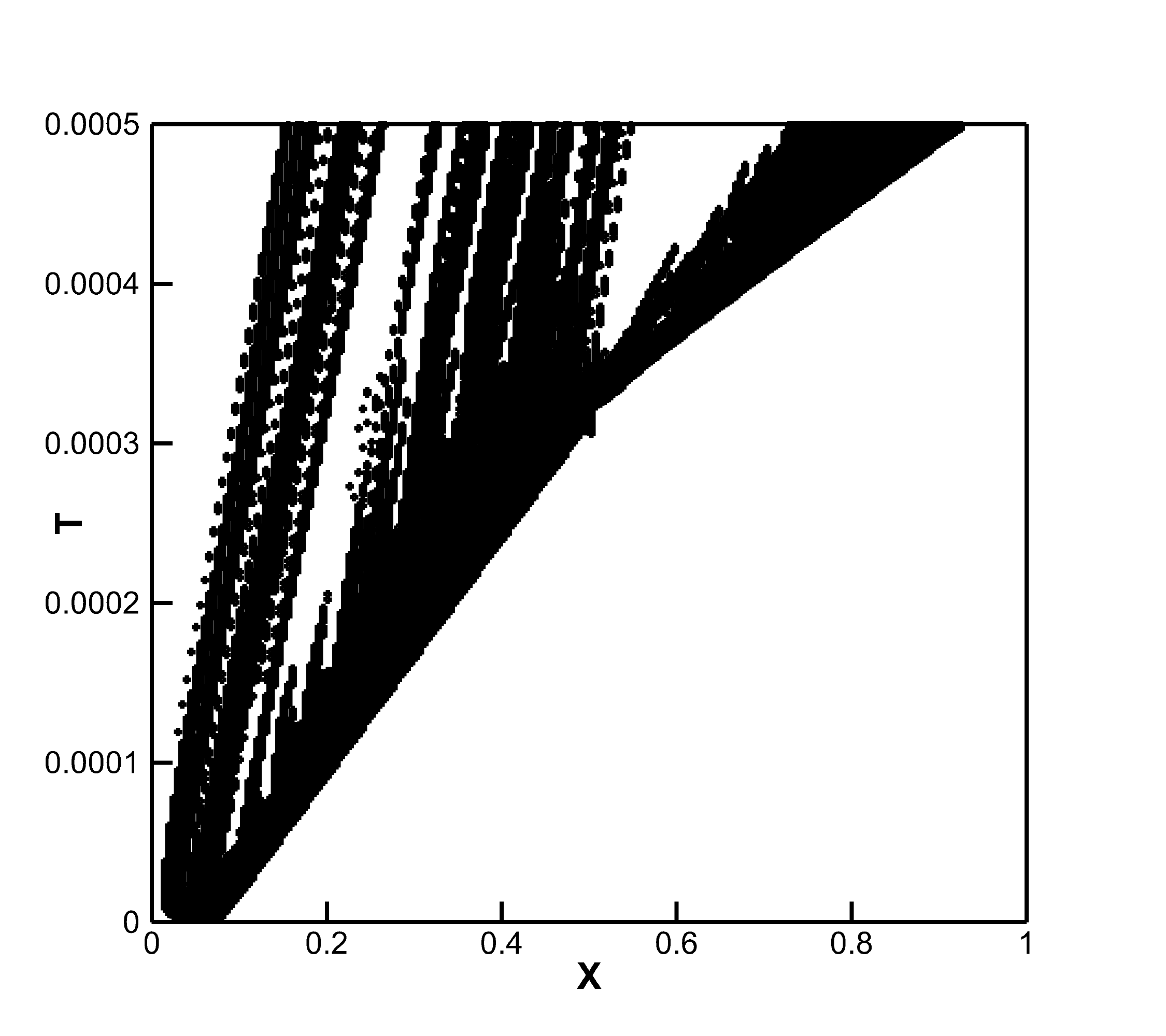,width=2 in}
 \psfig{file= 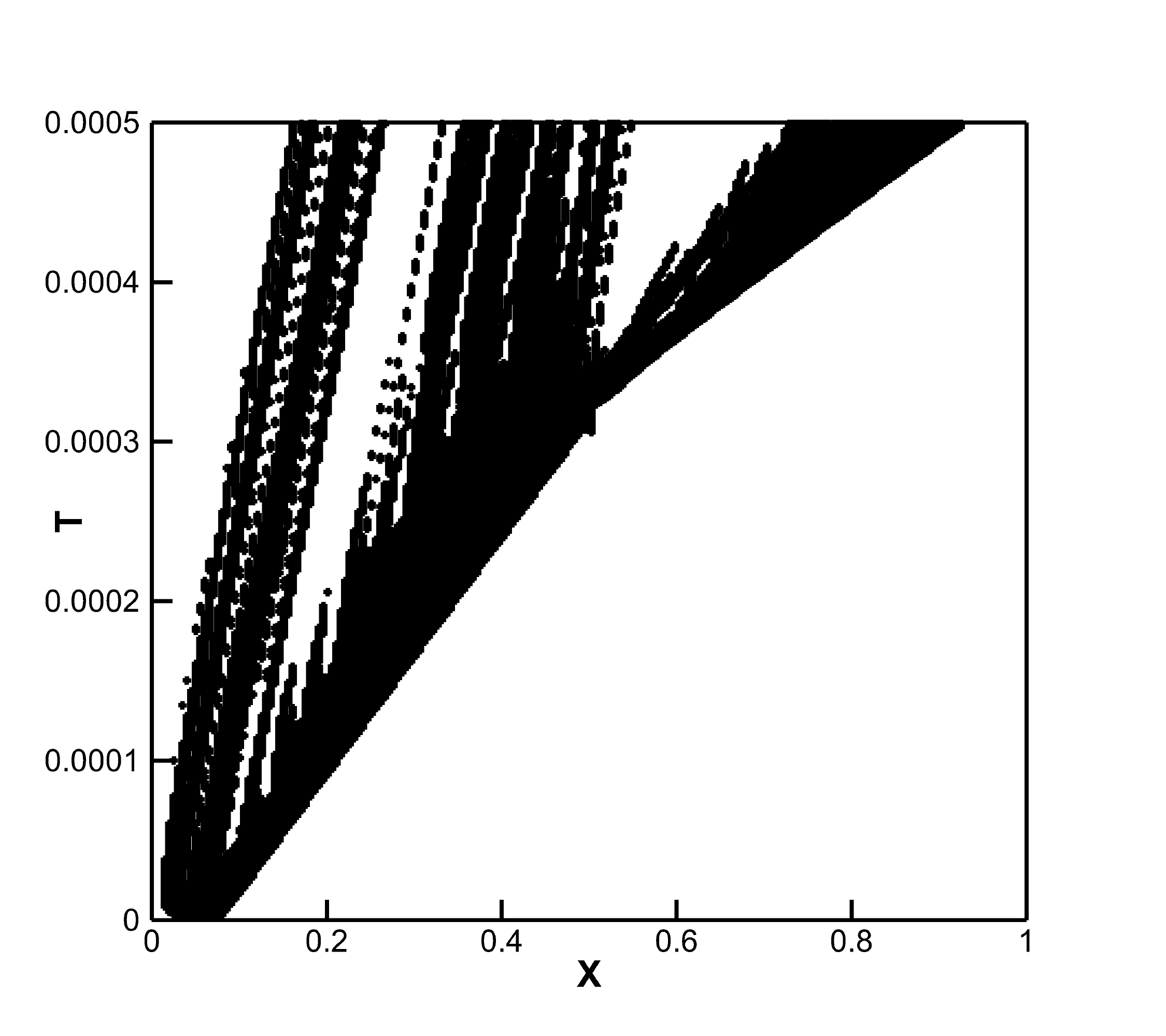,width=2 in}}
  \centerline{\psfig{file=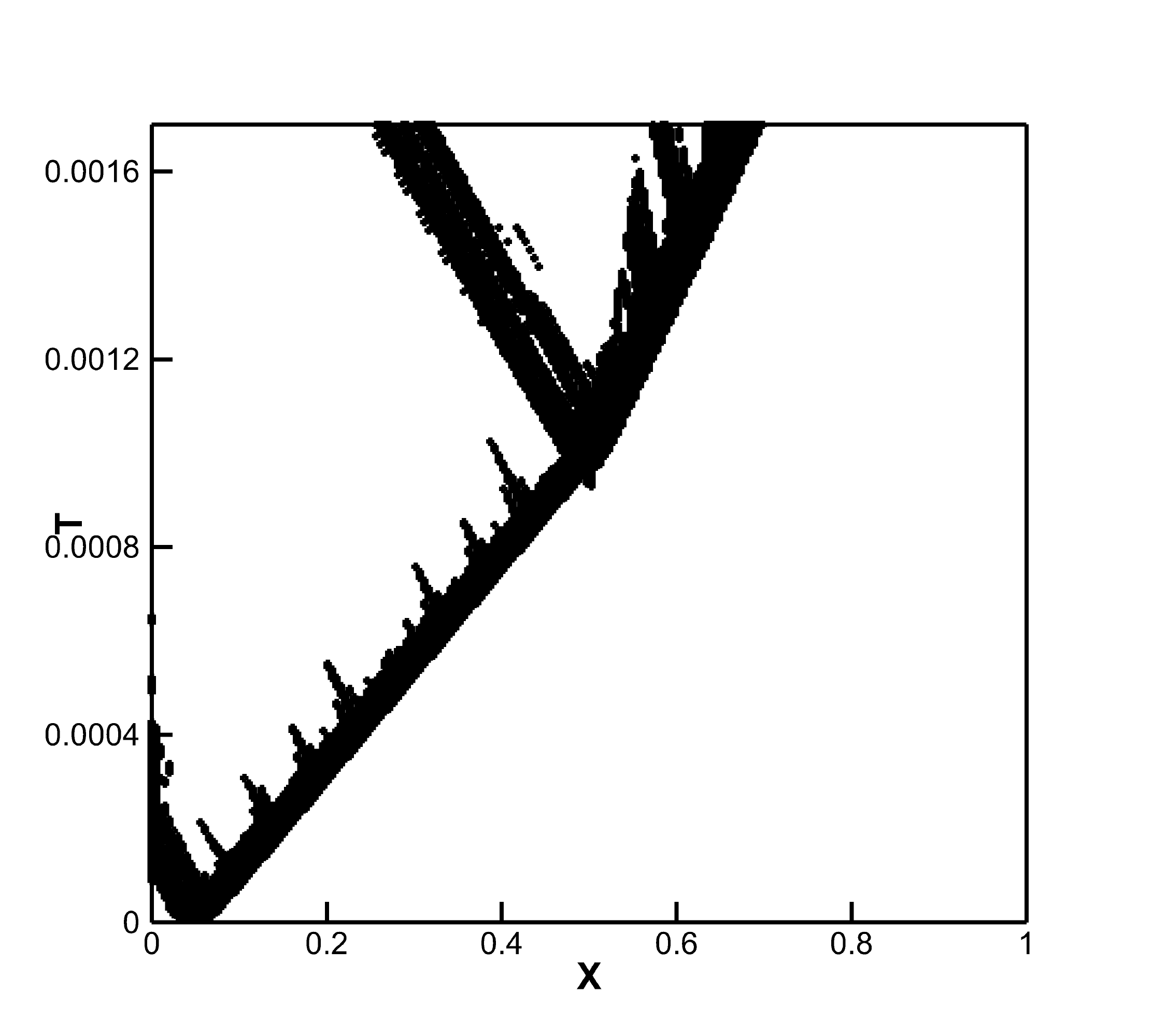,width=2 in}
 \psfig{file= 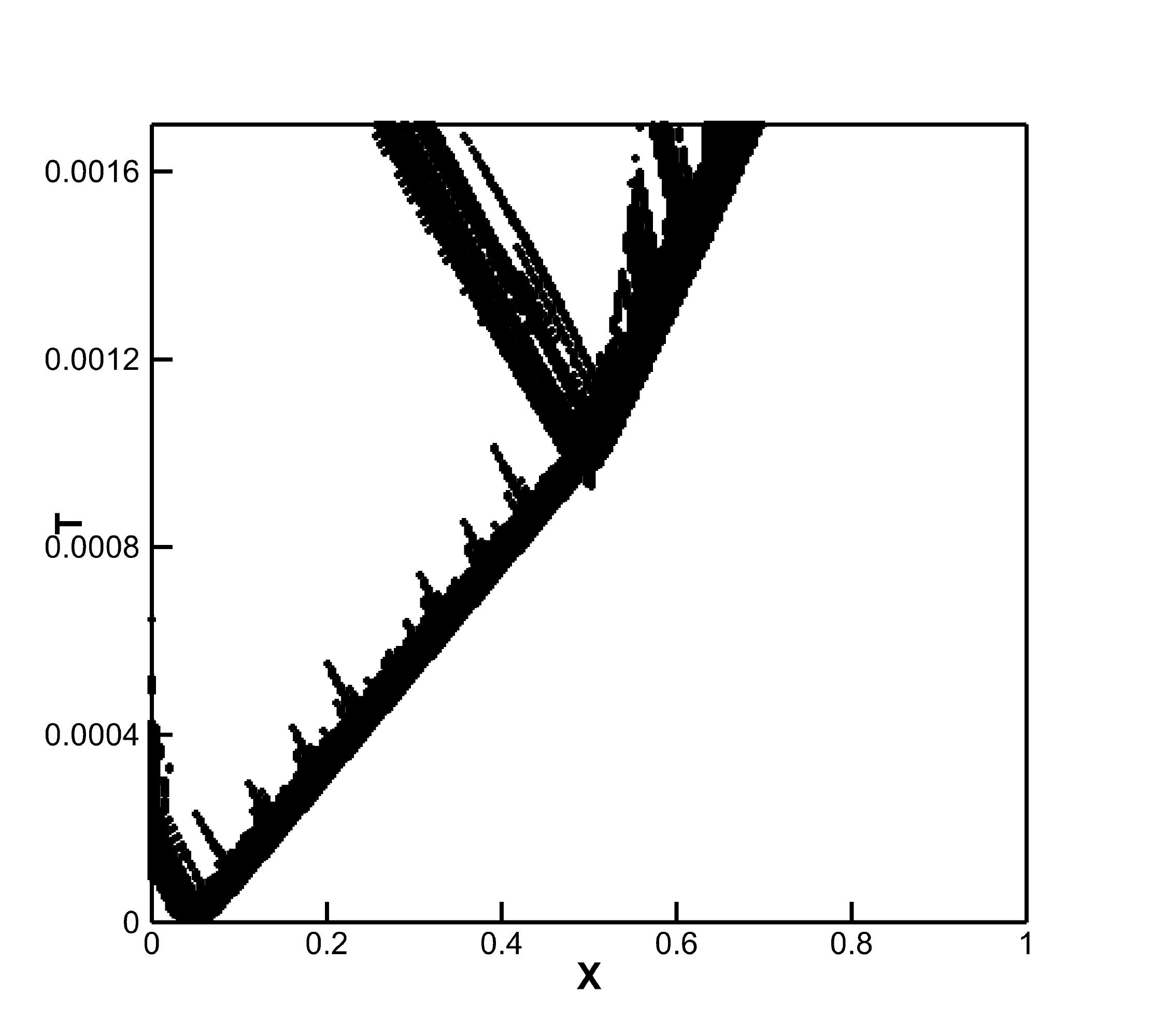,width=2 in}}
\caption{The points where the WENO procedures are performed for Examples 3.1 to 3.3. From left to right: the results of Old hybrid WENO method; the results of New/simplified hybrid WENO method.}
\label{Exlim13}
\end{figure}

\noindent{\bf Example 3.1.} This problem was taken from \cite{FAMO}, and  the initial conditions are
\begin{equation*}
(\rho,\mu,p,\gamma)= \left\{
\begin{array}{ll}
(1, 0, 1\times10^5,1.4),& x \in [0, 0.5),\\
(0.125, 0, 1\times10^4,1.2),& x \in [0.5,1].
\end{array}
\right.
\end{equation*}
In flow and out flow  boundary conditions are applied here, and the final computed time $t$ is 0.0007. We present the computed density $\rho$, velocity $\mu$ and pressure $p$ by New/Simplified hybrid WENO and Classical WENO methods against the exact solution in Figure \ref{Ex1}. We can find that the two methods both capture the location of the material interface correctly. These two schemes also have similar numerical  results, and the overall results  are comparable to analysis, but New/simplified hybrid WENO method can achieve higher efficiency than Classical WENO method for saving 22.93\% computation time. In addition, we find New/simplified hybrid WENO method can save 9.25\% CPU time than Old hybrid WENO method  by calculation, meanwhile, there are only 15.13\% and 15.14\% points where the WENO procedures are computed in New/simplified hybrid WENO and Old hybrid WENO methods, respectively, and the time history of the locations of WENO reconstruction for two methods are given in the top of Figure \ref{Exlim13}. These results show that the new identification skill in New/simplified hybrid WENO method can identify the regions  of the extreme points correctly as the old one in Old hybrid WENO method, but the new one has higher efficiency. The new identification technique  is also simpler as it only needs to solve the zero points of a quadratic polynomial, while the old one has to calculate the roots of a cubic polynomial.
\smallskip

\begin{figure}
 \centerline{\psfig{file=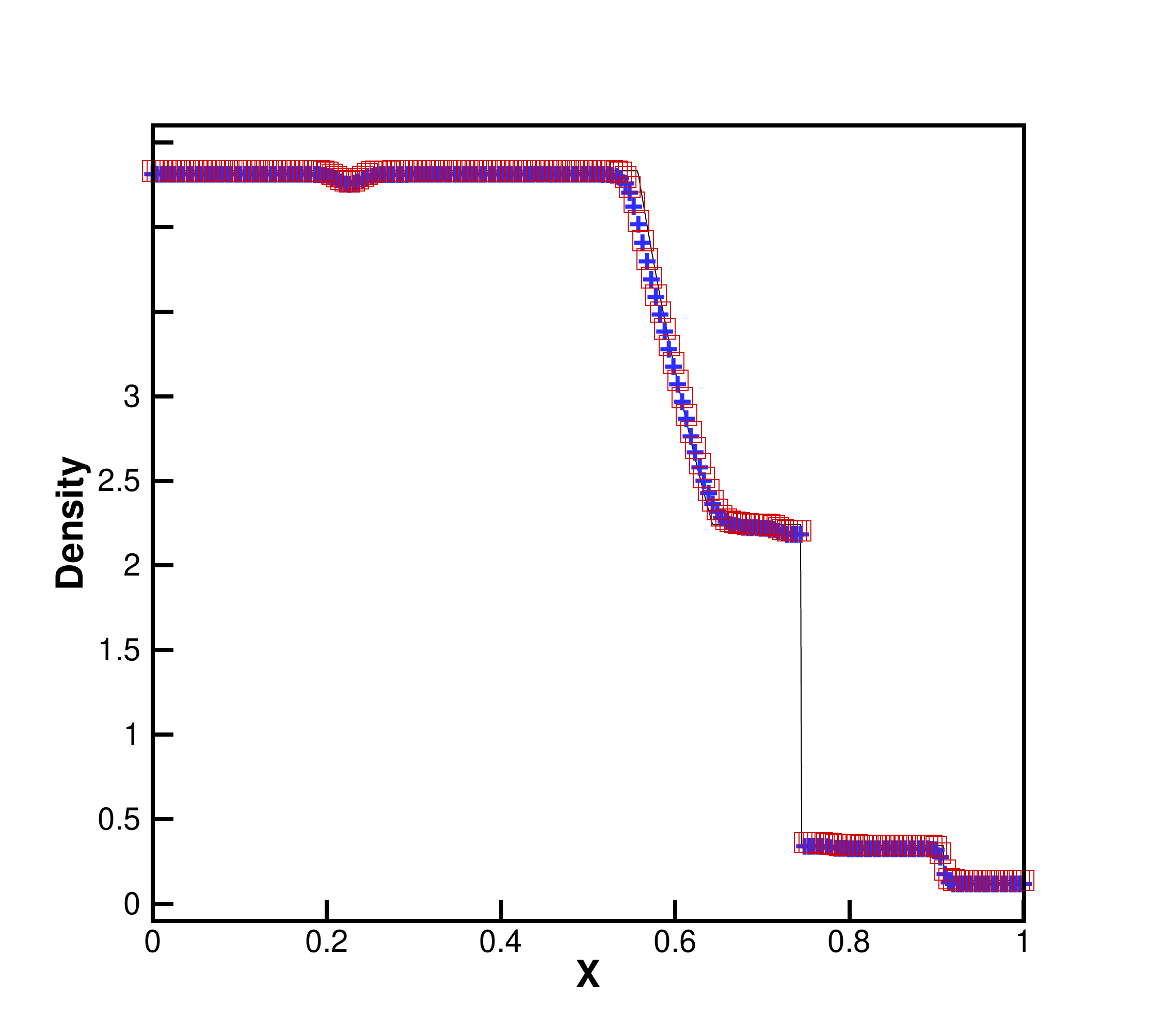,width=2 in} \psfig{file=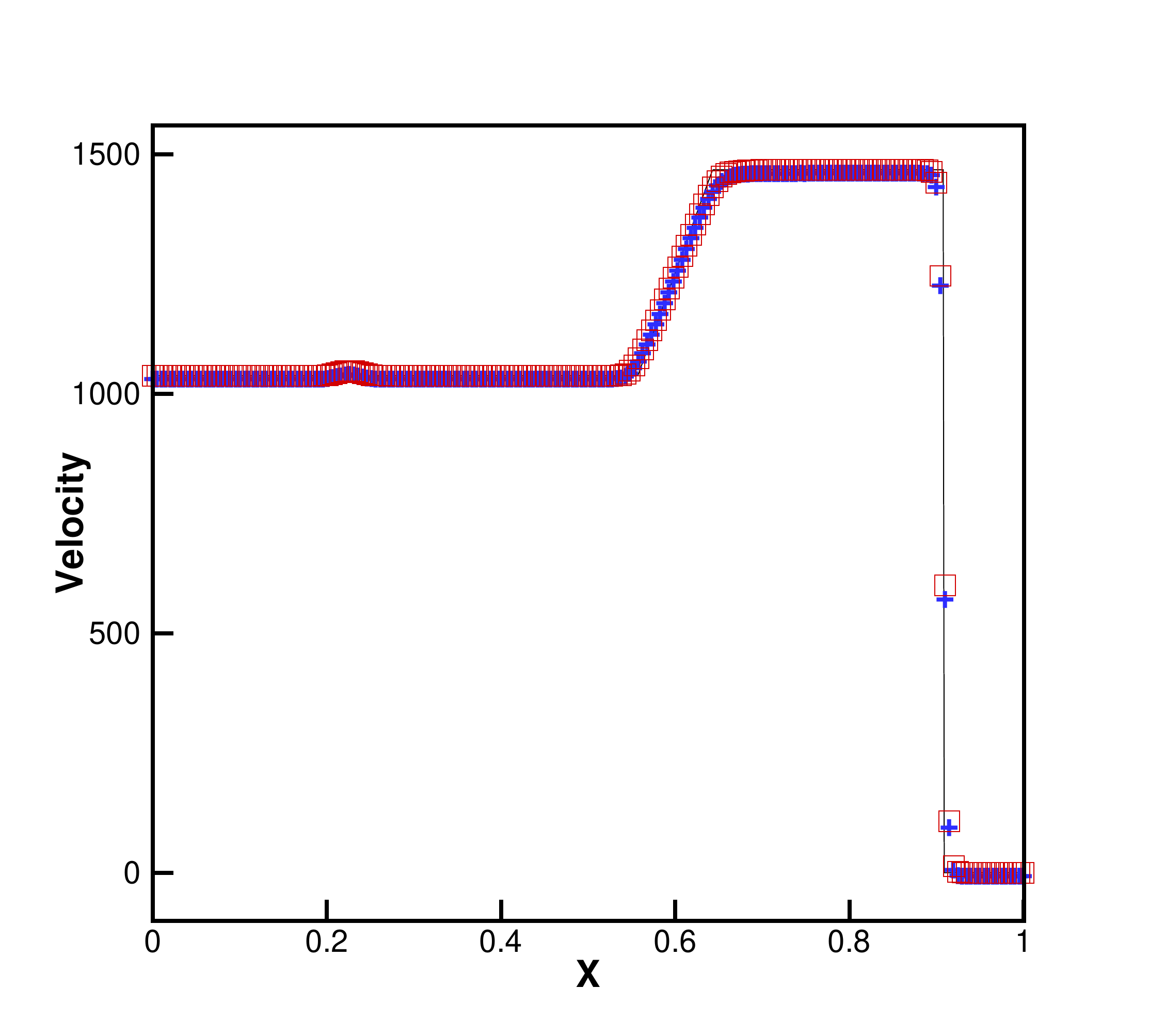,width=2 in}
 \psfig{file=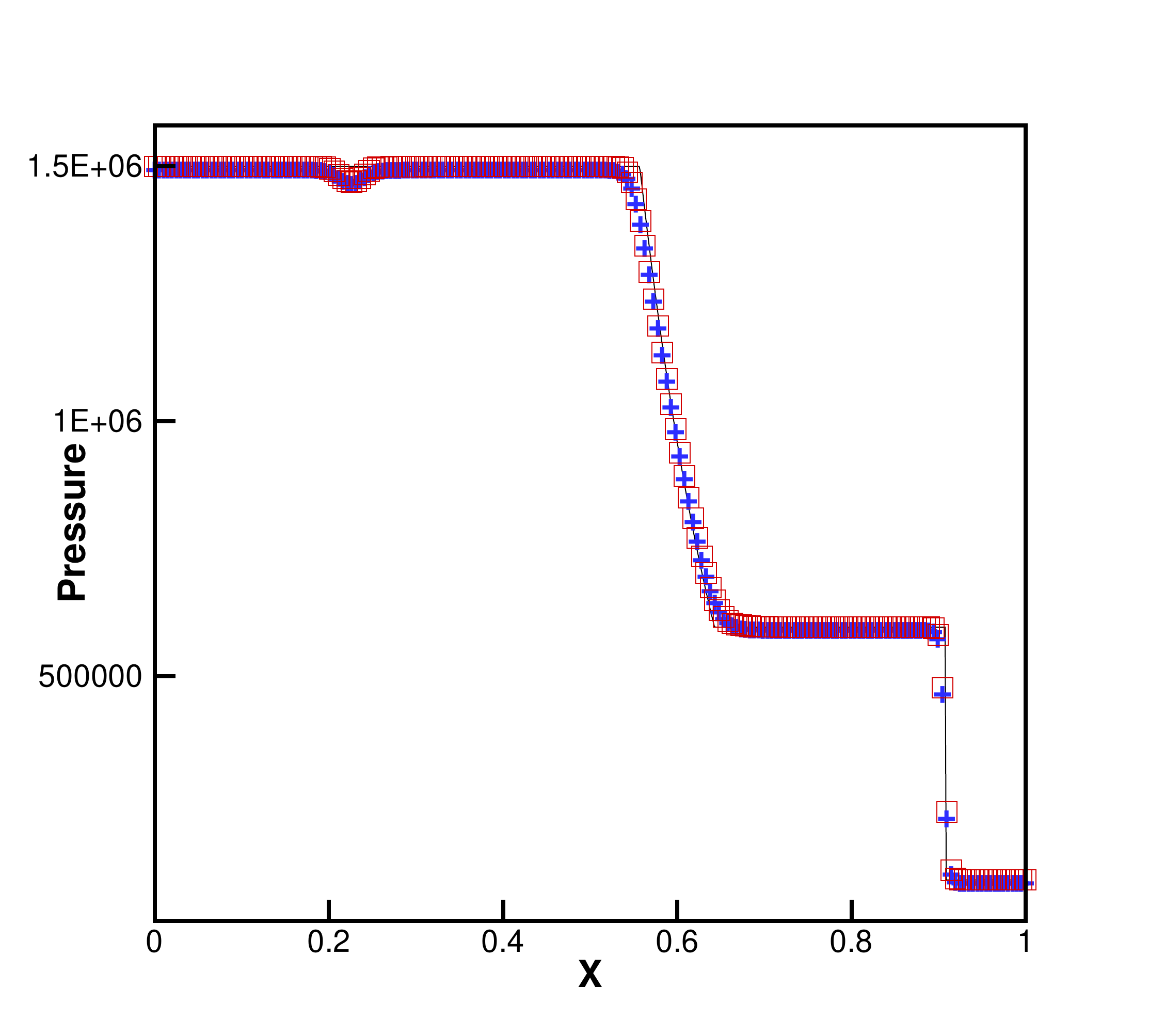,width=2 in}}
\caption{Example 3.2. t=0.0005. From left to right: density; velocity; pressure. Solid line: the exact solution; plus signs:  the results of Classical WENO method; squares:  the results of New/simplified hybrid WENO method. Grid points: 200.}
\label{Ex2}
\end{figure}

\noindent{\bf Example 3.2.} This problem is also taken from \cite{FAMO}, which contains  a right going shock refracting at an air-helium interface with a reflected rarefaction wave, and the initial conditions are given as
\begin{equation*}
 (\rho,\mu,p,\gamma)= \left\{
\begin{array}{ll}
(4.3333, 3.2817\sqrt{10^5}, 1.5\times10^6,1.4),& x \in [0, 0.05),\\
(1, 0, 1\times10^5,1.4),& x \in [0.05,0.5),\\
(0.1379, 0, 1\times10^5,5/3),& x \in [0.5,1],\\
\end{array}
\right.
\end{equation*}
with inflow/outflow  boundary conditions. The initial strength of the shock is $p_l/p_R=15$ at $x=0.05$, and the interface of air and helium is located at $x=0.5$. We ran the code to a final time of 0.0005, and the computed density $\rho$, velocity $\mu$ and pressure $p$ by New/simplified hybrid WENO and Classical WENO methods against the exact solution are shown in Figure \ref{Ex2}. We can see the contact discontinuity is located in the correct cell, and two methods have similar results, but New/simplified hybrid WENO method can achieve higher efficiency than Classical WENO method for saving 16.41\% CPU time. In addition, we find New/simplified  hybrid WENO method can save 9.54\% computation time than  Old hybrid WENO method by calculation, meanwhile, there are 25.07\%  and 24.50\% points where the WENO procedures are computed in New/simplified  hybrid WENO  and Old hybrid WENO methods, respectively, and the time history of the locations of WENO reconstruction are shown in the middle of Figure \ref{Exlim13}, which illustrate that the new identification skill in New/simplified  hybrid WENO method catchs the regions of the extreme points as the old one in Old hybrid WENO method. However, the new identification technique has higher efficiency, and it is also simpler for it only needs to calculate the roots of a quadratic polynomial.
\smallskip

\begin{figure}
 \centerline{\psfig{file=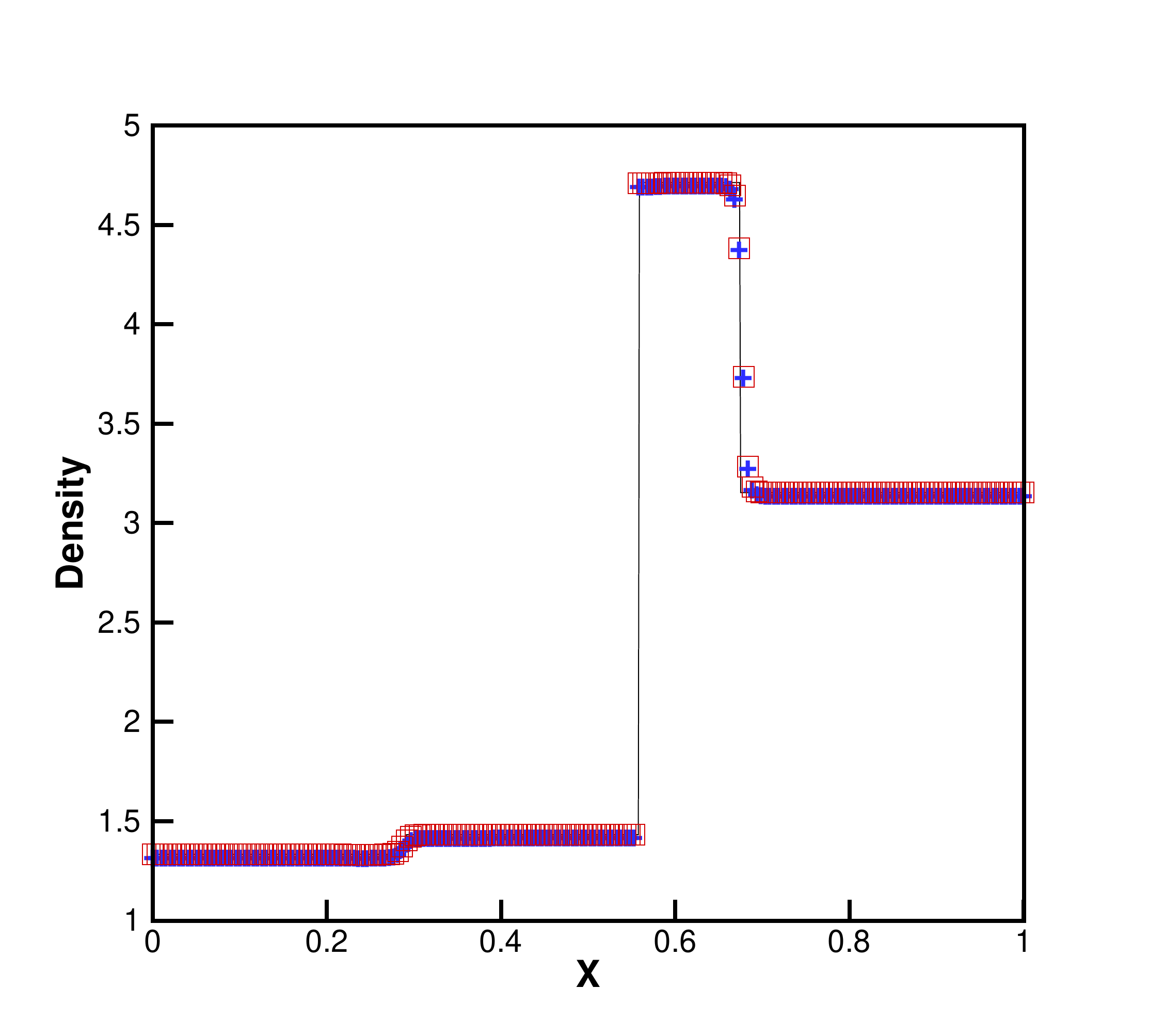,width=2 in} \psfig{file=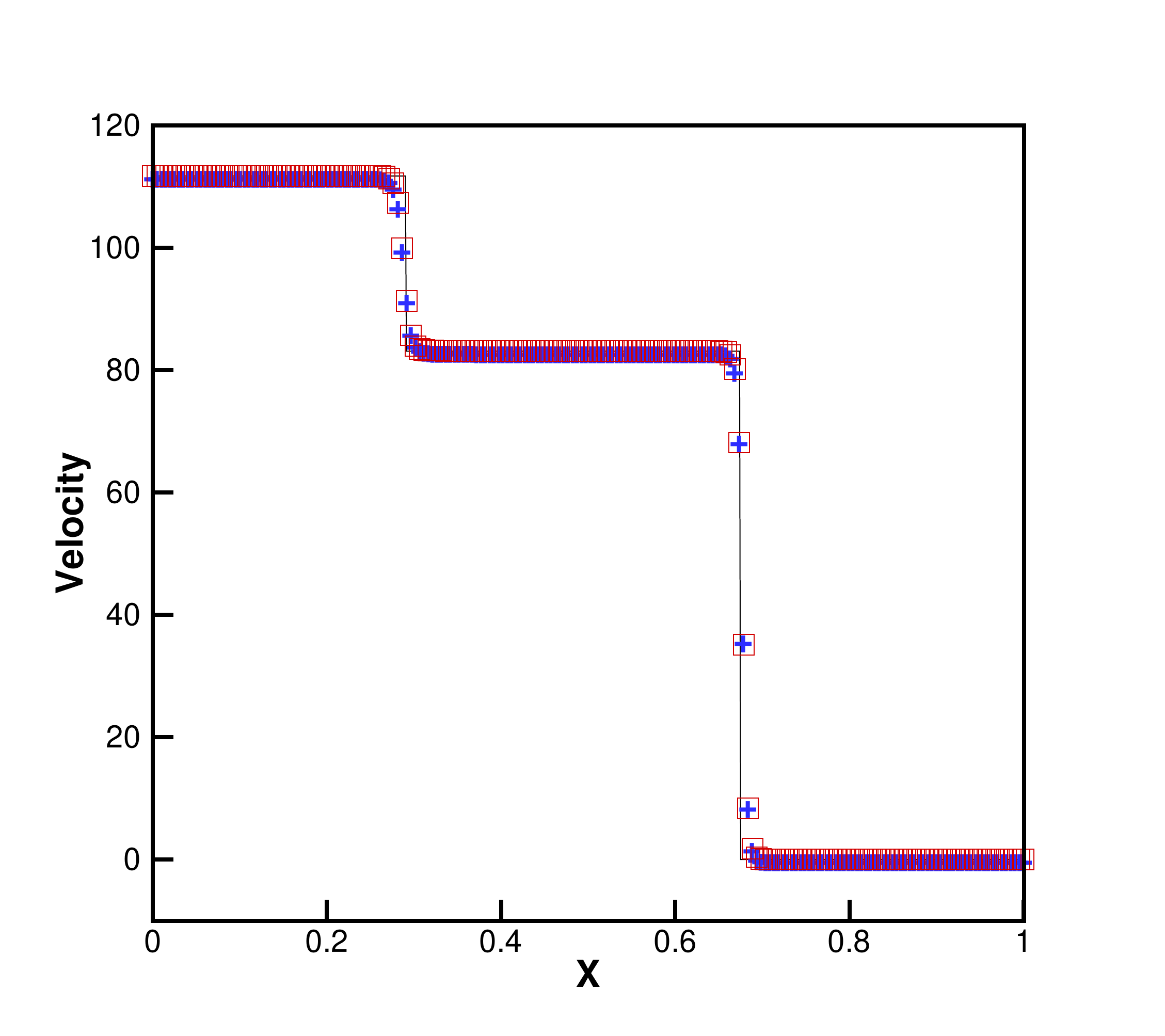,width=2 in}
 \psfig{file=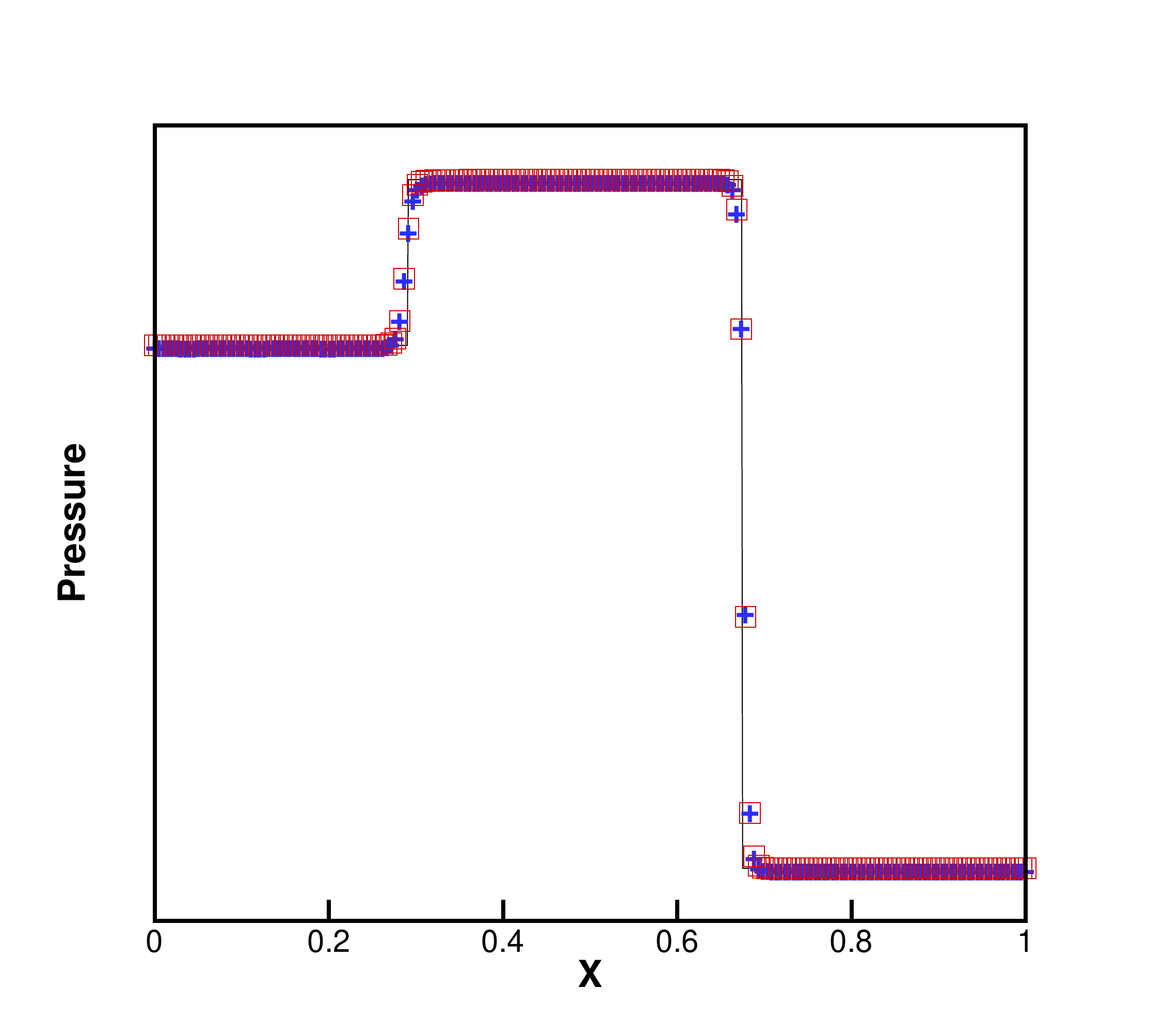,width=2 in}}
\caption{Example 3.3. T=0.0017. From left to right: density; velocity; pressure. Solid line: the exact solution; plus signs:  the results of Classical WENO method; squares:  the results of New/simplified hybrid WENO method. Grid points: 200.}
\label{Ex3}
\end{figure}

\noindent{\bf Example 3.3.} We solve the governing equations (\ref{EQ0}) for one dimensional Euler equations with the following Riemann initial conditions
\begin{equation*}
 (\rho,\mu,p,\gamma)= \left\{
\begin{array}{ll}
(1.3333, 0.3535\sqrt{10^5}, 1.5\times10^5,1.4),& x \in [0, 0.05),\\
(1, 0, 1\times10^5,1.4),& x \in [0.05,0.5),\\
(3.1538, 0, 1\times10^5,1.249),& x \in [0.5,1],\\
\end{array}
\right.
\end{equation*}
The final computed time $t$ is up to 0.0017. This example is also taken from \cite{FAMO}, and the computed density $\rho$, velocity $\mu$ and pressure $p$ by New/simplified hybrid WENO and Classical WENO methods against the exact solution are given in Figure \ref{Ex3}. The numerical results illustrate two schemes capture the contact discontinuity correctly, with non-oscillations and similar comparable results, meanwhile, New/simplified hybrid WENO method saves almost 25.49\% computation time comparing with Classical WENO method. In addition, we find New/simplified hybrid WENO method with the new identification skill  can save 8.31\% CPU time than Old hybrid WENO method by calculation, meanwhile, there are only 8.48\% and 8.52\% points where the WENO procedures are computed in
New/simplified hybrid WENO and Old hybrid WENO methods, respectively. The time history of the locations of WENO reconstruction by the two hybrid methods are seen in the bottom of Figure \ref{Exlim13}. These results illustrate that the new identification technique in New/simplified hybrid WENO  method catches the regions of the extreme points as the old one in  Old hybrid WENO method, but New/simplified hybrid WENO  method with the new one has higher efficiency, and the new identification technique is also simpler.
\smallskip

\begin{figure}
 \centerline{\psfig{file=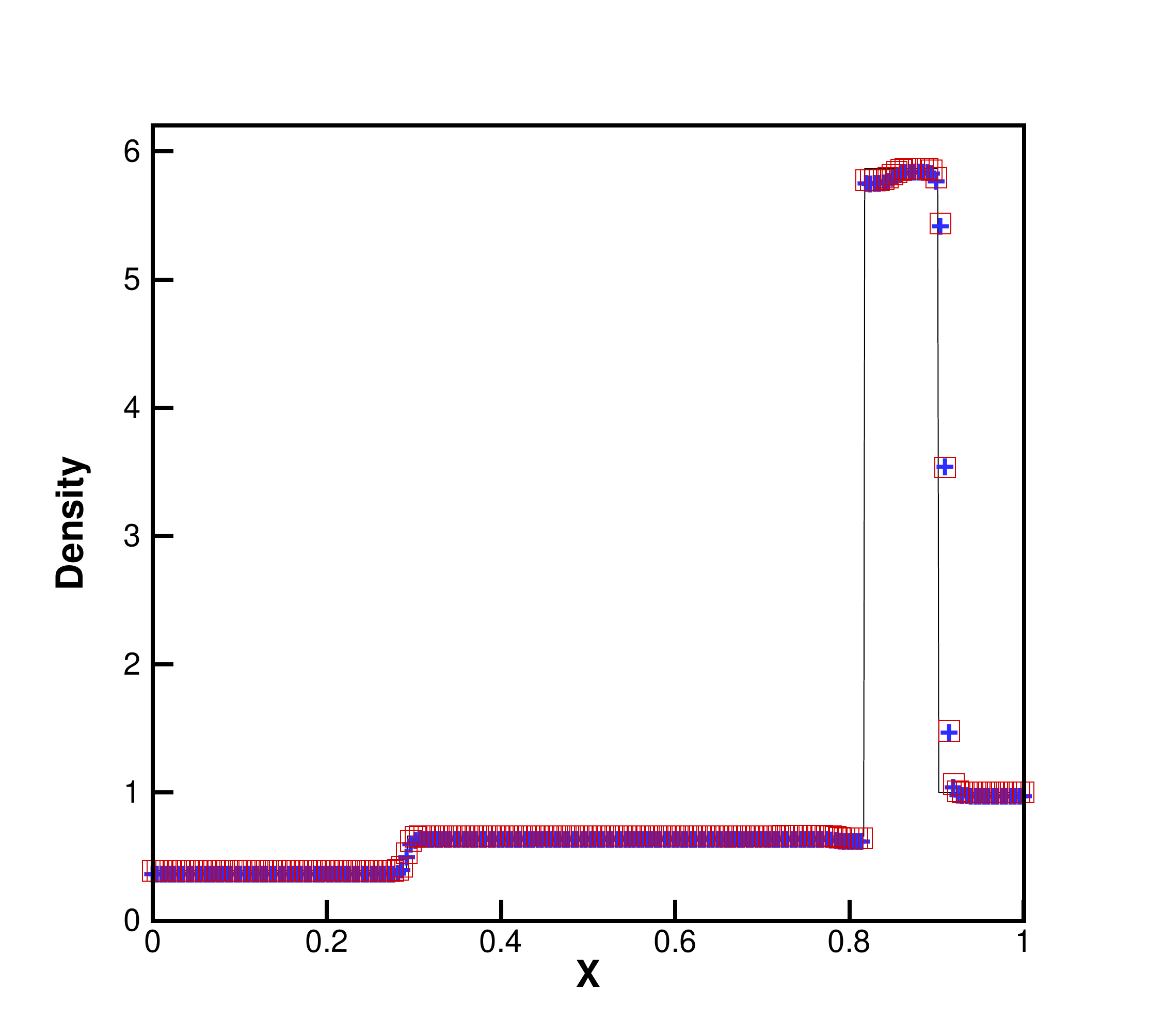,width=2 in} \psfig{file=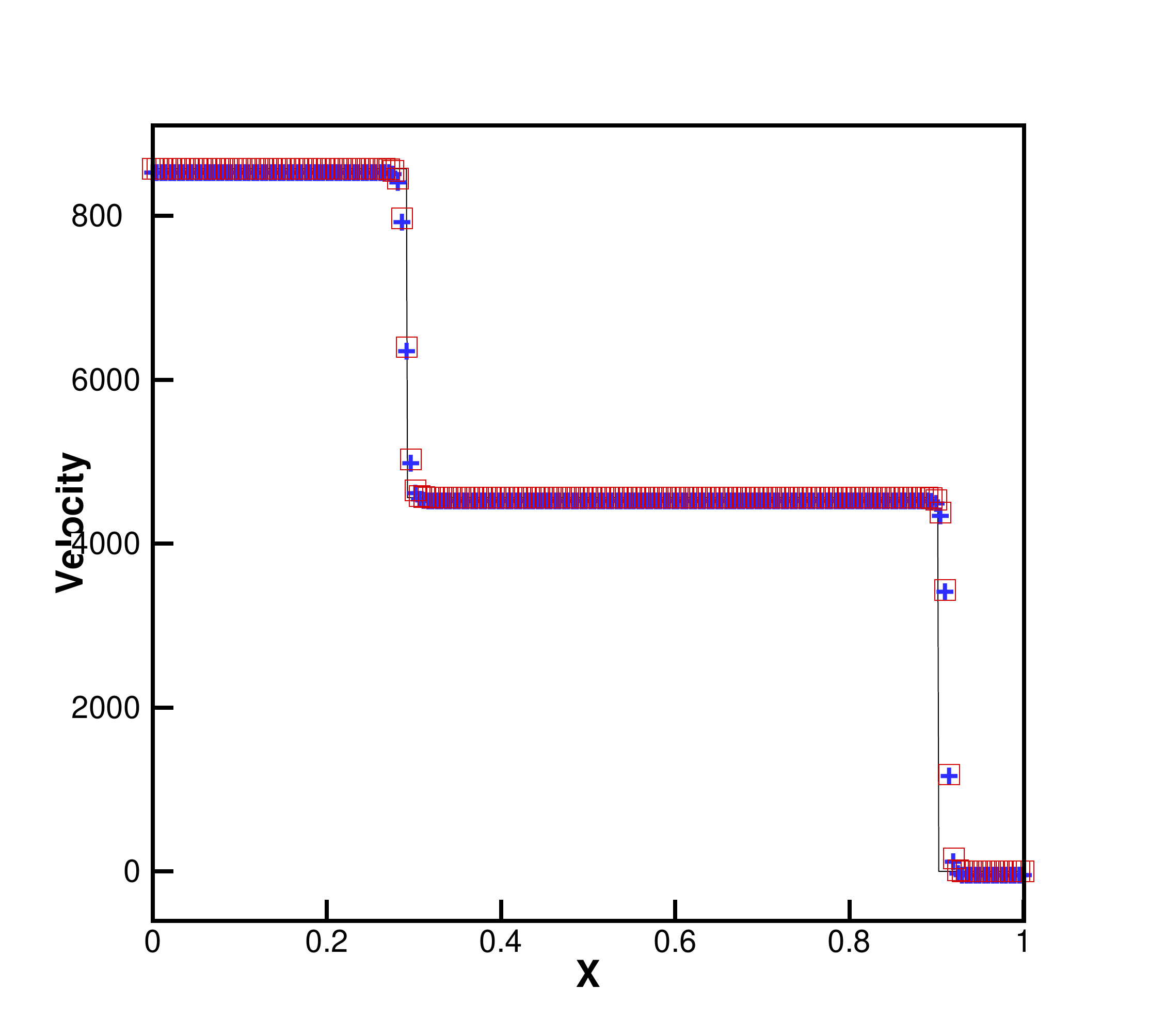,width=2 in}
 \psfig{file=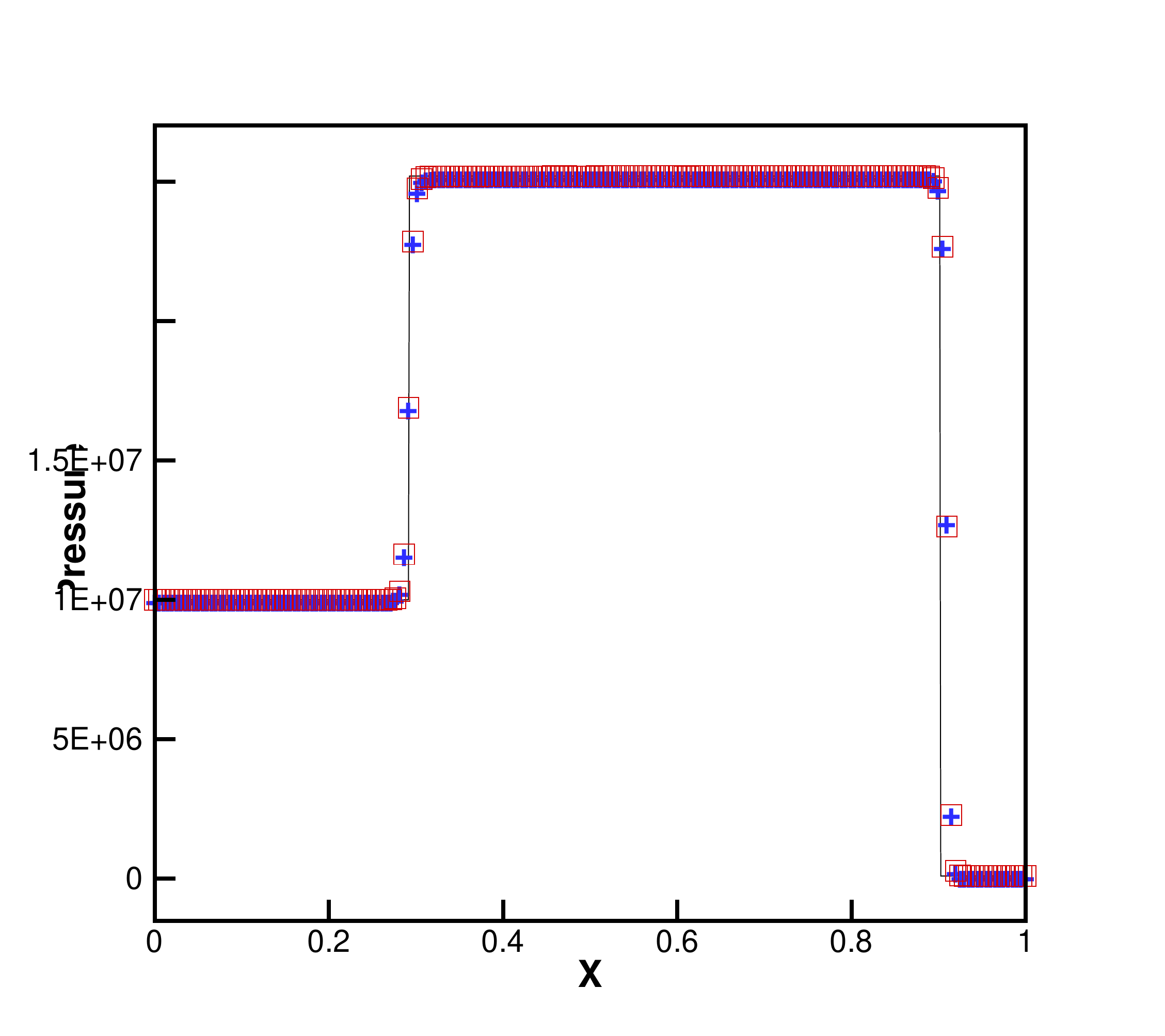,width=2 in}}
\caption{Example 3.4. t=0.0001. From left to right: density; velocity; pressure. Solid line: the exact solution; plus signs:  the results of Classical WENO method; squares:  the results of New/simplified hybrid WENO method. Grid points: 200.}
\label{Ex4}
\end{figure}

\begin{figure}
  \centerline{\psfig{file=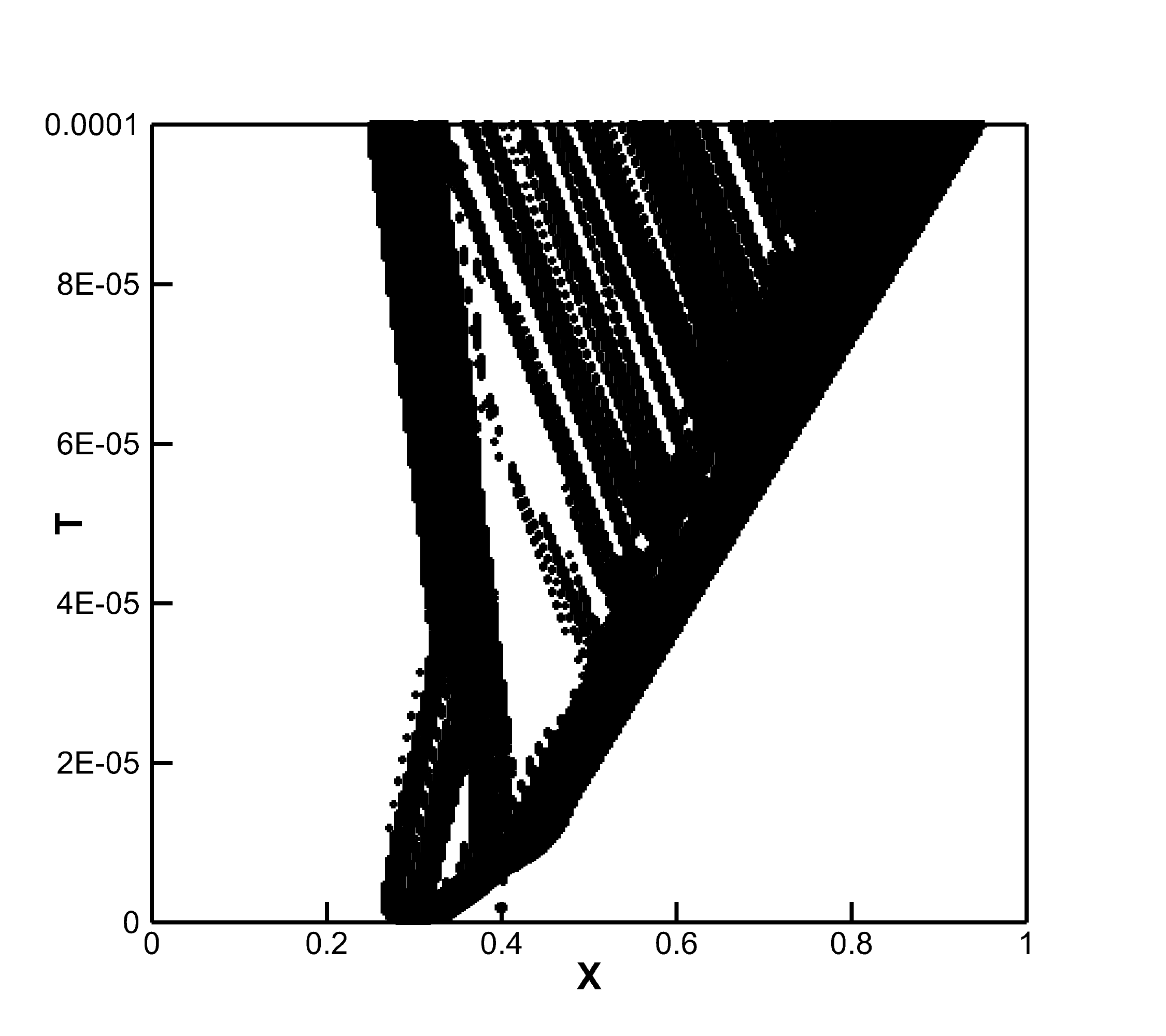,width=2 in}
 \psfig{file= 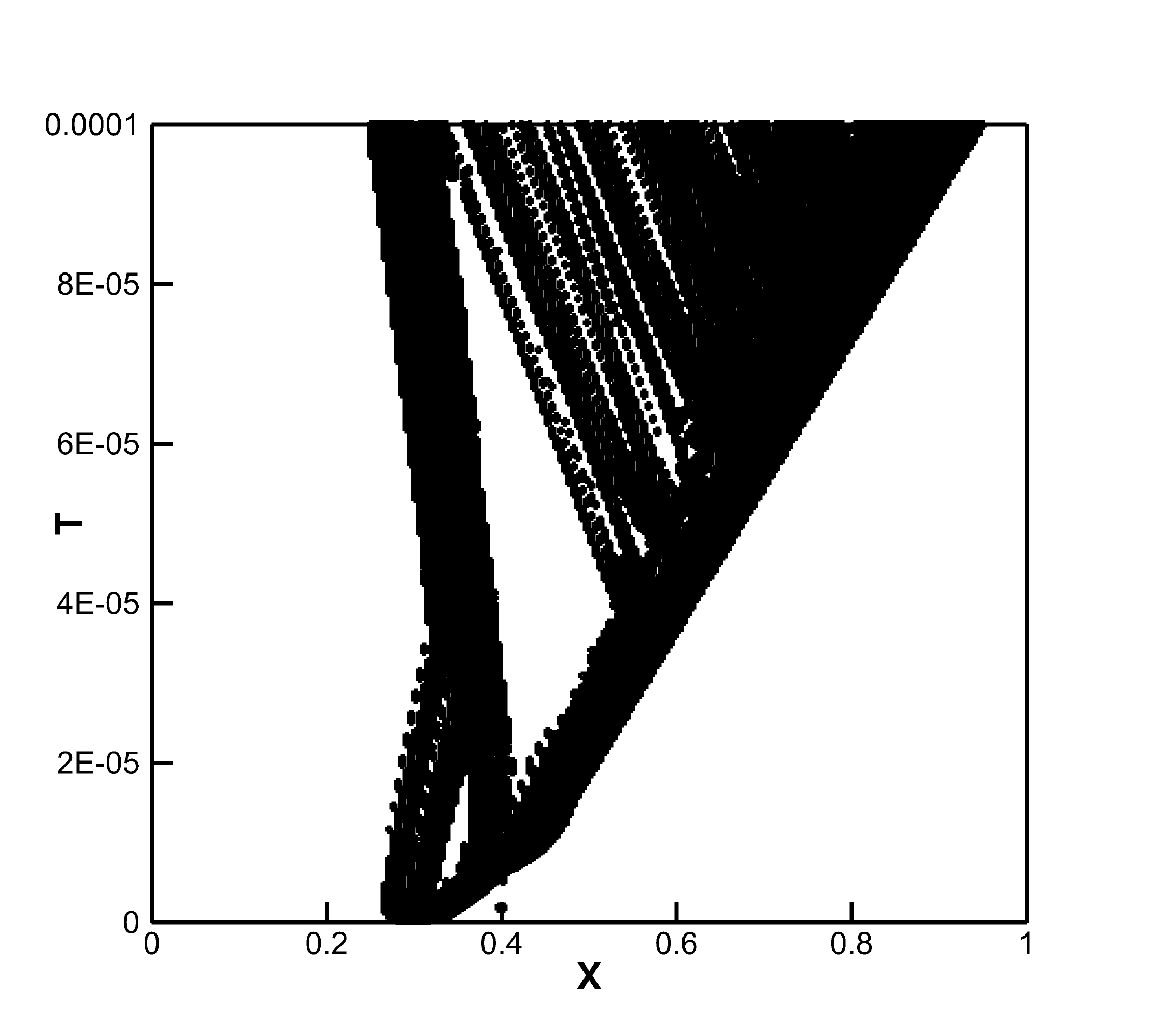,width=2 in}}
  \centerline{\psfig{file=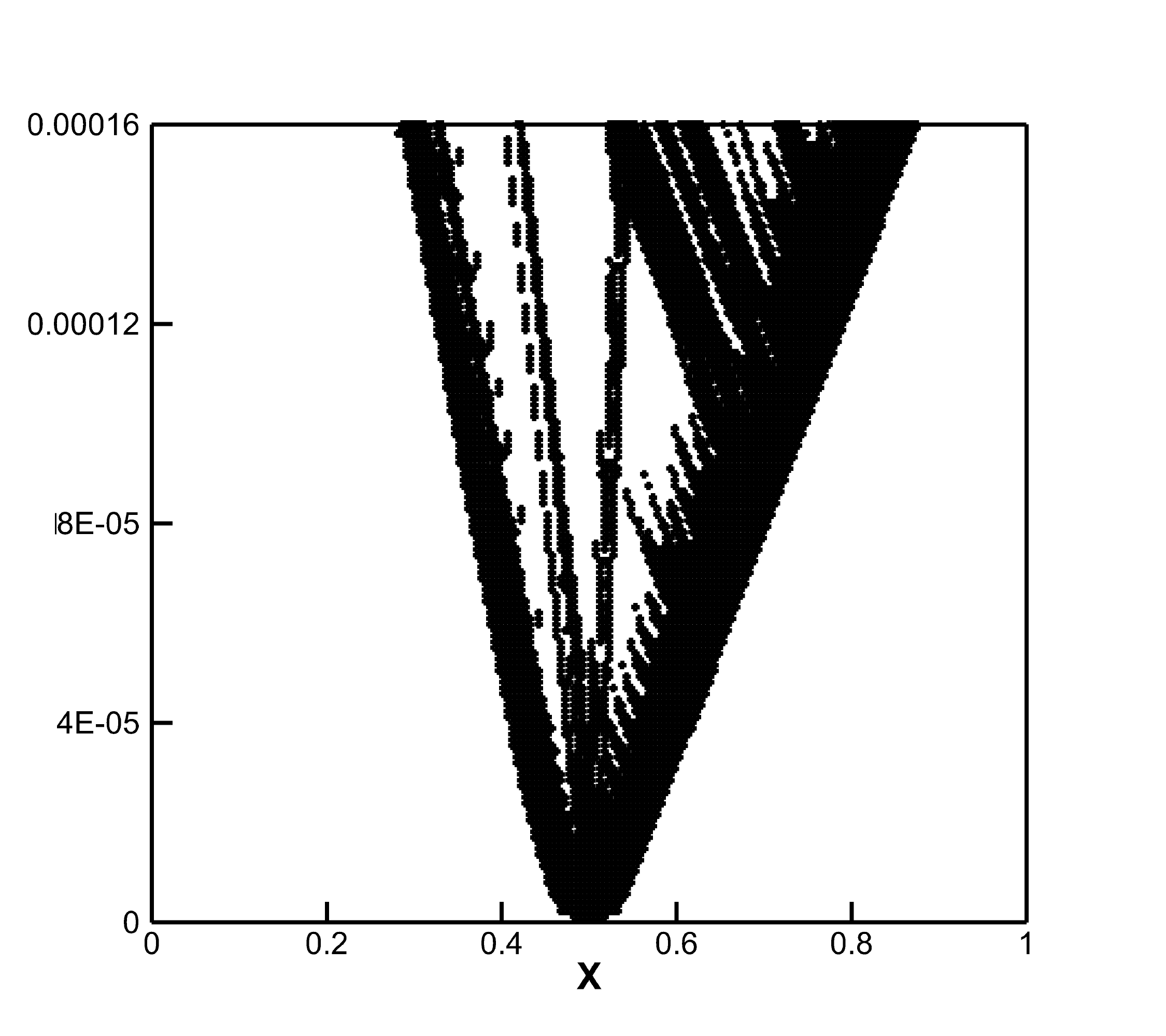,width=2 in}
 \psfig{file= 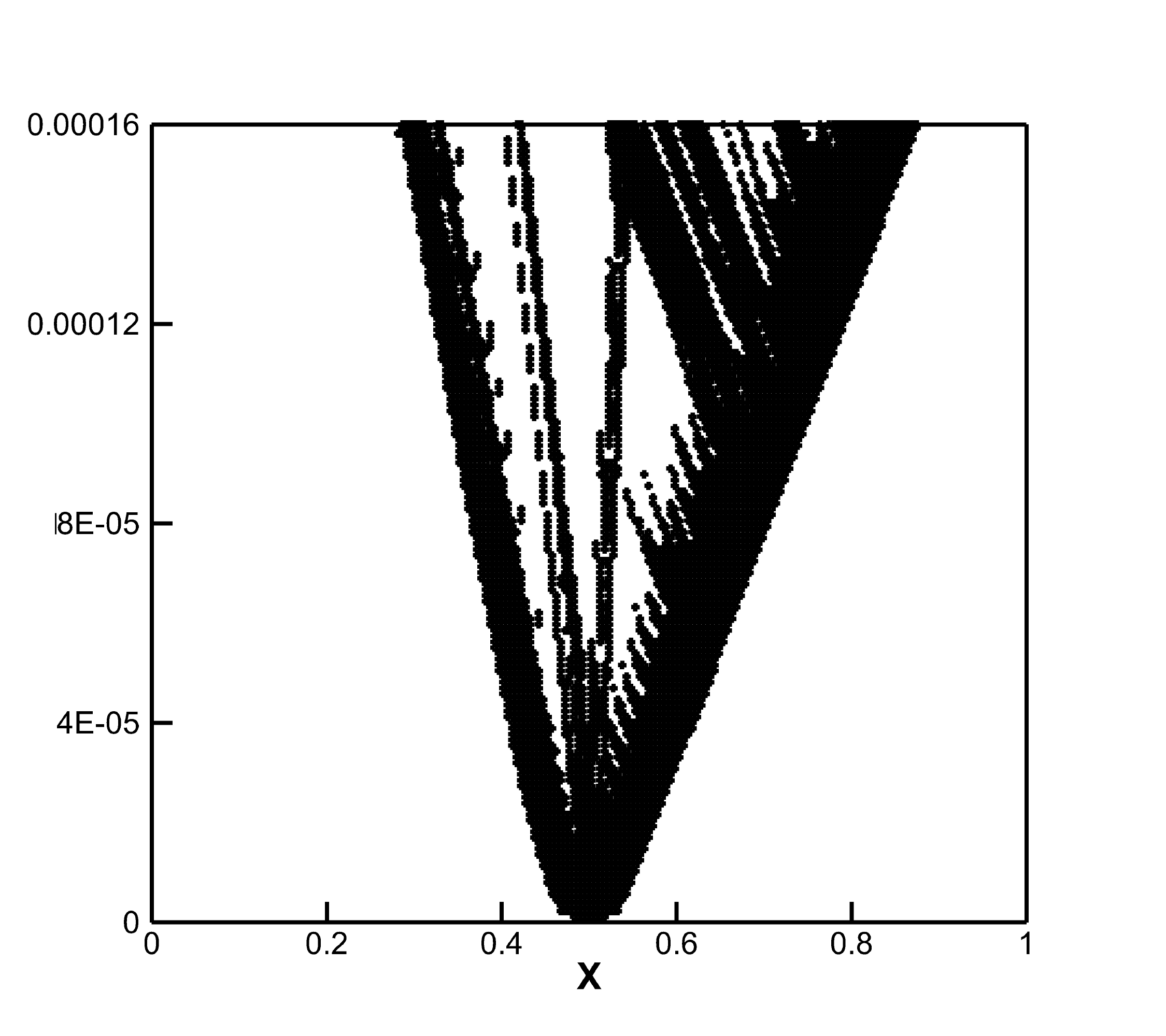,width=2 in}}
  \centerline{\psfig{file=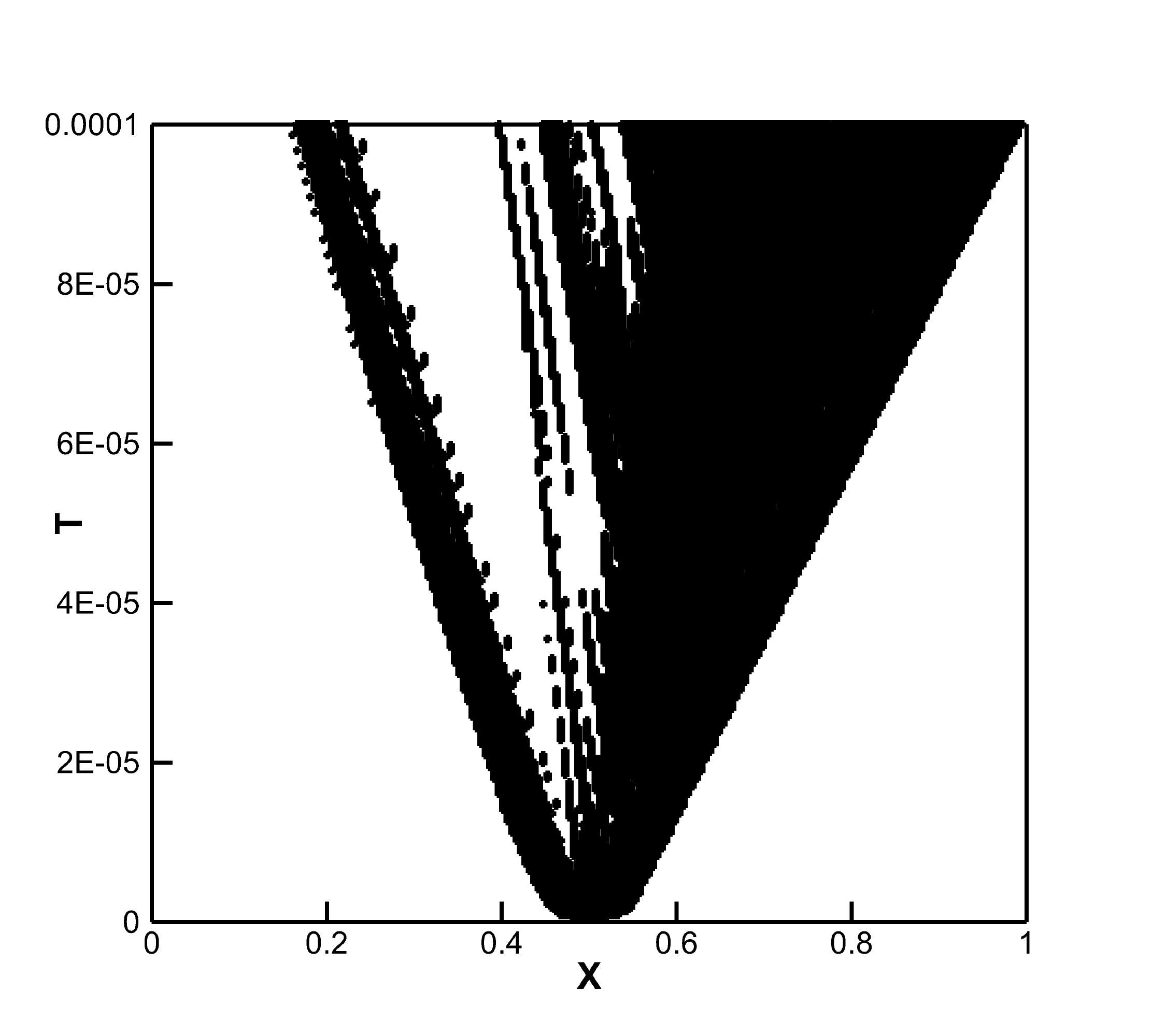,width=2 in}
 \psfig{file= 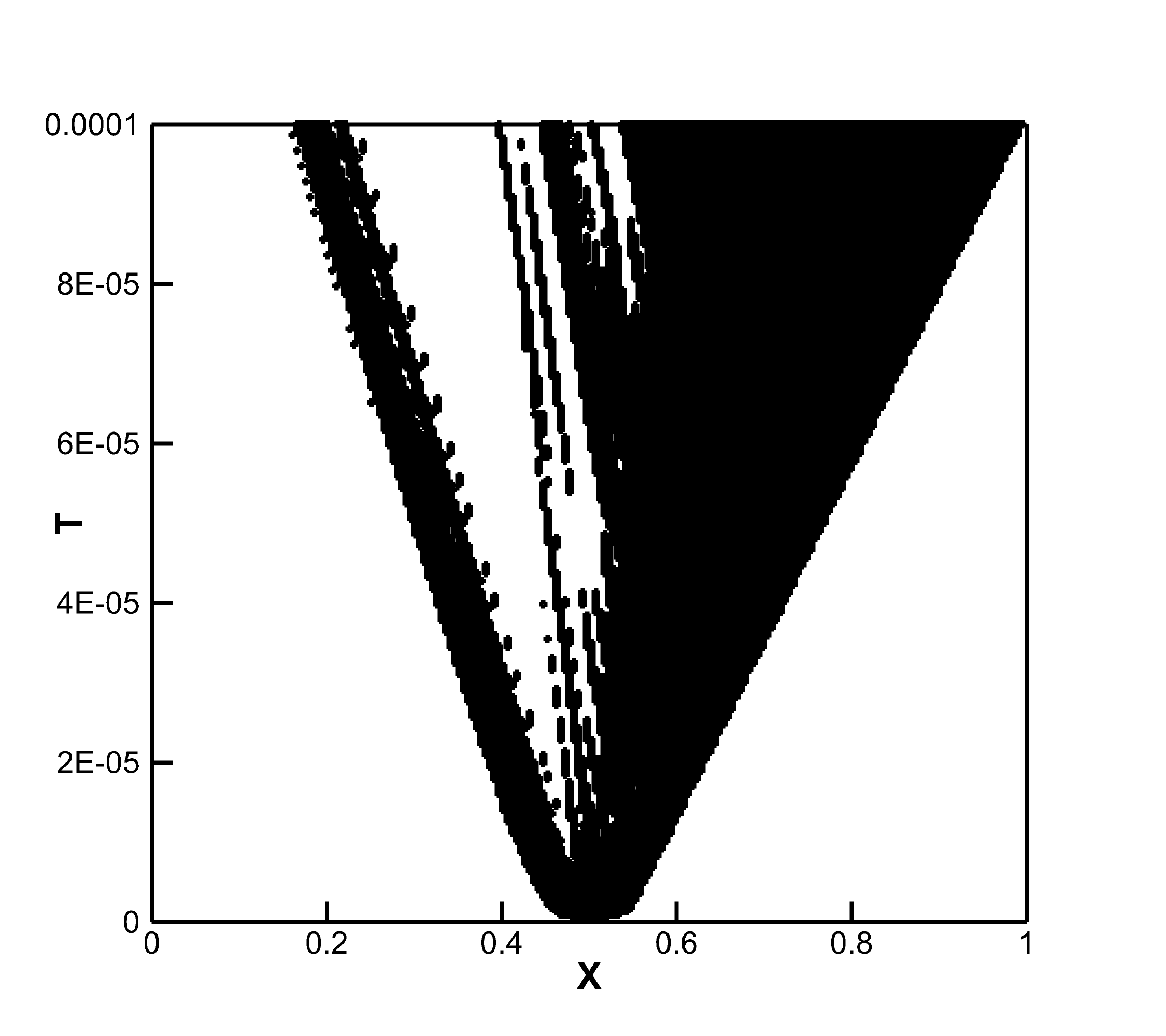,width=2 in}}
\caption{The points where the WENO procedures are performed for Examples 3.4 to 3.6. From left to right: the results of Old hybrid WENO method; the results of New/simplified hybrid WENO method.}
\label{Exlim46}
\end{figure}

\noindent{\bf Example 3.4.} This problem is  taken from \cite{LKY1}, having a strong shock on a gas-gas interface, and the strength of the right shock wave is up to $p_L/p_R=100$. The initial conditions are given as follows
\begin{equation*}
 (\rho,\mu,p,\gamma)= \left\{
\begin{array}{ll}
(0.3884, 27.1123\sqrt{10^5}, 1.0\times10^7,5/3),& x \in [0, 0.3),\\
(0.1, 0, 1\times10^5,5/3),& x \in [0.3,0.4),\\
(1, 0, 1\times10^5,1.4),& x \in [0.4,1].\\
\end{array}
\right.
\end{equation*}
In Figure \ref{Ex4}, we present the computed density $\rho$, velocity $\mu$ and pressure $p$ by New/simplified hybrid WENO and Classical WENO methods against the exact solution at the final time $0.0001$. We can find that two methods work well for simulating this two-phase flow problem, and they capture the correct location of the interface between two gases. Comparing with Classical WENO method, New/simplified hybrid WENO method saves almost 15.98\% computation time. In addition, we find New/simplified hybrid WENO method with the new identification skill  can save 14.52\% CPU time than Old hybrid WENO method by calculation, meanwhile, there are  26.76\% and 25.52\% points where the WENO procedures are performed in New/simplified hybrid WENO and Old hybrid WENO methods, respectively, and the time history of the locations of WENO reconstruction by two hybrid methods are given in the top of Figure \ref{Exlim46}, which illustrate the new identification skill in New/simplified hybrid WENO method  identifies the regions  of the extreme points as the old one in Old hybrid WENO method, but New/simplified hybrid WENO method  with the new one has higher efficiency. The new identification technique in New/simplified hybrid WENO method is also simpler as it only needs to solve the roots of a quadratic polynomial, while the old one in Old hybrid WENO method  has to calculate the zero points  of a cubic polynomial.
\smallskip

\begin{figure}
 \centerline{\psfig{file=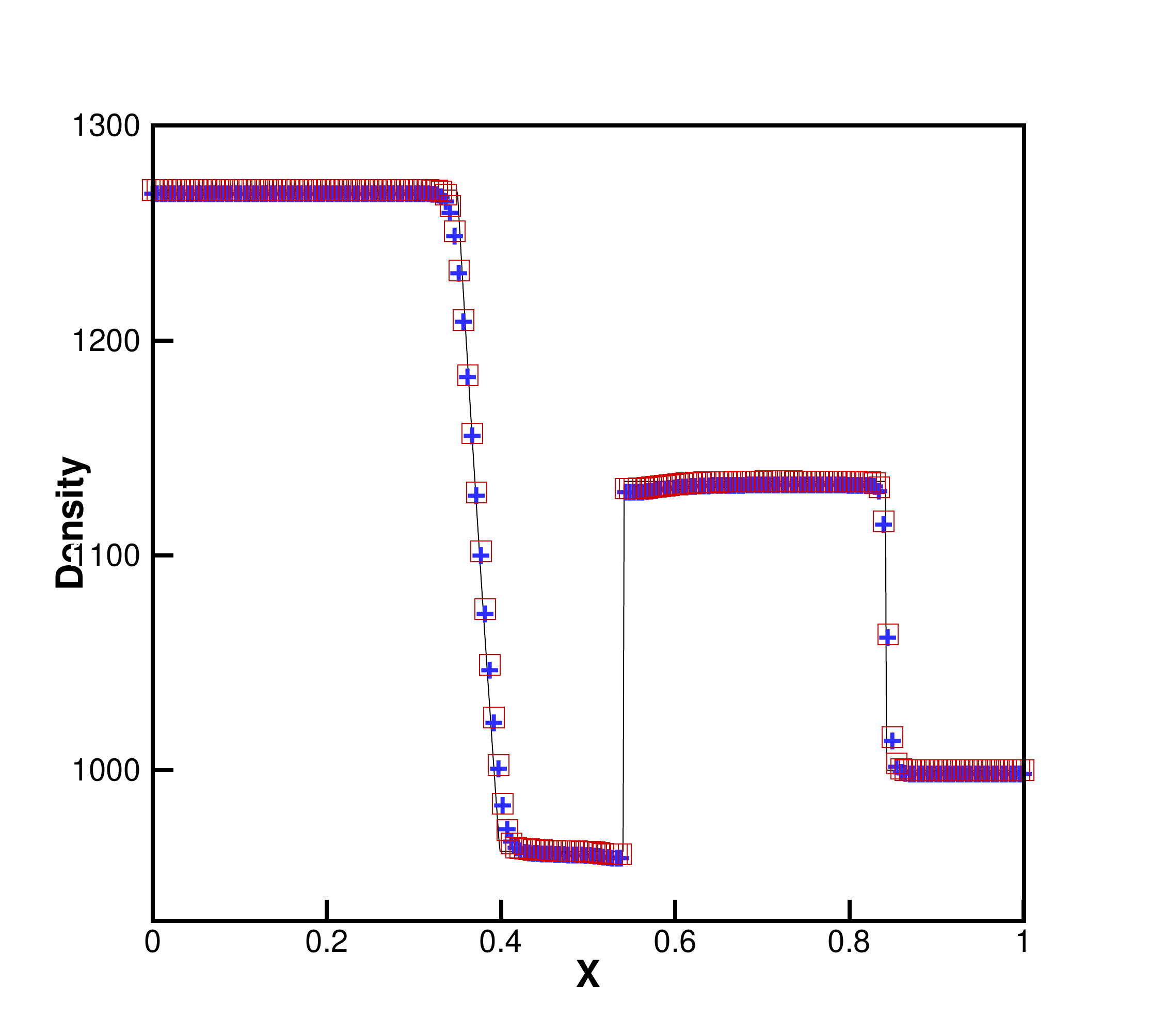,width=2 in} \psfig{file=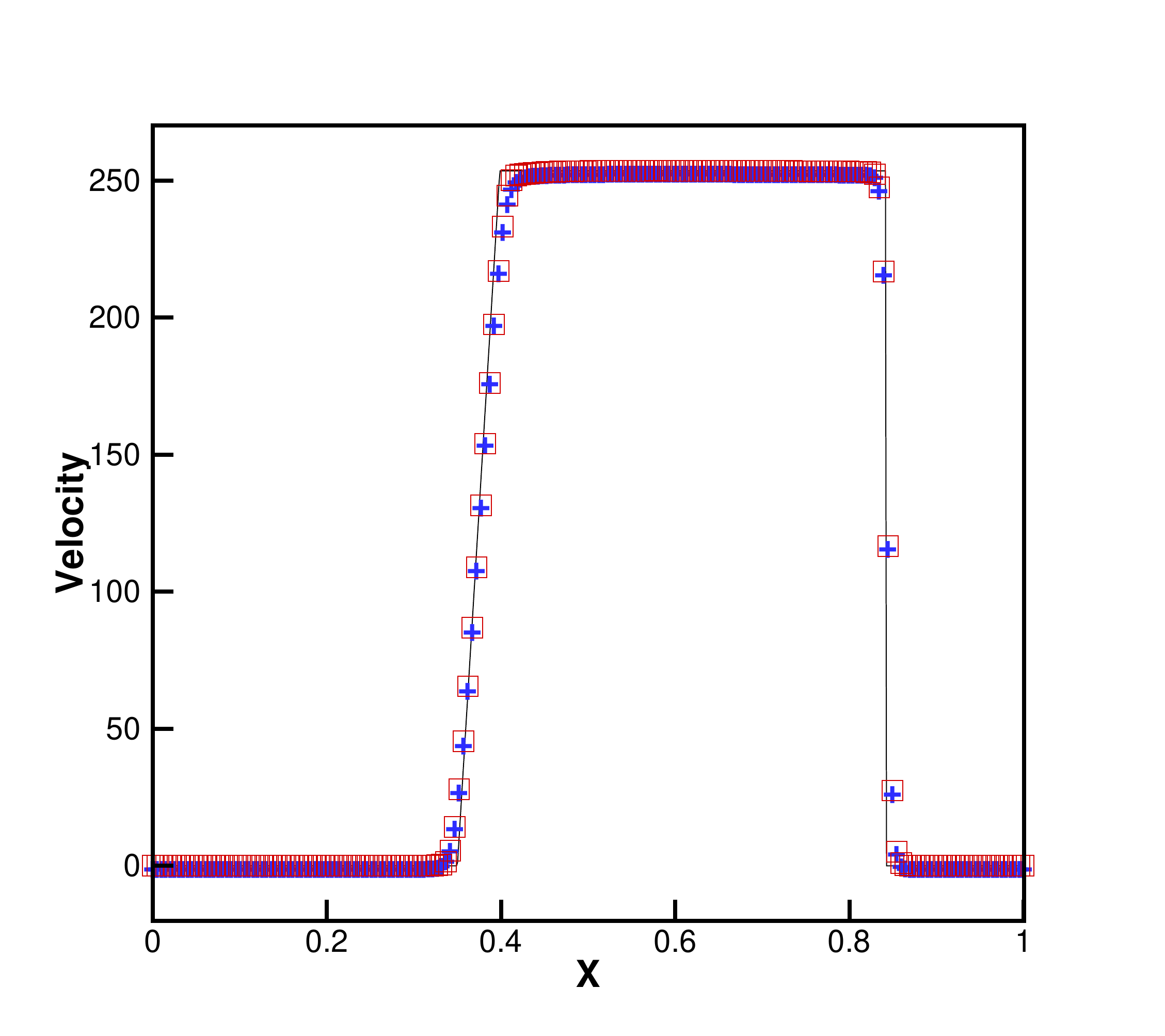,width=2 in}
 \psfig{file=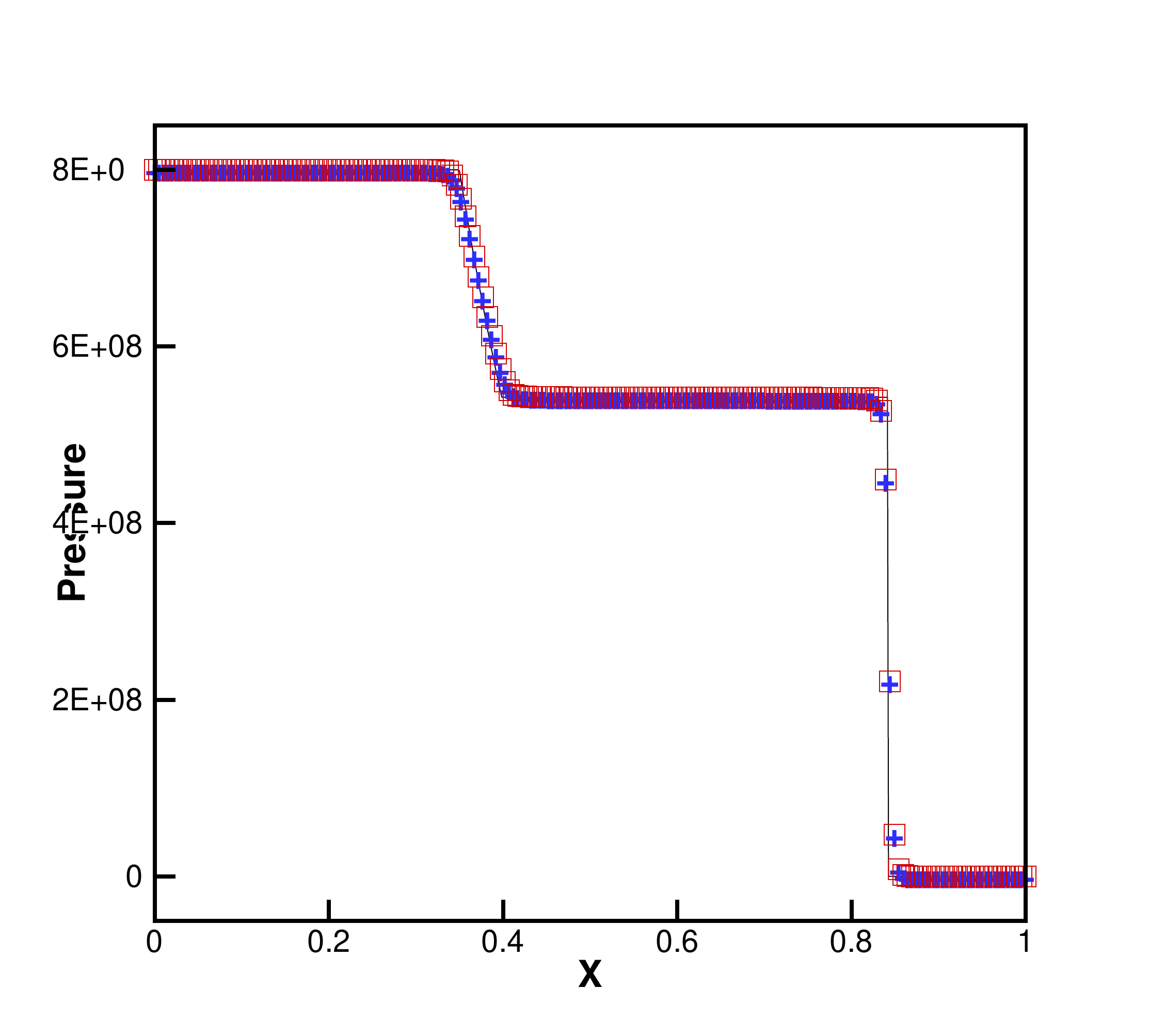,width=2 in}}
\caption{Example 3.5. t=00016. From left to right: density; velocity; pressure. Solid line: the exact solution; plus signs:  the results of Classical WENO method; squares:  the results of New/simplified hybrid WENO method. Grid points: 200.}
\label{Ex5}
\end{figure}

\noindent{\bf Example 3.5.} We consider the  gas-water shock tube problem taken from \cite{QLH}, and the initial condition are given as
\begin{equation*}
(\rho,\mu,p,\gamma)^T= \left\{
\begin{array}{ll}
(1270, 0, 8\times10^8,1.4)^T,& x \in [0, 0.5),\\
(1000, 0, 1\times10^5,7.15)^T,& x \in [0.5,1].
\end{array}
\right.
\end{equation*}
This underwater explosion problem has extremely high pressure in the gas medium, therefore, there is a very strong shock in the water. The final computation time is $0.00016$. We present the computed density $\rho$, velocity $\mu$ and pressure $p$ by New/simplified hybrid WENO and Classical WENO methods against the exact solution in Figure \ref{Ex5}, which illustrates two schemes capture the location of the material interface correctly, and they have good performance in the smooth and discontinuous regions. In addition, New/simplified hybrid WENO method saves about 16.53\% CPU time. Moreover, we find New/simplified hybrid WENO method with the new identification skill  can save 13.93\% computation time than the Old hybrid WENO method with the old identification technique  by calculation, meanwhile, there are both 20.68\% points where the WENO procedures are performed in New/simplified hybrid WENO  and Old hybrid WENO methods, respectively, and the time history of the locations of WENO reconstruction in two hybrid methods are given in the middle of Figure \ref{Exlim46}. These results show that the new identification skill in New/simplified hybrid WENO can catch the regions of the extreme points as the old one in Old hybrid WENO method. However, New/simplified hybrid WENO method with the new identification skill has higher efficiency, and the procedure of the new one is also simpler.
\smallskip

\begin{figure}
 \centerline{\psfig{file=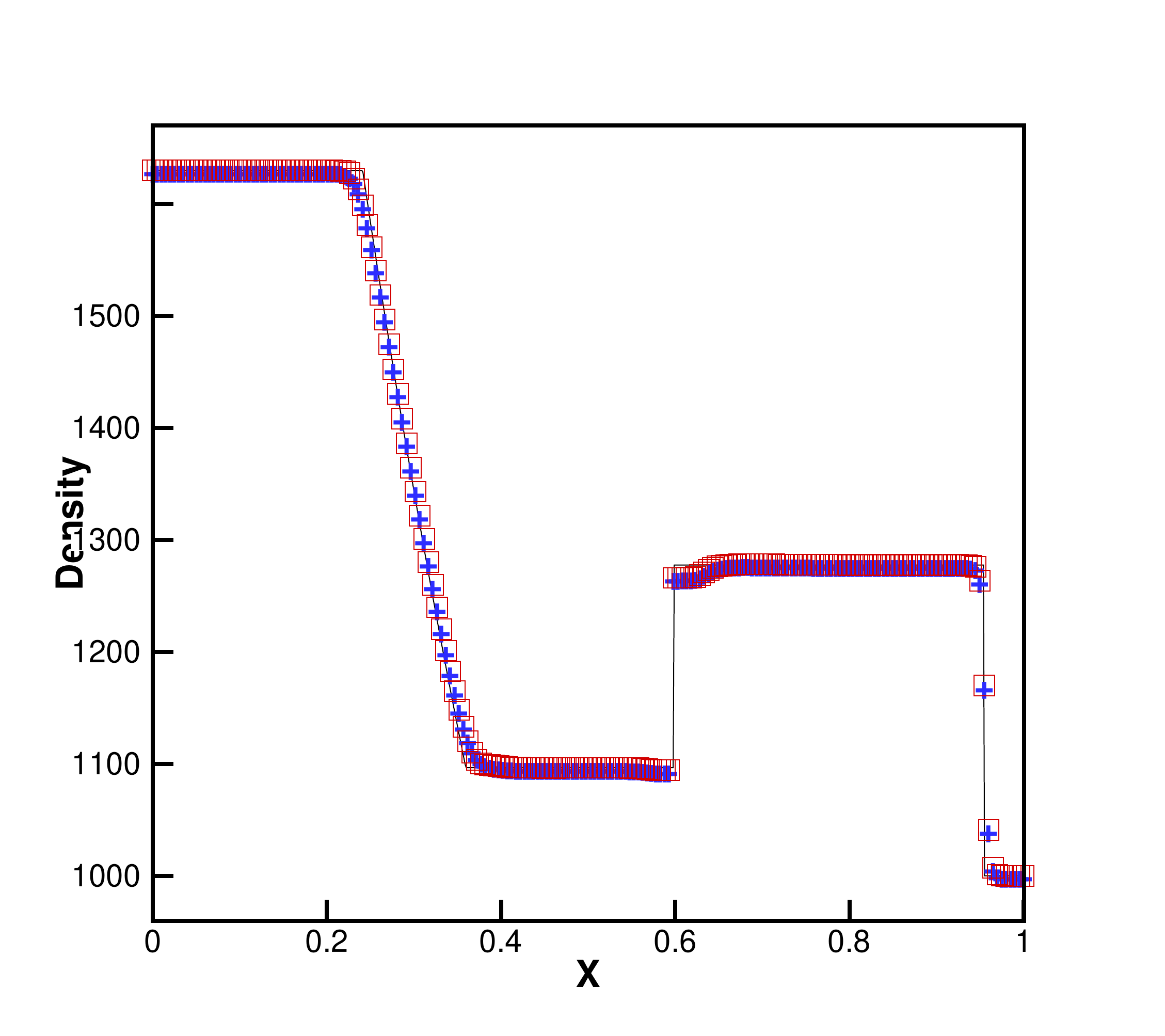,width=2 in} \psfig{file=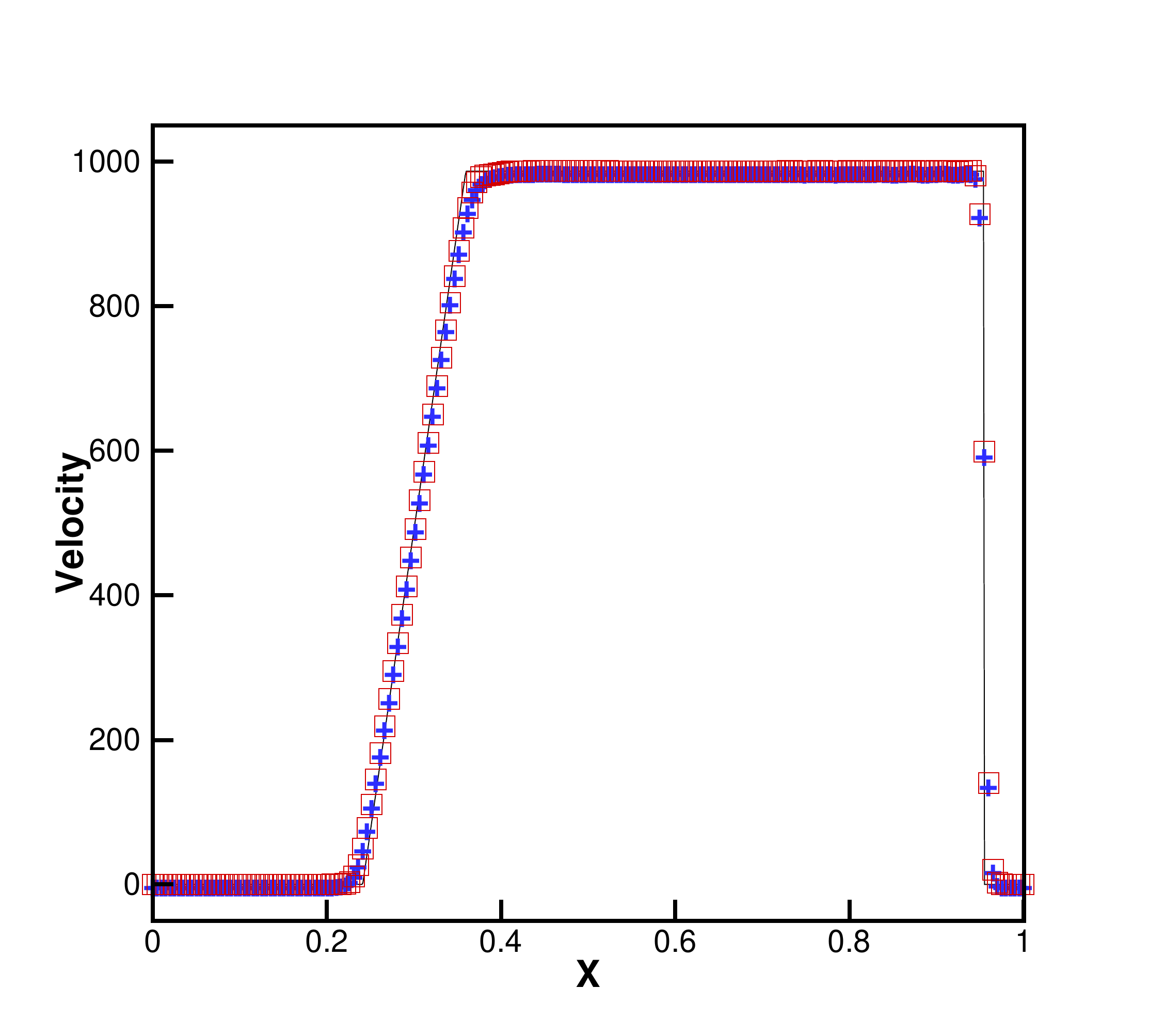,width=2 in}
 \psfig{file=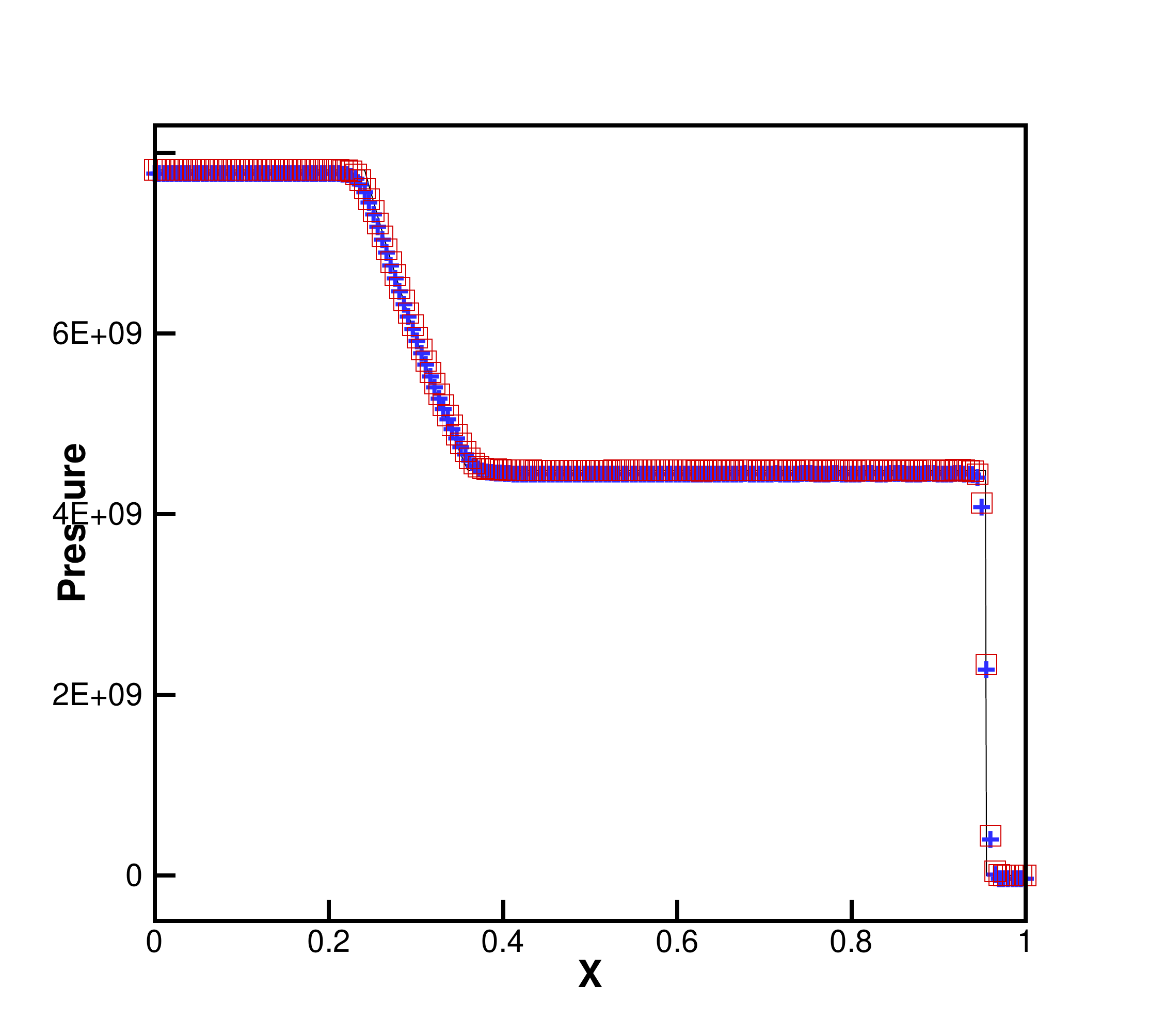,width=2 in}}
\caption{Example 3.6. t=0.0001. From left to right: density; velocity; pressure. Solid line: the exact solution; plus signs:  the results of Classical WENO method; squares:  the results of New/simplified hybrid WENO method. Grid points: 200.}
\label{Ex6}
\end{figure}

\noindent{\bf Example 3.6.} This gas-water shock tube problem is taken from \cite{FAMO}, which has higher energy of the explosive gaseous medium than the problem given in Example 3.5, and the initial conditions are
\begin{equation*}
(\rho,\mu,p,\gamma)^T= \left\{
\begin{array}{ll}
(1630, 0, 7.81\times10^9,1.4)^T,& x \in [0, 0.5),\\
(1000, 0, 1\times10^5,7.15)^T,& x \in [0.5,1].
\end{array}
\right.
\end{equation*}
We ran the code to the final time 0.0001, then, we give the computed density $\rho$, velocity $\mu$ and pressure $p$ by New/simplified hybrid WENO and Classical WENO methods against the exact solution in Figure \ref{Ex6}, which shows two schemes work well for this tough gas-water problem with non-oscillation in the non-smooth regions, and they also capture the right interface between two mediums. In addition, New/simplified hybrid WENO  method has higher efficiency for saving almost 15.12\% CPU time. Moreover, we find New/simplified hybrid WENO  method with the new identification skill  can save 13.52\% computation time than the Old hybrid WENO  method by calculation, meanwhile, there are both 32.88\% points where the WENO procedures are computed in two hybrid WENO method, and the time history of the locations of WENO reconstruction in New/simplified hybrid WENO and Old hybrid WENO methods are given in the bottom of Figure \ref{Exlim46}, which show that the new identification skill New/simplified hybrid WENO method can catch the regions of the extreme points correctly as the old one, but New/simplified hybrid WENO method with the new one has higher efficiency, and the new identification technique is also simpler.
\smallskip

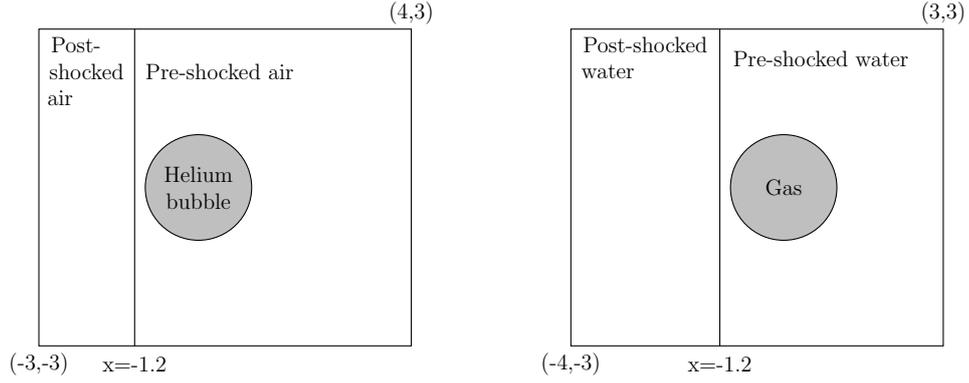
\begin{figure}
\tikzset{global scale/.style={
 scale=#1,every node/.append style={scale=#1}}}
  \centering
  \begin{tikzpicture}[global scale = 0.7]
\definecolor{grey}{rgb}{0.75,0.75,0.75}
\draw (-3,-3) rectangle(4,3);
\draw (-1.2,-3) -- (-1.2,3);
\draw (-3,-3-0.35) node{(-3,-3)};
\draw (4,3+0.3) node{(4,3)};
\draw (-1.2,-3-0.35) node{x=-1.2};
\draw(-2.3,2.7)node{Post-};
\draw(-2.1,2.2)node{shocked};
\draw(0.4,2.2)node{Pre-shocked air};
\draw(-2.6,1.7)node{air};
\draw[fill=grey](0,0)circle(1);
\draw(0,0.25)node{Helium};
\draw(0,-0.25)node{bubble};
\draw (7,-3) rectangle(14,3);
\draw (9.8,-3) -- (9.8,3);
\draw (7,-3-0.35) node{(-4,-3)};
\draw (14,3+0.3) node{(3,3)};
\draw (9.8,-3-0.35) node{x=-1.2};
\draw(8.4,2.7)node{Post-shocked};
\draw(7.7,2.2)node{water};
\draw(11.7,2.45)node{Pre-shocked water};
\draw[fill=grey](11,0)circle(1);
\draw(11,0)node{Gas};
\end{tikzpicture}
\caption{Physical domain for Example 3.7 (left) and Example 3.8 (right)}
\label{phy2d}
\end{figure}

\noindent{\bf Example 3.7.} This problem is a Mach 1.22 air shock acting on a helium bubble, then, we solve the governing equations (\ref{EQ0}) for two dimensional Euler equations, and its physical initial schematic diagram is shown in the left of Figure \ref{phy2d}. The reflective conditions are applied in the upper and lower boundary, while the inflow/outflow conditions are given in the left and right boundary, respectively. The non-dimensionalized initial conditions are given as follows
\begin{equation*}
\begin{array}{rll}
(\rho,\mu,\nu,p,\gamma)&=&\left\{
\begin{array}{ll}
(1, 0, 0,1,1.4),& \text {pre-shocked air},\\
(1.3764,0.394,0,1.5698,1.4),& \text{post-shocked air},\\
(0.138, 0, 0, 5/3),& \text{helium},\\
\end{array}\right.\\
\phi&=&\sqrt{x^2+y^2}-1, \quad\quad\quad\quad\quad\quad\quad\quad\ \text{level set},
\end{array}
\end{equation*}
in which $\phi\leq0$ represents helium and $\phi>0$ represents the air. In addition, the region for $x<1.2$ is the post-shocked air state.

\begin{figure}
 \centerline{\psfig{file=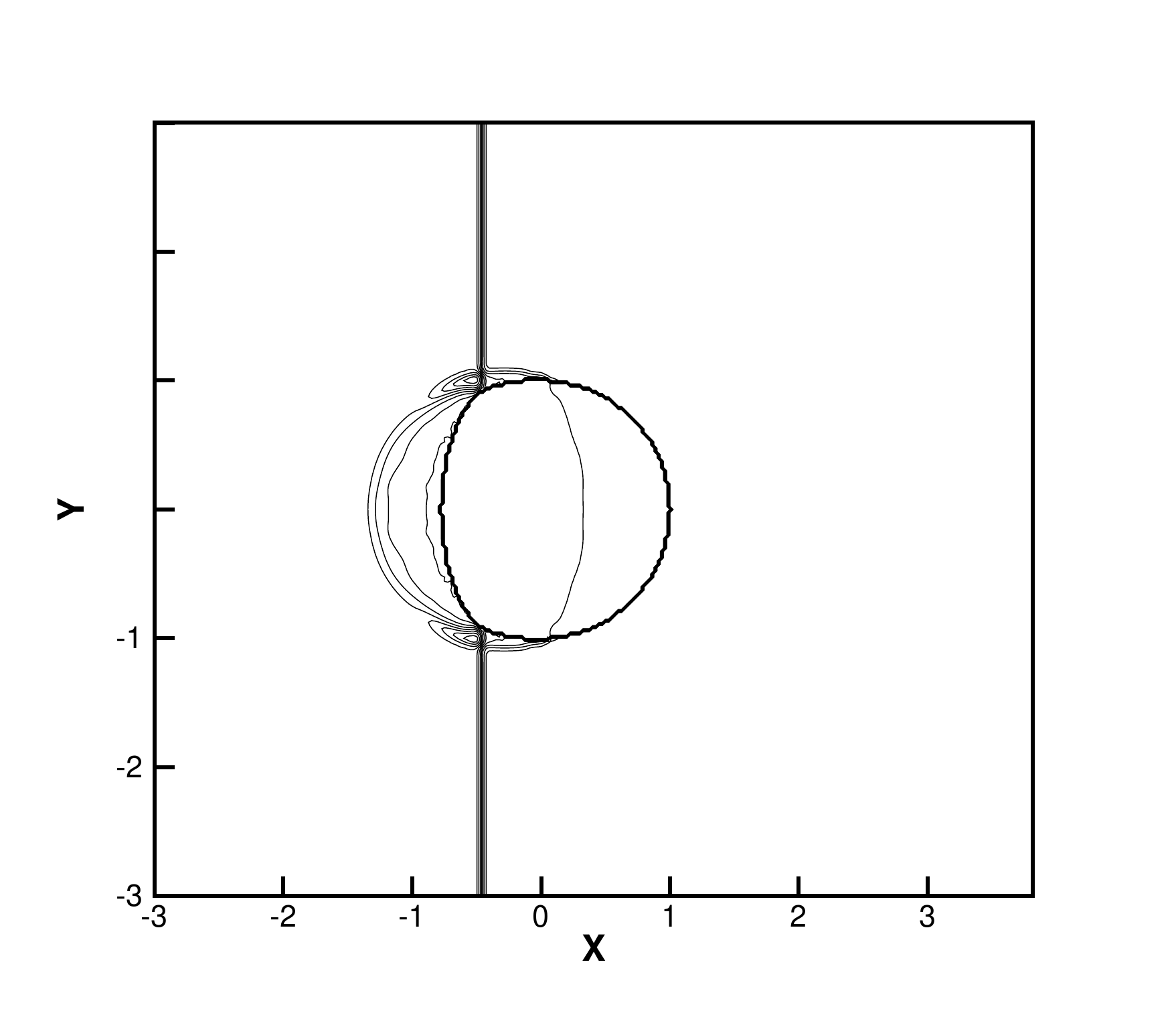,width=2.3 in}
 \psfig{file=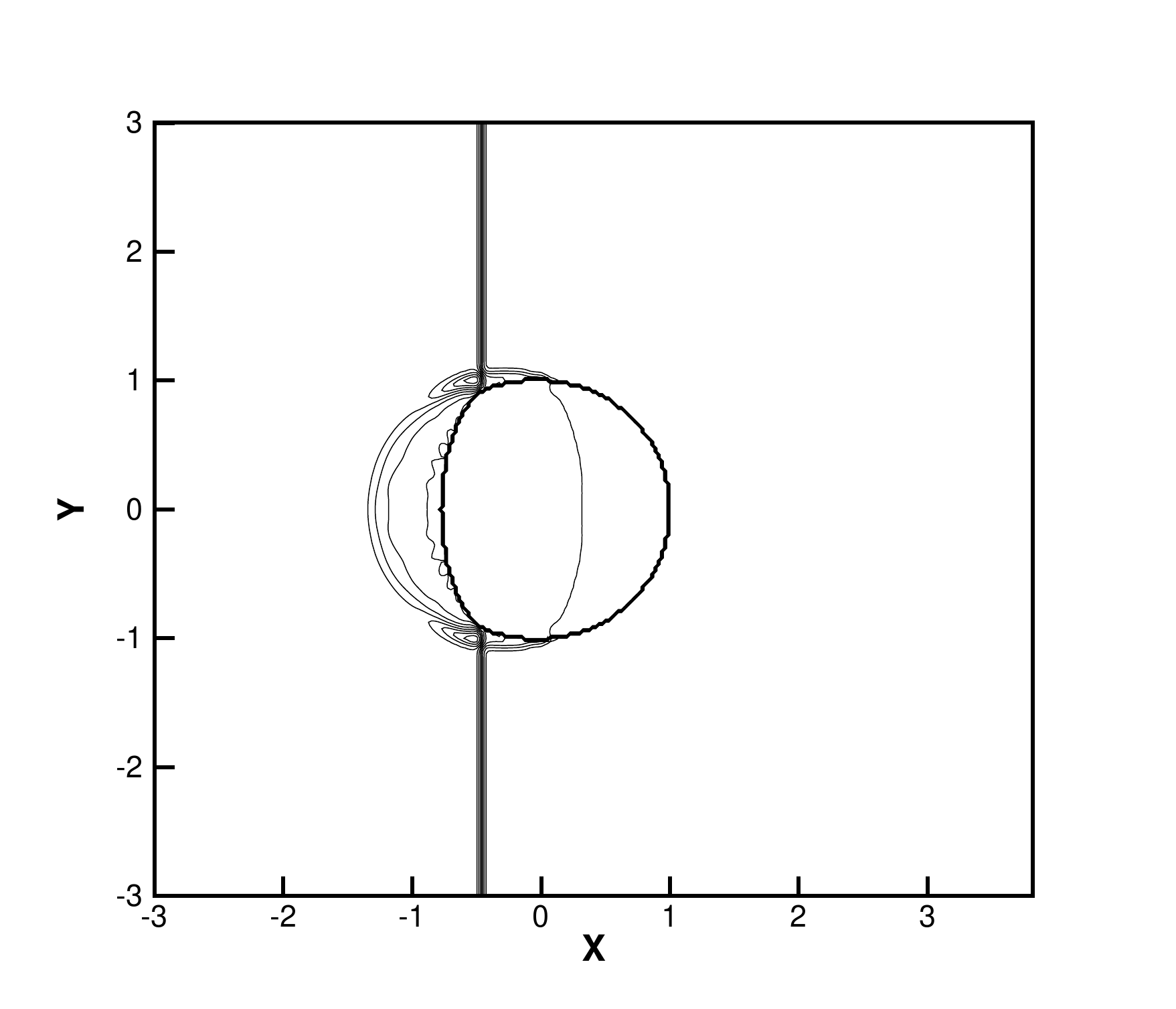,width=2.3 in}}
\centerline{\psfig{file=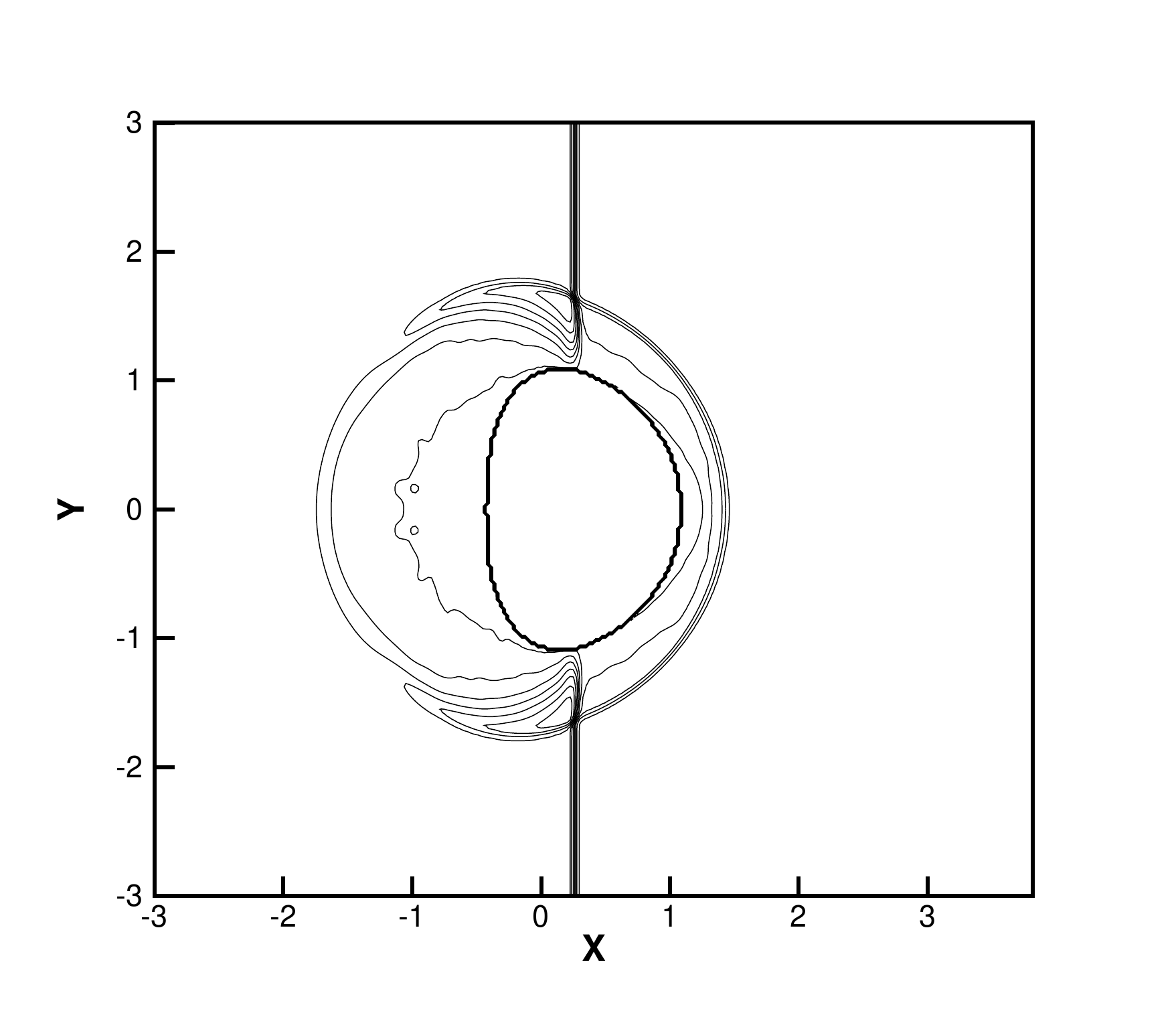,width=2.3 in}
 \psfig{file=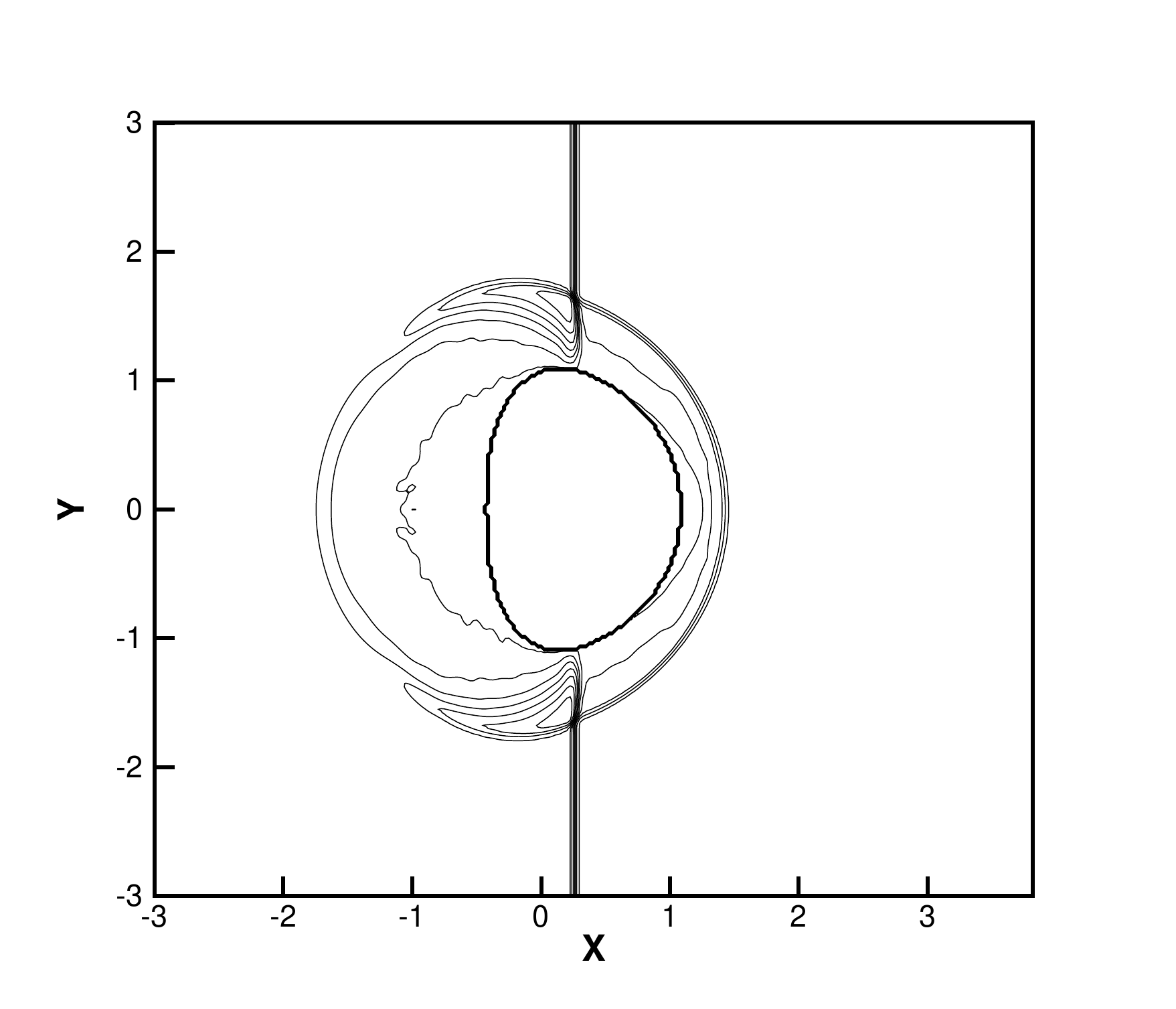,width=2.3 in}}
 \centerline{\psfig{file=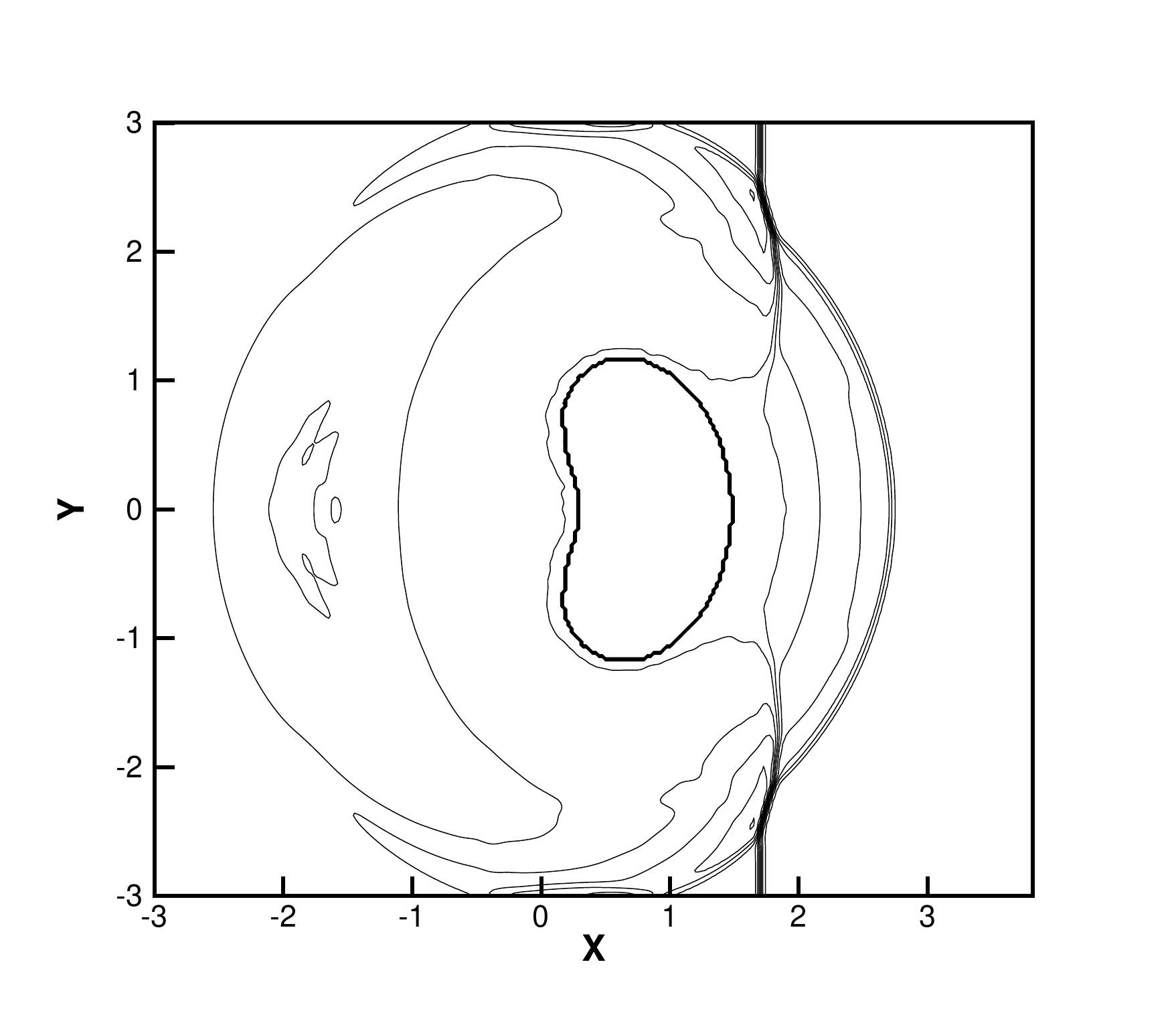,width=2.3 in}
 \psfig{file=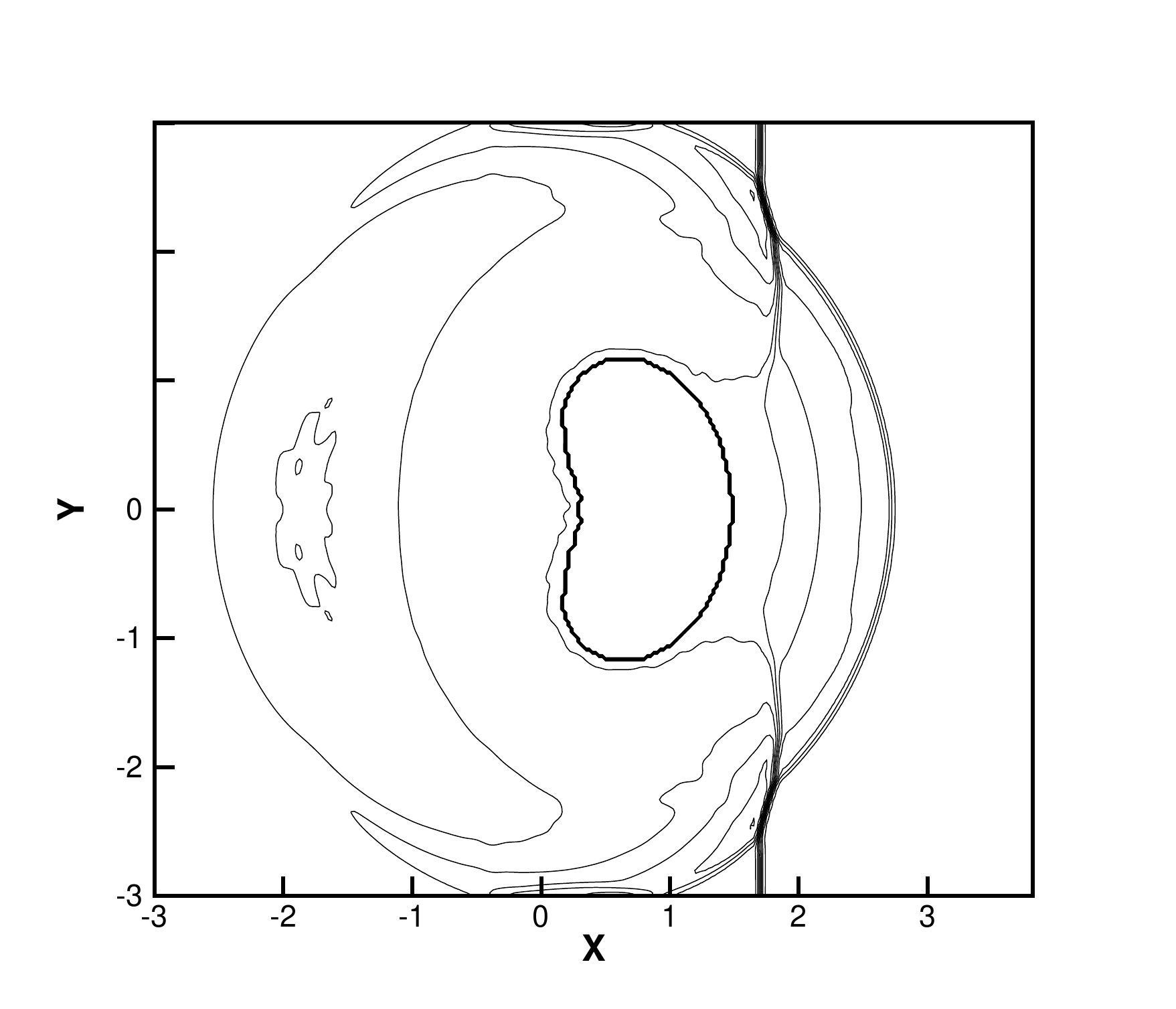,width=2.3 in}}
 \centerline{\psfig{file=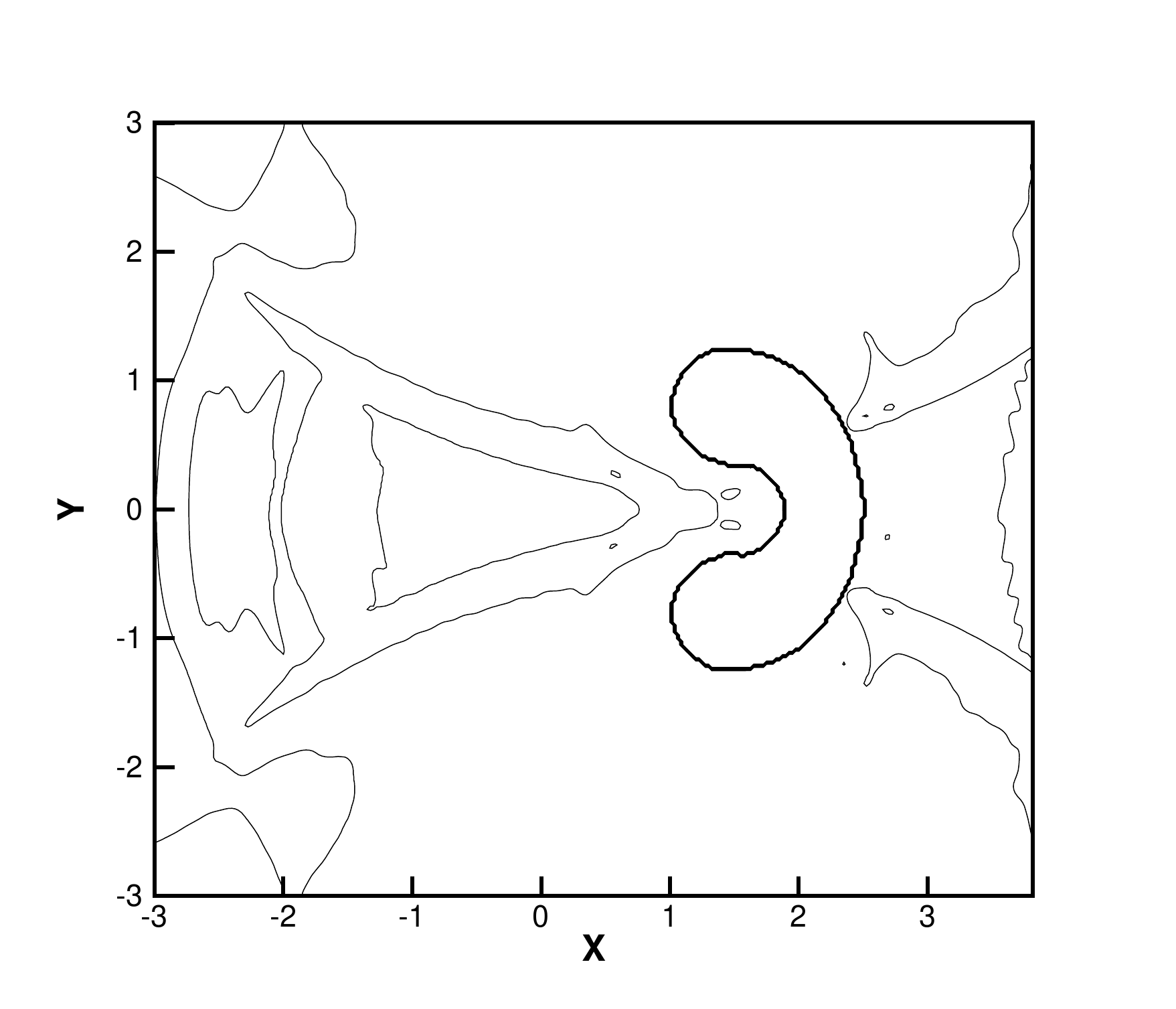,width=2.3 in}
 \psfig{file=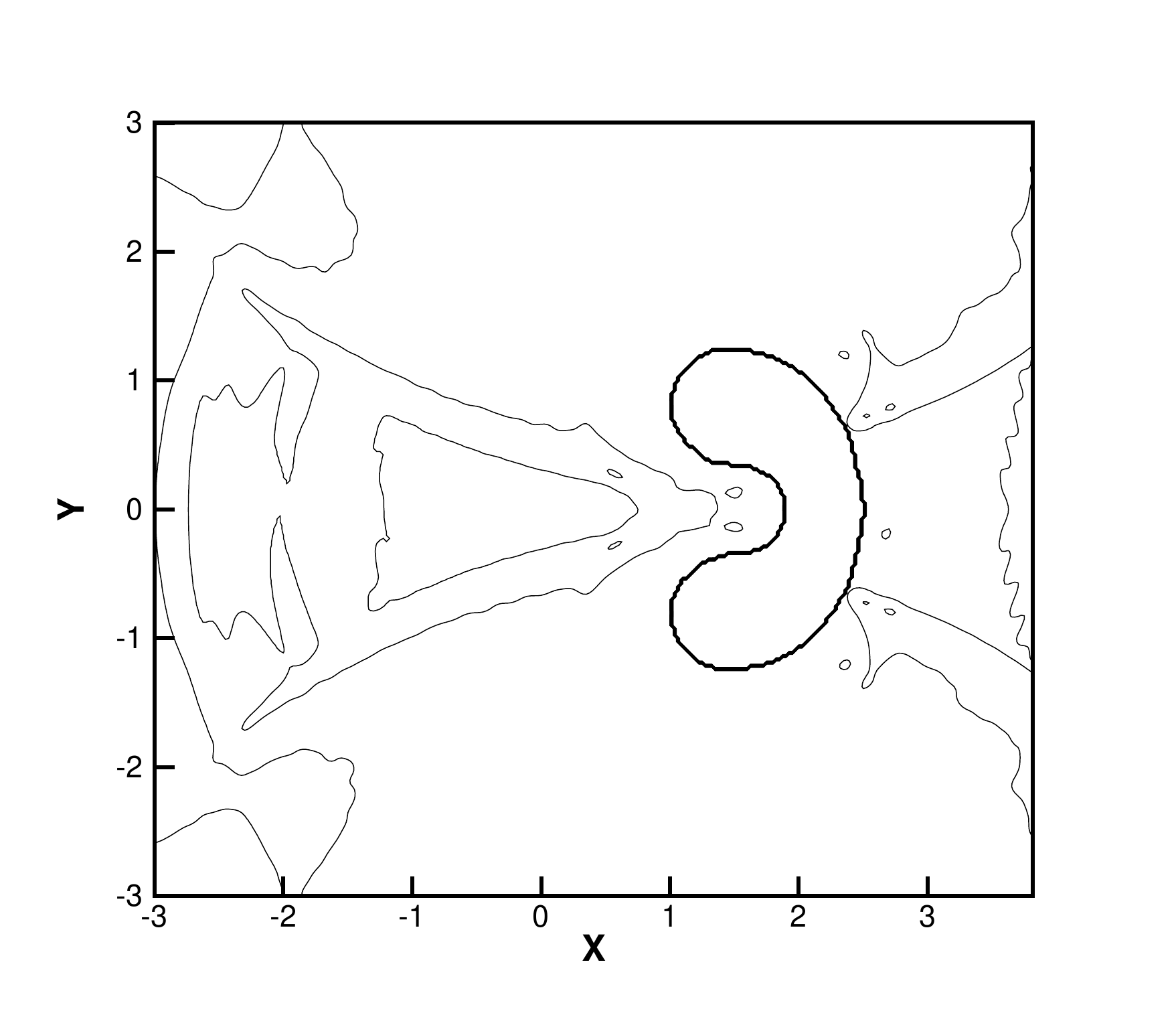,width=2.3 in}}
\caption{Example 3.7. The results computed by  Classical WENO method (left) and New/simplified hybrid WENO method (right). 30 equally spaced density contours from 0.1 to 1.6. From top to bottom are $T=0.5$, $T=1.0$, $T=2.0$ and $T=4.0$, respectively. Grid points: $280\times 240$.}
\label{Ex7}
\end{figure}

This shock impacting on a helium bubble problem had been experimentally studied in \cite{JFHBS}, and we present our computed results for density at time $0.5$, $1.0$, $2.0$ and $4.0$ with $280\times240$ uniform points in Figure \ref{Ex7}, then, we can know the numerical results are comparable to the results given in \cite{JFHBS}, and our computation for this example stops at time $t=4.0$ before the generation of the strong re-entrant jet, which is a complex physical phenomenon, and it might need to employ quite fine meshes or adaptive refinement technique seen in \cite{JC,QK}.

Then, we would give some descriptions for the numerical results. At first, a initial shock impacts on the helium bubble, then, a part of the incident shock refracts into the helium bubble, and other part of the shock reflects from the surface and backs into the air. At $t=0.5$, we can see that the initial regular shock becomes irregular having bifurcation of the shock on the bubble surface for the sound speed in helium is faster than that in air, which was also illustrated in the experimental results given in \cite{JFHBS}. At $t=1.0$, the refracted shock inside the bubble has interacted with the rear of the bubble and enters into the air, but the incident shock just went through the top of the helium bubble, then, the whole bubble starts moving to the right. At $t=2.0$, the incident shock has gone through the whole bubble, and the shape of the bubble begins to misshape. After this, a re-entrant jet begins to form. At $t=4.0$, the re-entrant jet actually has been formed, and the interface would be instable, when the re-entrant jet becomes stronger and stronger, which would affect
the rear side of the bubble and cause the bubble to collapse, and the quite fine meshes or adaptive refinement technique might be needed seen in \cite{JC,QK}. Hence, our computation stops at the non-dimensional time $t=4.0$.

\begin{figure}
  \centerline{\psfig{file=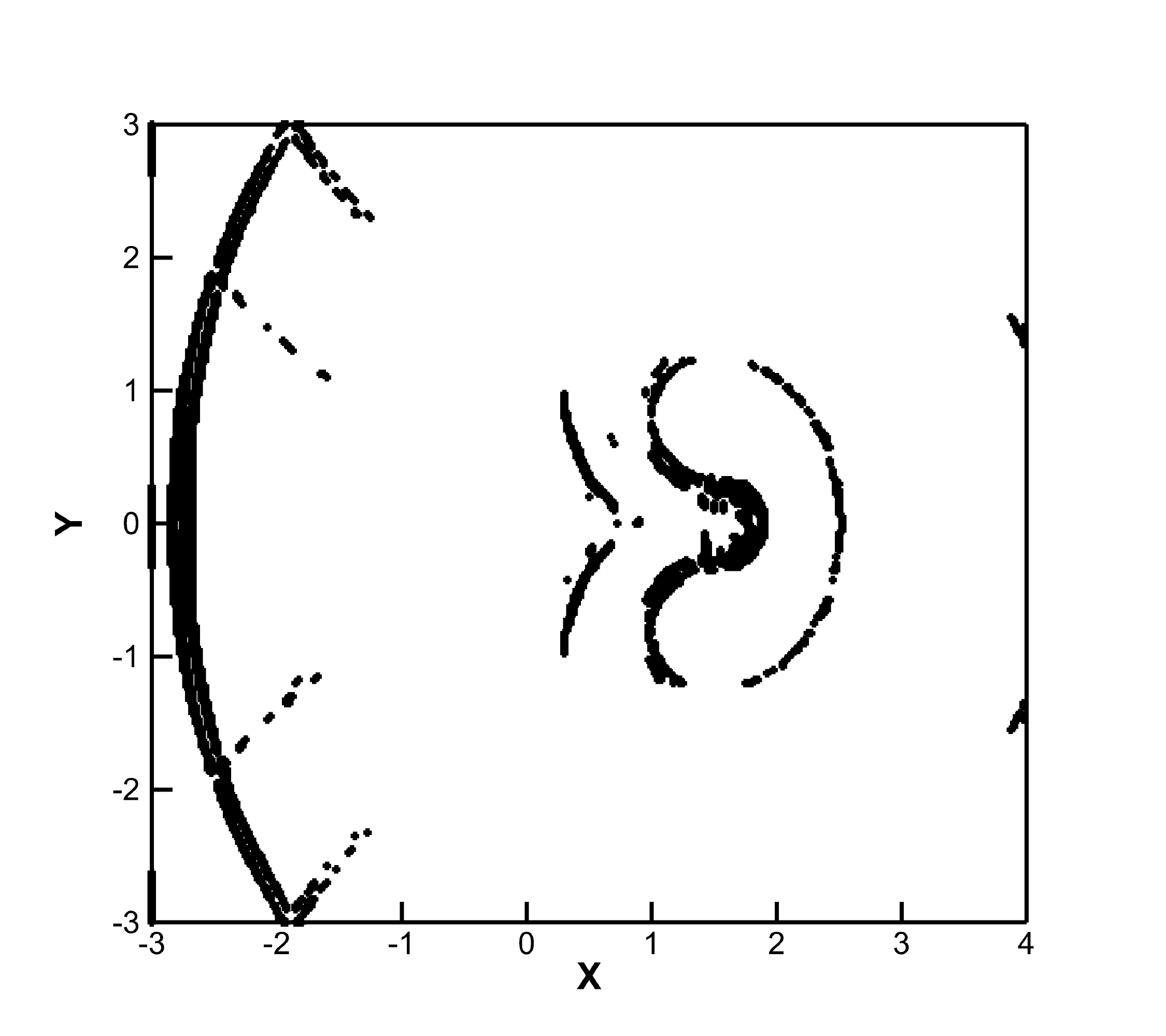,width=2 in}
 \psfig{file= 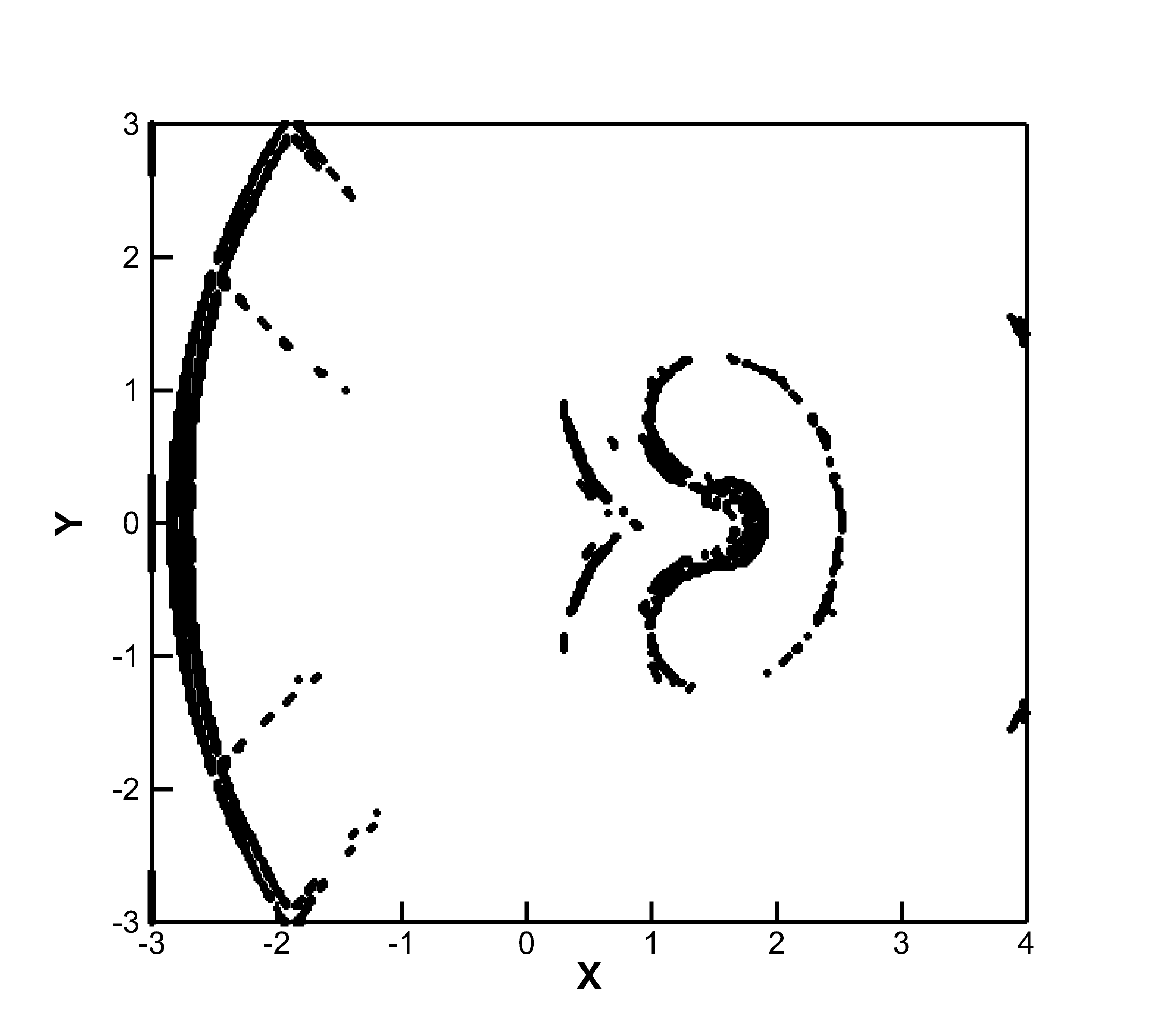,width=2 in}}
  \centerline{\psfig{file=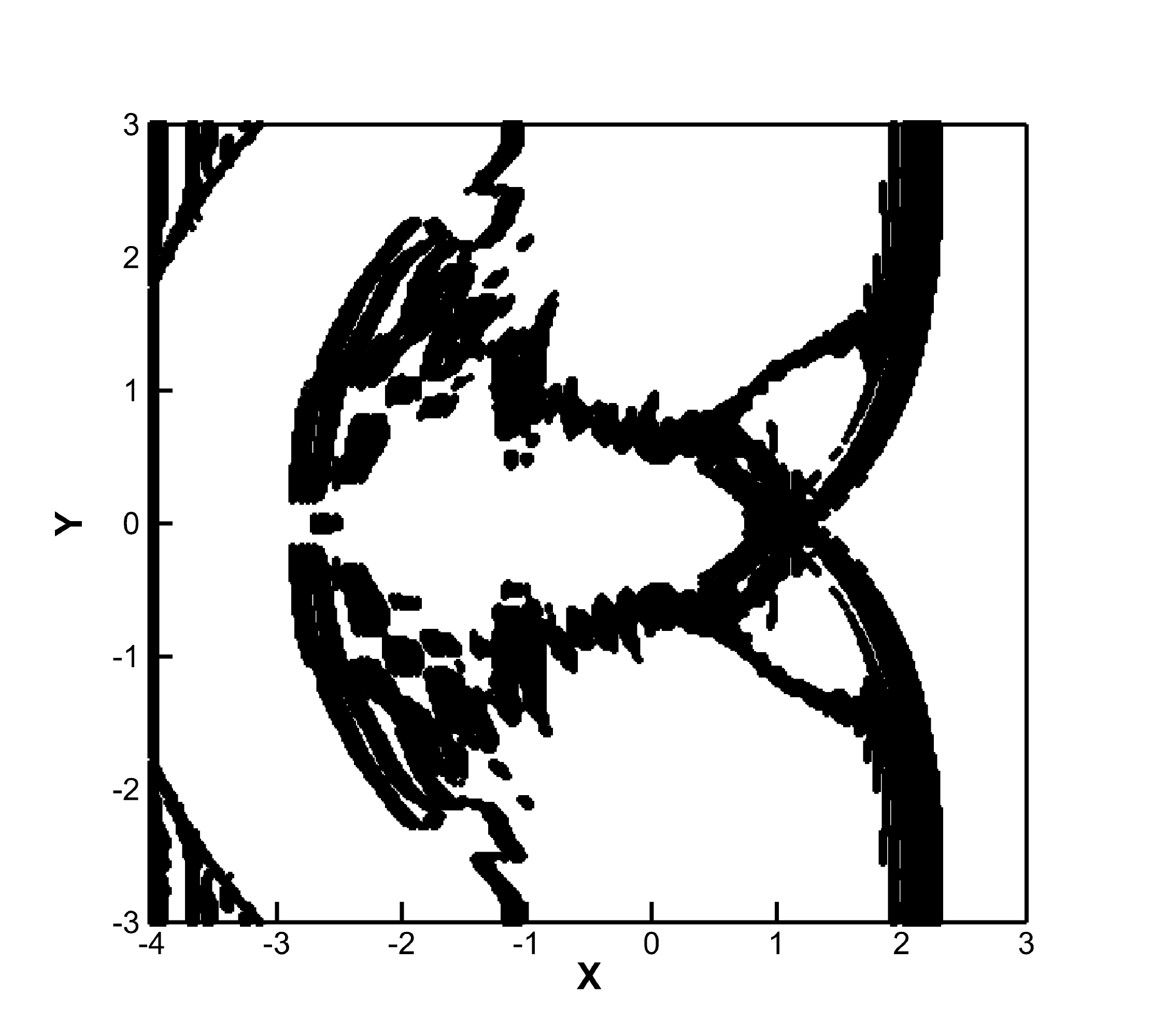,width=2 in}
 \psfig{file= 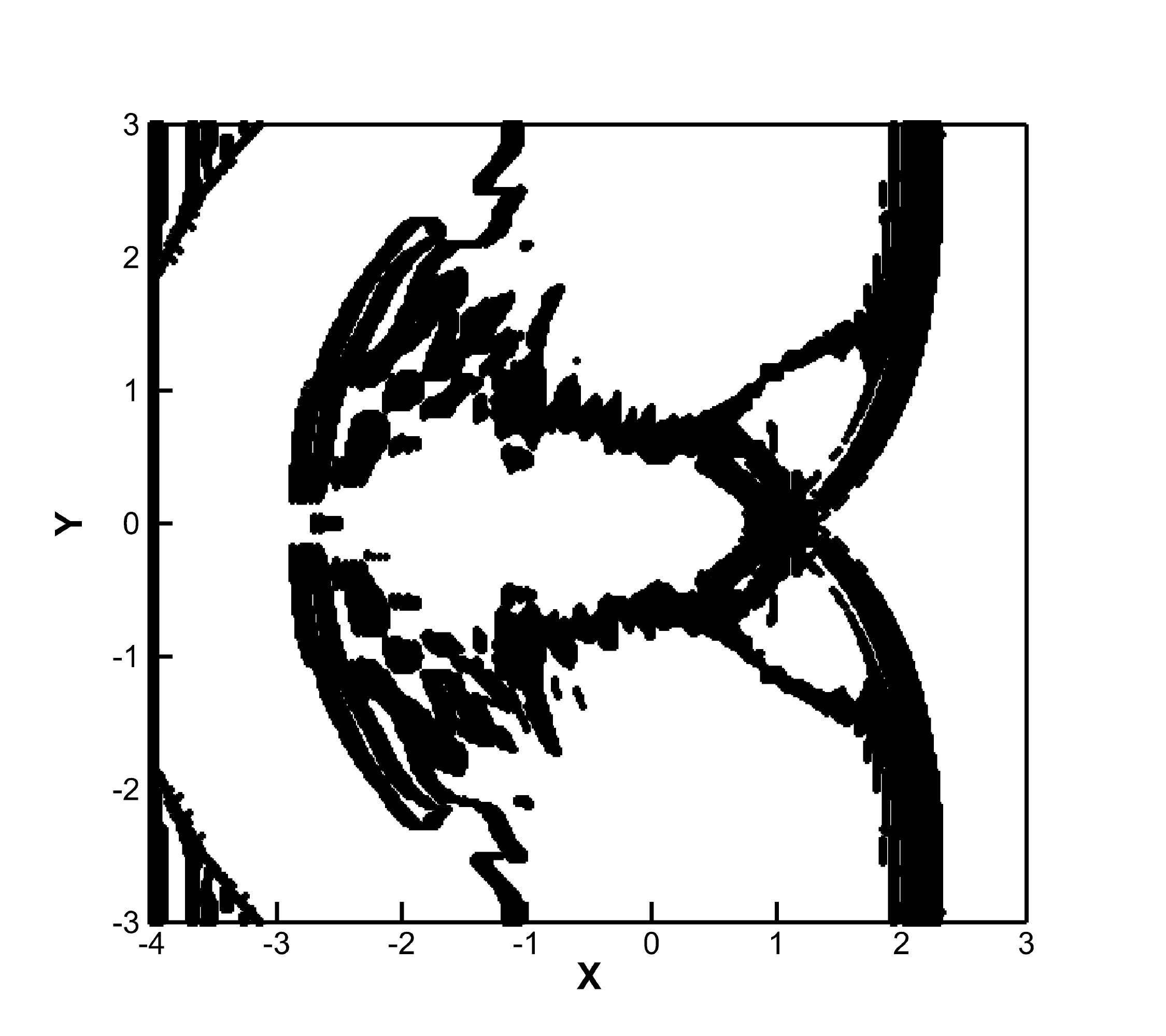,width=2 in}}
\caption{The points where the WENO procedures are performed at the final time step  for Examples 3.7 and 3.8. From left to right: the results of Old hybrid WENO method; the results of New/simplified hybrid WENO method.}
\label{Exlim78}
\end{figure}

Finally, we also find that the results computed by New/simplified hybrid WENO and Classical WENO methods are similar, but New/simplified hybrid WENO method saves almost 35.49\% computation time as we use linear approximation directly for the governing equation, the level set function and its re-initialization in the smooth regions. In addition, we find New/simplified hybrid WENO method with the new identification skill  can save 9.41\% computation time than Old hybrid WENO method by calculation, meanwhile, there are only 2.84\% and 2.75\%  points where the WENO procedures are computed in New/simplified hybrid WENO and Old hybrid methods  at the final time step, respectively, and the locations of WENO reconstruction computed by the two hybrid WENO methods at the final time step are given in the top of Figure \ref{Exlim78}, which illustrate that the new identification skill has similar ability as the old one in Old hybrid WENO method, but New/simplified hybrid WENO method with the new one has higher efficiency, and the new one is simpler as it only needs to solve the roots of a quadratic polynomial, while the old one has to calculate the zero points of a cubic polynomial.
\smallskip

\noindent{\bf Example 3.8.} The final example is a initial Mach 1.653 planar underwater shock interacting with a gas bubble in an open domain taken from \cite{QLH}, then we solve the governing equations (\ref{EQ0}) for two dimensional Euler equations with the next non-dimensionalized initial conditions:
\begin{equation*}
\begin{array}{rll}
(\rho,\mu,\nu,p,\gamma)&=&\left\{
\begin{array}{ll}
(1000, 0, 0,1,7.15),& \text {pre-shocked water},\\
(1176.3333,1.1692,0,9120,7.15),& \text{post-shocked water},\\
(1, 0, 0,1.4),& \text{gas},\\
\end{array}\right.\\
\phi&=&\sqrt{x^2+y^2}-1, \quad\quad\quad\quad\quad\quad\quad\quad\ \text{level set},
\end{array}
\end{equation*}
where $\phi\leq0$ represents gas and $\phi>0$ represents the water. In addition, the region for $x<1.2$ is the post-shocked water state. The physical initial schematic diagram is given in the right of Figure \ref{phy2d}. Reflective boundary conditions are applied in the upper and lower boundary. In flow and out flow boundary conditions are given in the left and right boundary, respectively. We present the computed density at $t=0.06$, $t=0.19$, $t=0.357$ and $=0.481$.
The detailed physical analysis can be seen in \cite{LKY2} for the earlier stage, while for the late time, one can be found in \cite{NDT}. Our numerical results are similar with the computed results by Qiu et al. \cite{QLH}, where they solved this problem by the discontinuous Galerkin finite element methods with MGFM. Again, our computation stops at time $=0.481$ before the form of the strong re-entrant jet, and the bubble doesn't collapse at this time.

\begin{figure}
 \centerline{\psfig{file=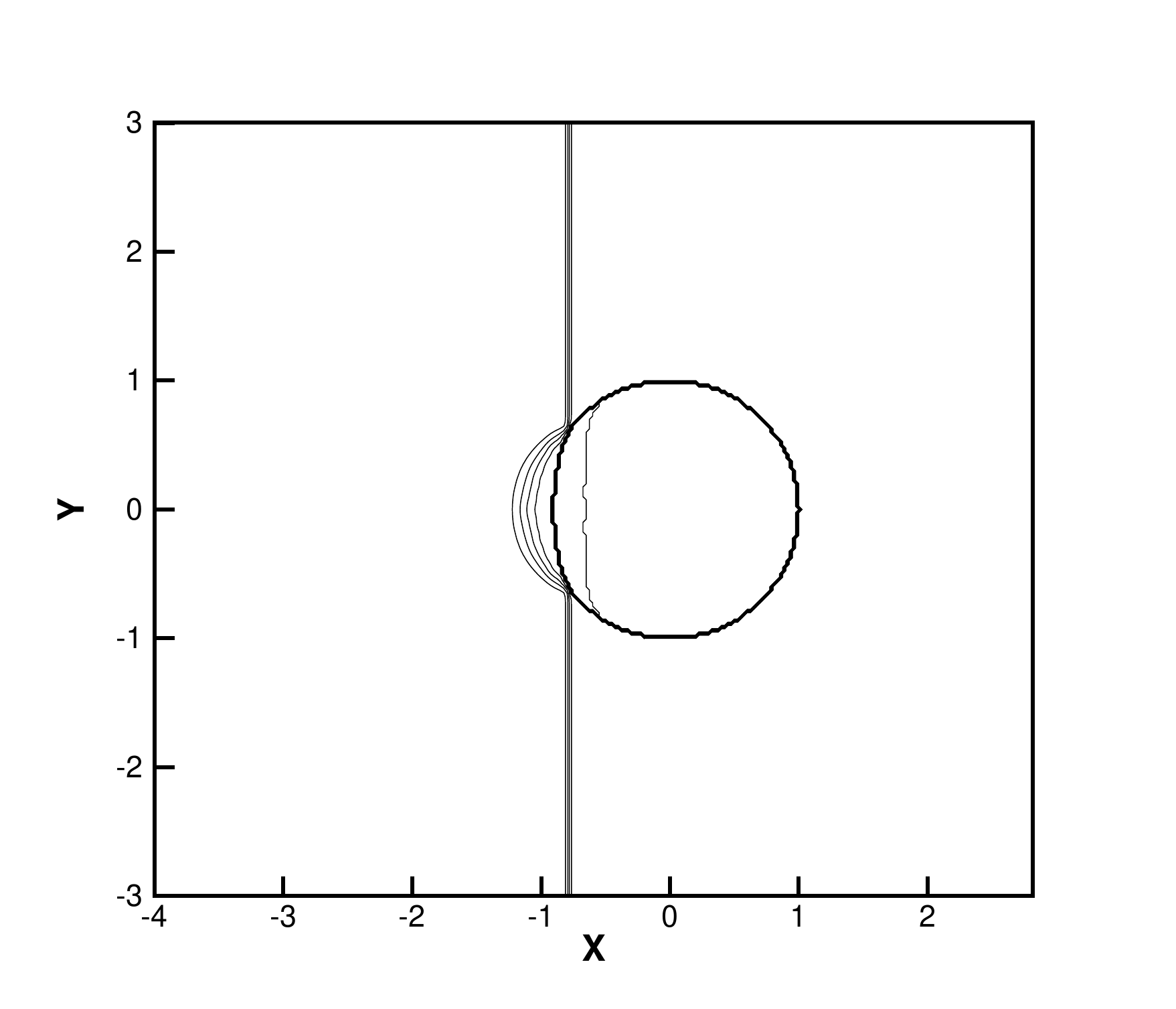,width=2.3 in}
 \psfig{file=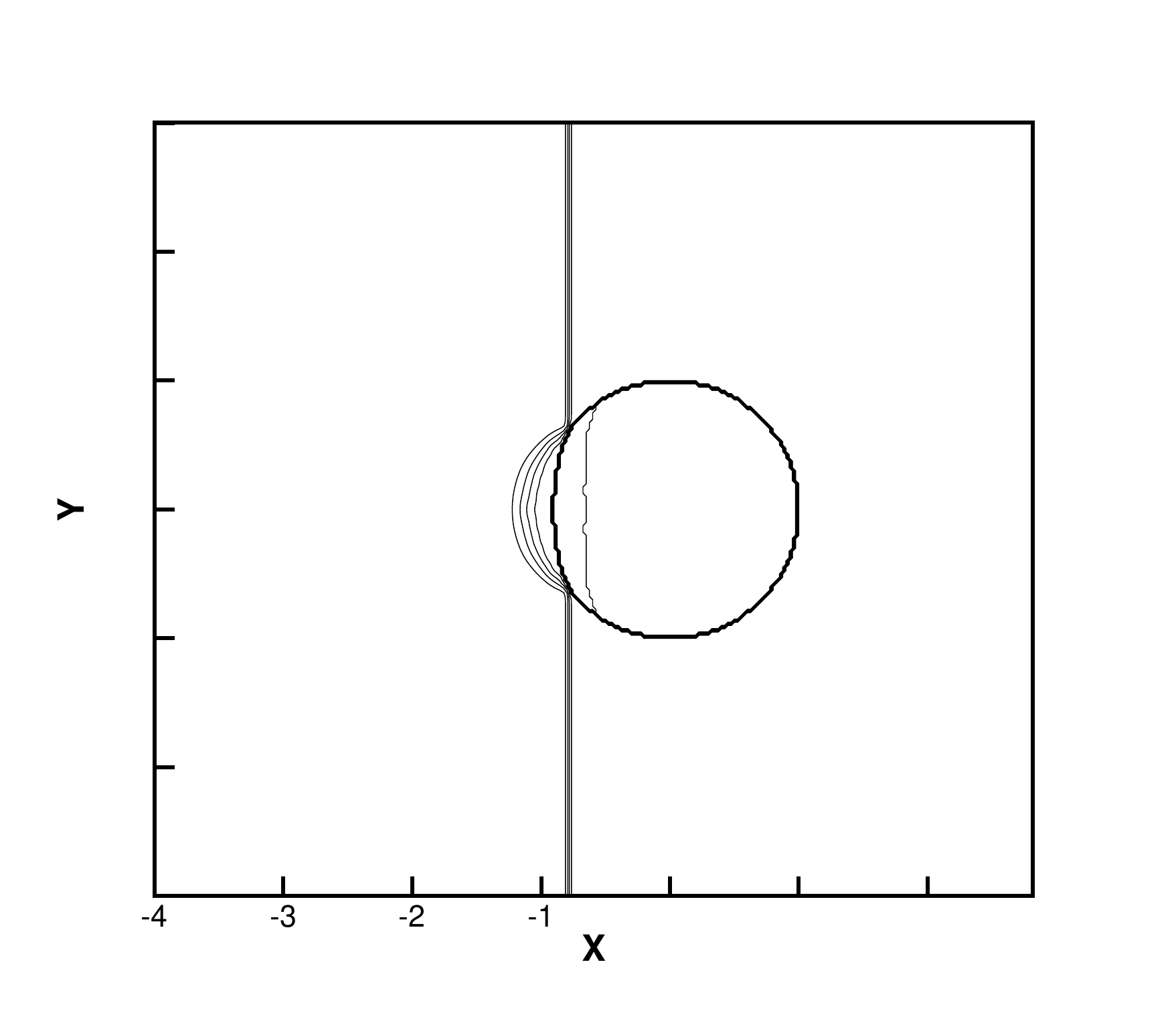,width=2.3 in}}
 \centerline{\psfig{file=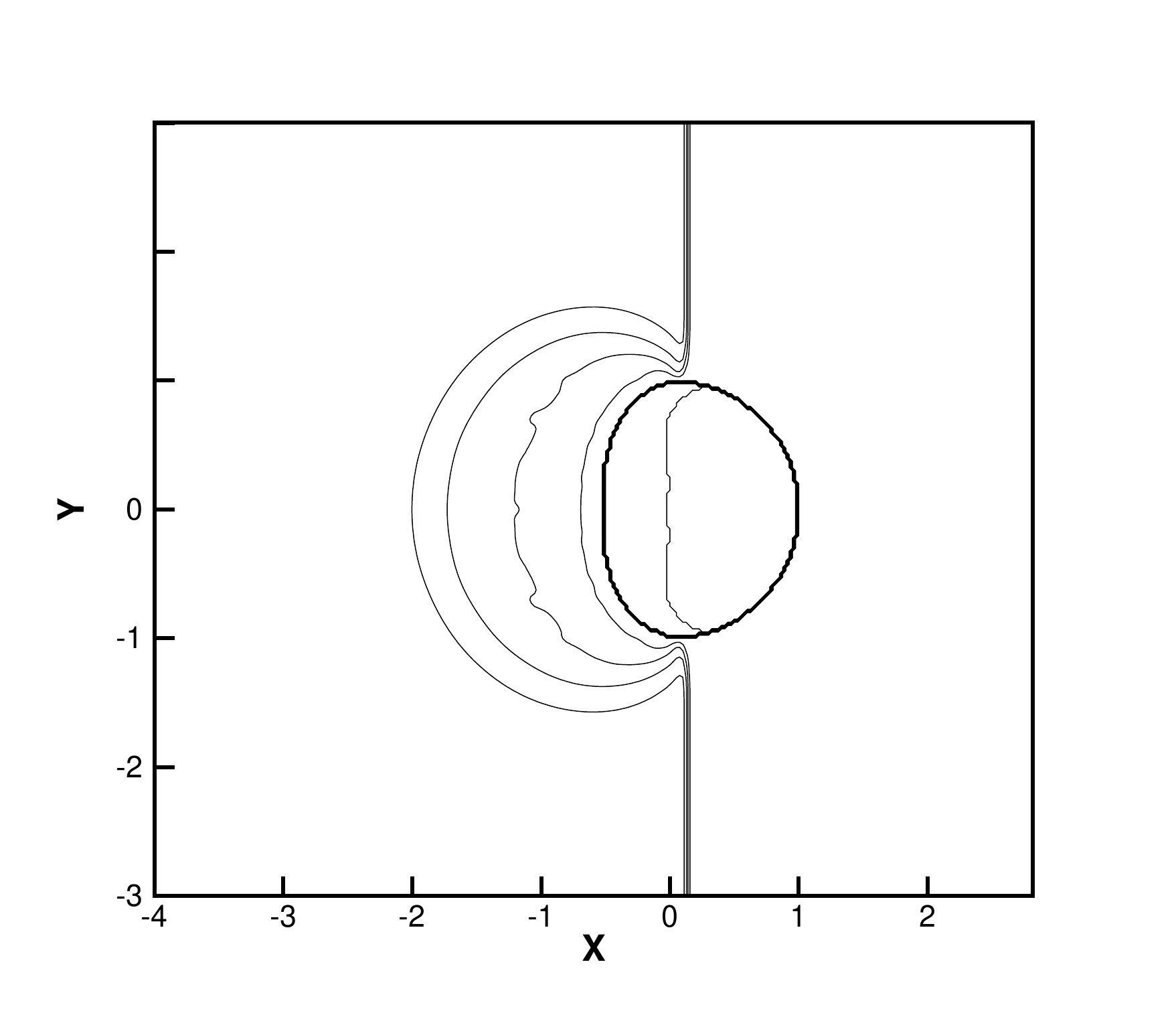,width=2.3 in}
 \psfig{file=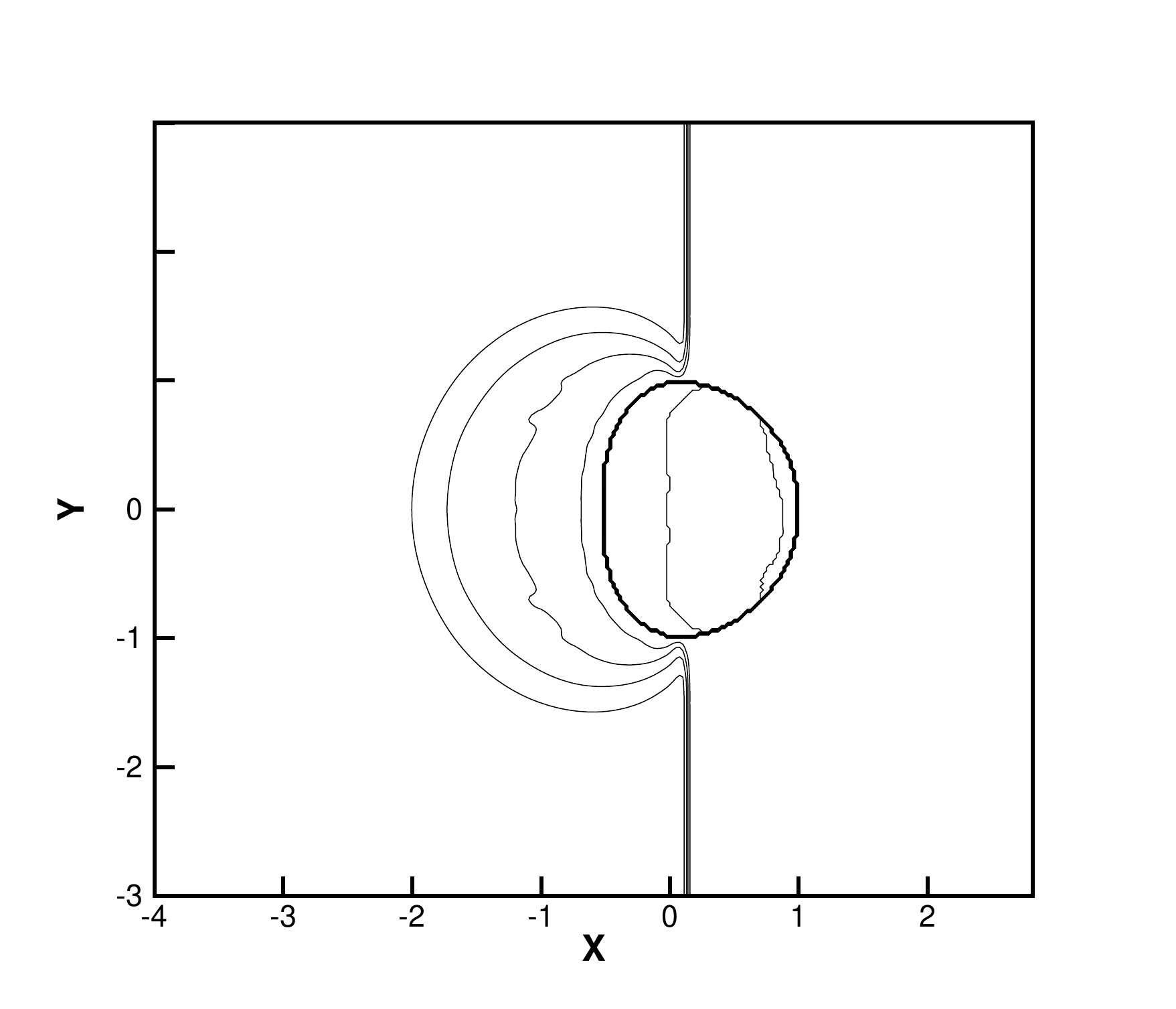,width=2.3 in}}
 \centerline{\psfig{file=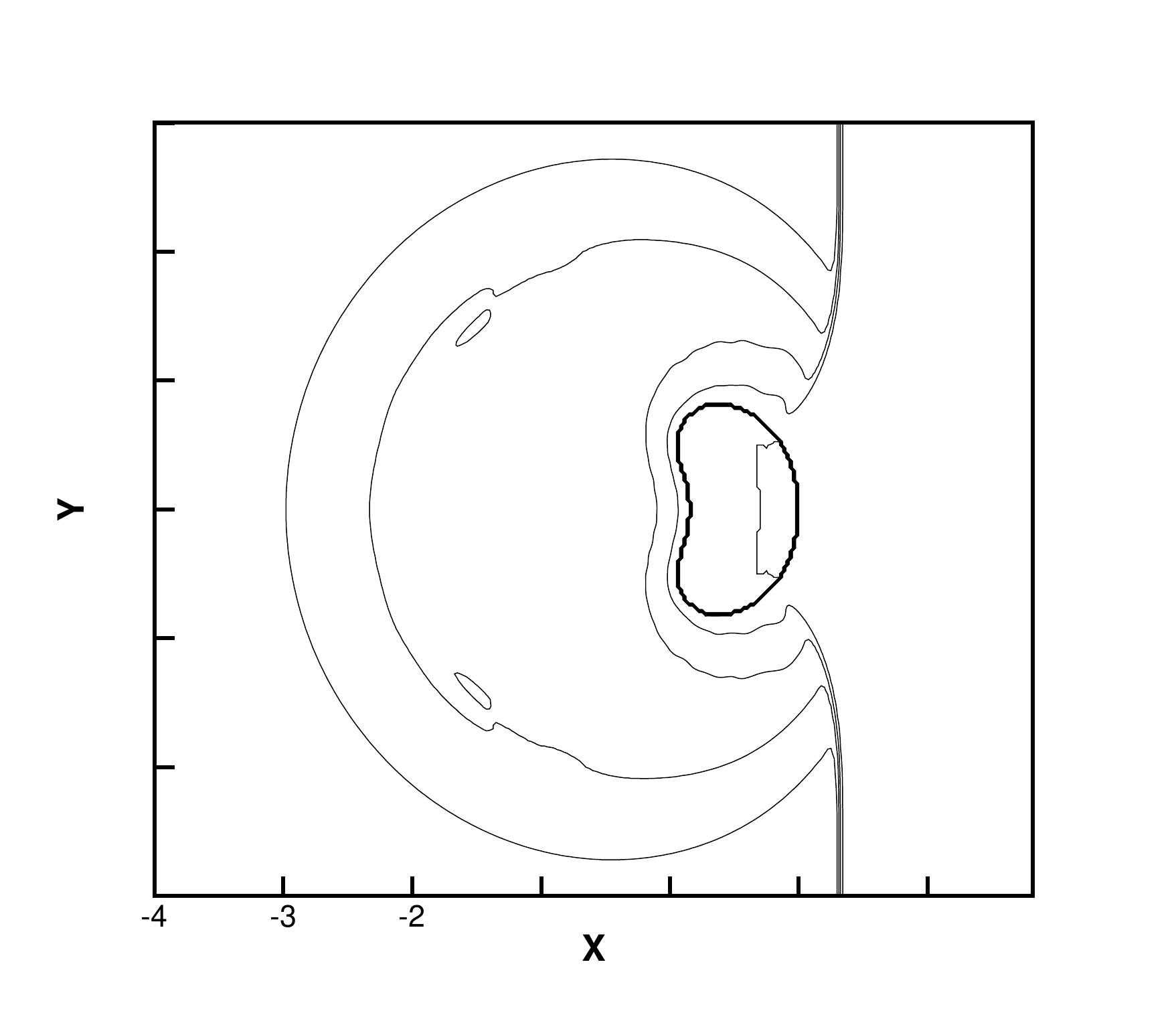,width=2.3 in}
 \psfig{file=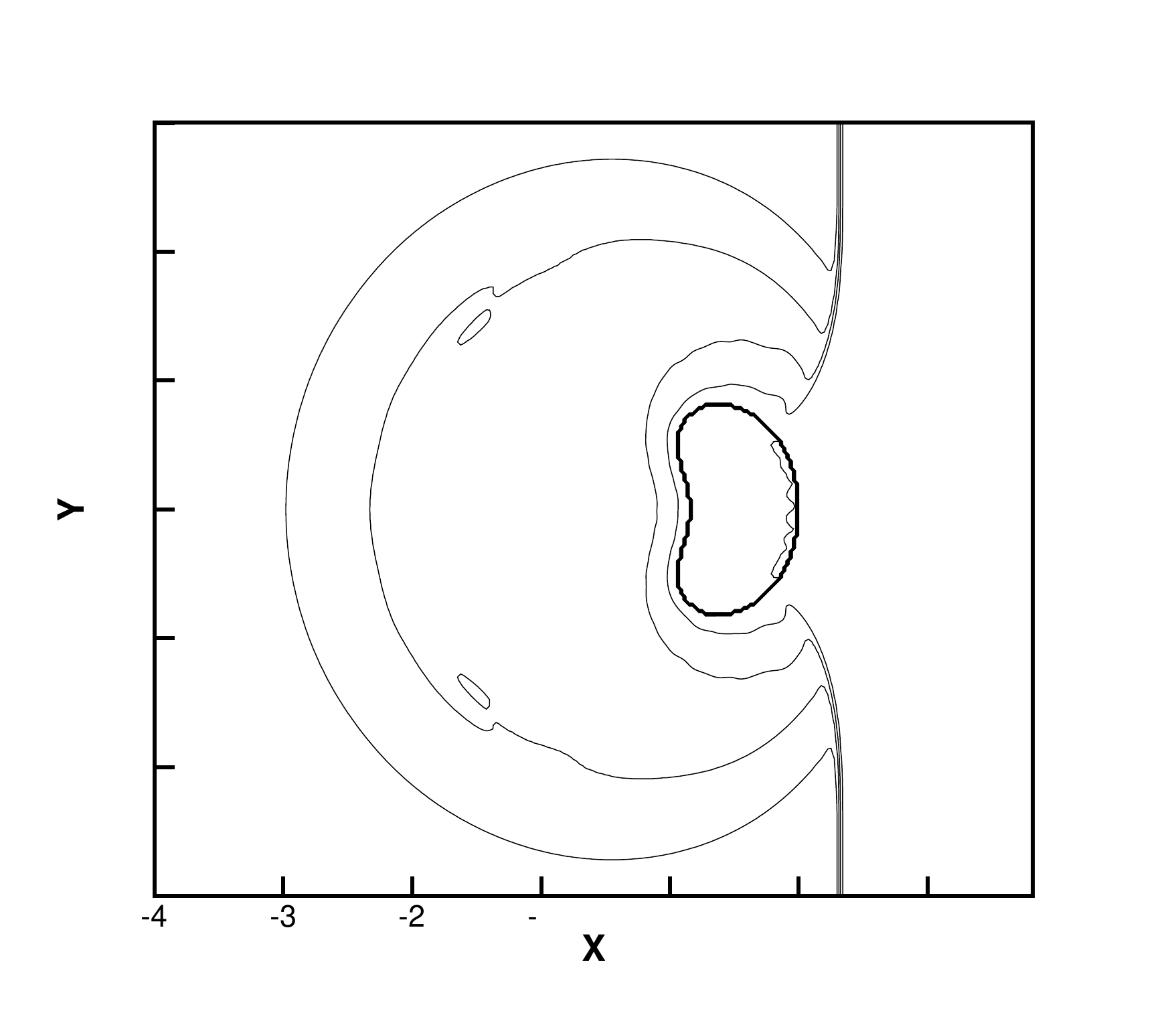,width=2.3 in}}
 \centerline{\psfig{file=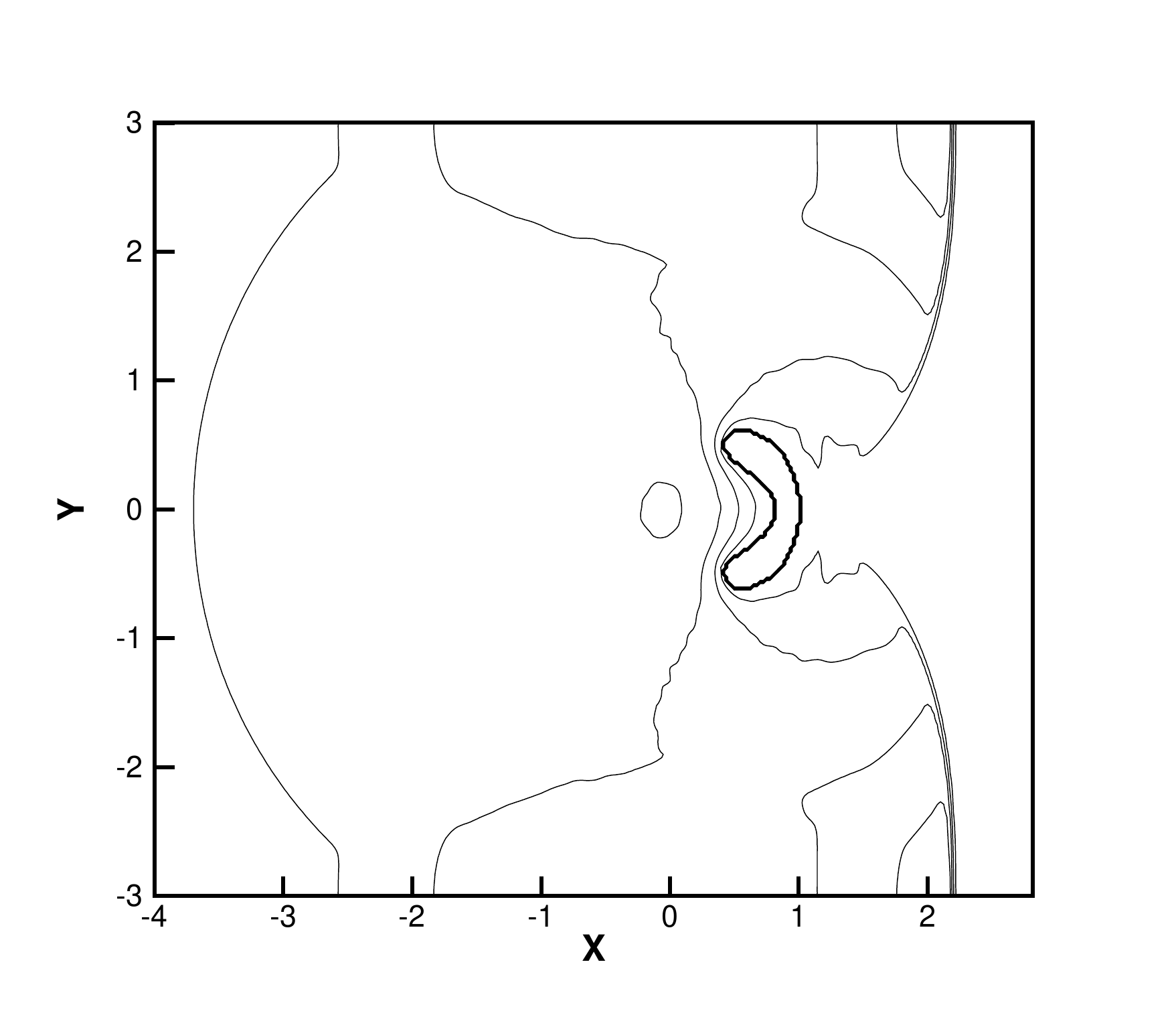,width=2.3 in}
 \psfig{file=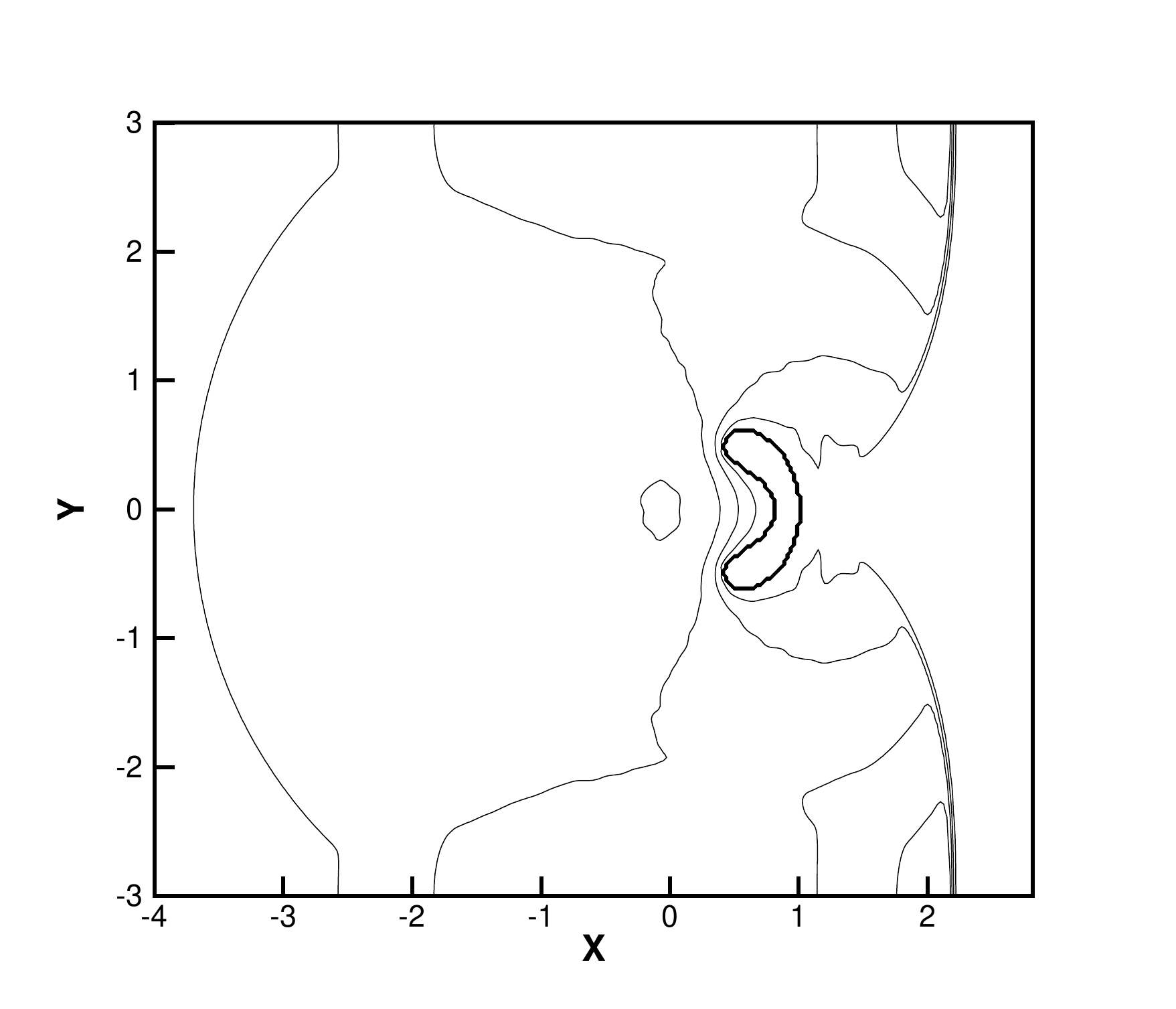,width=2.3 in}}
\caption{Example 3.8. The results computed by  Classical WENO method (left) and New/simplified hybrid WENO method (right). 30 equally spaced density contours from 1.0 to 1200. From top to bottom are $t=0.06$, $t=0.19$, $t=0.357$ and $t=0.481$, respectively. Grid points: $280\times 240$.}
\label{Ex8}
\end{figure}

From Figure \ref{Ex8}, we can see that the density computed by New/simplified hybrid WENO and Classical WENO methods  are similar, however, New/simplified hybrid WENO method  has higher efficiency than Classical WENO scheme for saving almost 27.13\% computation  time. In addition, we find New/simplified hybrid WENO method with the new identification skill  can save 12.00\% CPU time than the old one in Old hybrid WENO method by calculation, meanwhile, there are only 19.05\% and 19.21\% points where the WENO procedures are computed in the two hybrid WENO methods  at the final time step, respectively, and the locations of WENO reconstruction at the final time step are shown in the bottom of Figure \ref{Exlim78}, which illustrate that the new identification skill in New/simplified hybrid WENO method can identify the regions of the extreme points as the old one in Old hybrid WENO, but New/simplified hybrid WENO method with the new one has higher efficiency, and the new one has simpler implementation procedure.
\smallskip

\section{Concluding remarks}
\label{sec4}
\setcounter{equation}{0}
\setcounter{figure}{0}
\setcounter{table}{0}

In this paper, we combine the new simplified hybrid WENO method with the modified ghost fluid method \cite{LKY1} to simulate the compressible two-medium flow problems, which adapts between the linear upwind approximation and WENO reconstruction  automatically in terms of the regions of the extreme points for the big reconstruction polynomial, and we have an improvement about the identification technique for the regions of the extreme points of the big reconstruction polynomial. This new switch principle is not only simple for we doesn't need to adjust the parameters, but also it is effective as we just need to know the range of the extreme point for the big reconstruction polynomial, rather than solving the exact location of the extreme point as the old one in hybrid WENO schemes \cite{mzq,ZZCQ}. Comparing with the classical WENO scheme \cite{js}, the new simplified hybrid WENO scheme is more efficient with  less numerical errors in the smooth region and less computation costs, meanwhile, the new simplified hybrid WENO scheme with MGFM is robust and non-oscillatory to simulate these two-medium flow problems. In general, these numerical results all show the good performances of the new simplified hybrid WENO scheme with the modified ghost fluid method.

\section{Appendix A: The  finite difference hybrid WENO method for Hamilton-Jacobi equations}
\label{appendixA}

The next  finite difference hybrid WENO  method for Hamilton-Jacobi equations is mainly developed from the fifth order WENO scheme introduced by Jiang and Peng \cite{jp} to solve the Hamilton-Jacobi equations (\ref{Levelset1}), (\ref{Levelset2}) and (\ref{Leve_reinitial}) in Section \ref{sec22}.

We first consider one dimensional Hamilton-Jacobi equation
\begin{equation}
\label{EQHJ} \left\{
\begin{array}
{ll}
\phi_t+ H(x,t,\phi,\phi_x)=0, \\
\phi(x,0)=\phi_0(x). \\
\end{array}
\right.
\end{equation}
The computing domain is divided by uniform grid points $\{x_i\}$, and the semi-discrete form of (\ref{EQHJ}) is
\begin{equation}
\label{odeHJ1} \frac{d\phi_i(t)}{dt}= -\hat H(x_i,t,\phi_i,\phi_{x_i}^+,\phi_{x_i}^-),
\end{equation}
where $\phi_i(t)$ is represented as $\phi(x_i,t)$, and $\phi_{x_i}^\pm$ are linear or WENO approximations for $\frac{\partial\phi(x_i)}{\partial x}$. $\hat H $ is a numerical flux to approximate $H$, and we use the Lax-Friedrichs (LF) flux here as:
\begin{equation*}
\label{llfsplit11} \hat H(x,t,\phi,u^+,u^-) =H\left(x,t,\phi,\frac{u^++u^-}{2}\right) -\alpha(u^+,u^-)\frac{u^+-u^-}{2},
\end{equation*}
where $\alpha$ is $\max_{u}|H_1(u)|$, where $H_1$ is represented as the partial derivative of $H$ with respect to $\phi_x$.

Next, we only introduce the reconstruction procedures for $\phi_{x_i}^-$, and the reconstruction for $\phi_{x_i}^+$ is mirror symmetric with respect to $x_{i}$ of that for  $\phi_{x_i}^-$. Firstly, we can easily obtain the fourth degree polynomial $p_0(x)$ to approximate $\phi_x$ in terms of the requirements:
\begin{eqnarray*}
\frac 1 {\Delta x} \int_{x_{i-1+l}}^{x_{i+l}} p_0(x)dx= \frac 1 {\Delta x} \int_{x_{i-1+k}}^{x_{i+k}}  \phi_x dx= \frac{\phi_{x_{i+k}}-\phi_{x_{i-1+k}}}{\Delta x},\
k=-2,...,2.
\end{eqnarray*}

To increase the efficiency, if all extreme points of $p_0(x)$ are located outside of the big spatial stencil, $\phi_{x_i}^-$ is taken as $p_0(x_i)$ directly, otherwise we would use the next classical WENO procedures \cite{jp,js}, and the method to identify the regions of the extreme points for $p_0(x)$ has been detailedly introduced in Section \ref{sec22}.

Now, we would give a brief review of the WENO reconstruction for $\phi_{x_i}^-$. Similarly, we obtain three quadratic polynomials $p_l(x)$ to approximate $\phi_x$, satisfying
\begin{eqnarray*}
\frac 1 {\Delta x} \int_{x_{i-3+k+l}}^{x_{i-2+k+l}} p_l(x)dx=\frac 1 {\Delta x} \int_{x_{i-3+k+l}}^{x_{i-2+k+l}} \phi_xdx= \frac{\phi_{x_{i-2+k+l}}-\phi_{x_{i-3+k+l}}}{\Delta x},\ k=-1,0,1,\ l=1,2,3.
\end{eqnarray*}
For saving space, the explicit values of $p_l(x_i)$, the linear weights $\gamma_l$, the smoothness indicators $\beta_l$, and the  nonlinear weights $\omega_l$ are not present here, and these expressions can be seen in \cite{jp,js}. Finally, the WENO reconstruction of $\phi^-_{x_i}$ is  approximated by
\begin{equation*}
\phi^-_{x_i}=\sum_{l=1}^3\omega_lp_l(x_{i}).
\end{equation*}

For the two dimensional Hamilton-Jacobi equation
  \begin{equation}
\label{EQHJ2} \left\{
\begin{array}
{ll}
\phi_t+ H(x,y,t,\phi,\phi_x,\phi_y)=0, \\
\phi(x,y,0)=\phi_0(x,y). \\
\end{array}
\right.
\end{equation}
The computing domain is divided by uniform grid points $\{(x_i,y_j)\}$, and the semi-discrete form of (\ref{EQHJ2}) is
\begin{equation}
\label{odeHJ2} \frac{d\phi_{i,j}(t)}{dt}= -\hat H(x_i,y_j,t,\phi_{i,j},\phi_{x,i,j}^+,\phi_{x,i,j}^-,\phi_{y,i,j}^+,\phi_{y,i,j}^-),
\end{equation}
where $\phi_{i,j}(t)$ is represented as $\phi(x_i,y_j,t)$.
$\phi_{x,i,j}^\pm$ and $\phi_{y,i,j}^\pm$ are linear or WENO approximations for $\frac{\partial\phi(x_i,y_j)}{\partial x}$ and $\frac{\partial\phi(x_i,y_j)}{\partial y}$, respectively. $\hat H $ is a numerical flux to approximate $H$,  and we use the Lax-Friedrichs (LF) flux here as:
\begin{equation*}
\label{llfsplit2} \hat H(x_i,y_j,t,\phi_{i,j},u^+,u^-,v^+,v^-) =H(x_i,y_j,t,\phi_{i,j},\frac{u^++u^-}{2},\frac{v^++v^-}{2}) -\alpha\frac{u^+-u^-}{2}-\beta\frac{v^+-v^-}{2},
\end{equation*}
where $\alpha$ is $ \max_{u}|H_1(u,v)|$ and $\beta$ is  $\max_{v}|H_2(u,v)|$. $H_1$ and $H_2$ are represented as the partial derivative of $H$ with respect to $\phi_x$ and $\phi_y$, respectively. Finally, $\phi_{x,i,j}^\pm$ and $\phi_{y,i,j}^\pm$ are reconstructed by dimension by dimension as one dimension case.

For the time discretization, the semi-discrete schemes (\ref{odeHJ1}) and (\ref{odeHJ2}) are discretized by the third order Runge-Kutta method \cite{so1} in (\ref{RK}) of Section \ref{sec22}.


\end{document}